\documentclass[10pt,letterpaper]{article}
\usepackage[top=0.85in,left=2.75in,footskip=0.75in,marginparwidth=2in]{geometry}
\usepackage{amsthm,amsmath,amssymb}
\usepackage{booktabs}
\usepackage{float}
\usepackage{multirow}
\usepackage{ragged2e}
\usepackage{tabularx}
\usepackage{array}
\usepackage{hyperref}
\usepackage[utf8]{inputenc}
\usepackage{natbib}

\usepackage{tikz}
\usepackage{graphicx}
\usepackage{subcaption}
\usetikzlibrary{matrix,positioning,calc,arrows.meta}

\usepackage{nameref,hyperref}

\usepackage[right]{lineno}
\usepackage{xr-hyper}
\usepackage[expansion=false]{microtype}
\DisableLigatures[f]{encoding = *, family = * }

\raggedright
\usepackage{changepage}

\usepackage[aboveskip=1pt,labelfont=bf,labelsep=period,singlelinecheck=off]{caption}

\makeatletter
\renewcommand{\@biblabel}[1]{\quad#1.}
\makeatother

\usepackage{lastpage,fancyhdr,graphicx}
\usepackage{epstopdf}
\fancyheadoffset[L]{2.25in}
\fancyfootoffset[L]{2.25in}

\usepackage{color}

\definecolor{Gray}{gray}{.25}

\usepackage{graphicx}
\usepackage{subcaption}

\usepackage{sidecap}

\usepackage{wrapfig}
\usepackage[pscoord]{eso-pic}
\usepackage[fulladjust]{marginnote}
\reversemarginpar

\def\r{{\boldsymbol r}}
\def\x{{\boldsymbol x}}
\def\by{{\boldsymbol y}}
\def\u{{\boldsymbol u}}
\def\w{{\boldsymbol w}}

\def\K{{\boldsymbol K}}
\def\Y{{\boldsymbol Y}}

\def\bthe{{\boldsymbol \theta}}
\def\bdel{{\boldsymbol \delta}}
\def\balp{{\boldsymbol \alpha}}
\def\bbet{{\boldsymbol \beta}}
\def\bF{{\boldsymbol F}}

\def\be{\begin{eqnarray}}
\def\ee{\end{eqnarray}}
\def\non{\nonumber}
\renewcommand{\hat}{\widehat}

\begin{document}
\vspace*{0.35in}

\begin{flushleft}
{\Large
\textbf\newline{Differential Dynamic Causal Nets: Model Construction, Identification and Group Comparisons}
}
\newline
\\
Kang You\textsuperscript{1},
Gary Green\textsuperscript{2},
Jian Zhang\textsuperscript{1,*}
\\
\bigskip
\bf{1} School of Engineering, Mathematics and Physics, University of Kent, Canterbury, Kent CT2 7NF, UK.
\\
\bf{2} York Neuroimaging Centre, Department of Psychology, University of York, Heslington, York YO10 5DD, UK.
\\
\bigskip
* Corresponding author: Jian Zhang (J.Zhang-79@kent.ac.uk)

\end{flushleft}

\section*{Abstract}
Pathophysiological modelling of brain systems from microscale to macroscale remains difficult in group comparisons partly because of the infeasibility of modelling the interactions of thousands of neurons at the scales involved. Here, to address the challenge, we present a novel approach to construct differential causal networks directly from grouped electroencephalogram (EEG) data.   In the proposed modular network, each node represents a locally averaged regional neural system via a canonical Jansen-Rit model  while directed edges stand for node-wise input-output connections. The network is hierarchically structured in the sense that node and edge parameters are random, varying in subjects, but follow a mixed-effects model.  We show that unlike the classical Jansen-Rit models, the proposed node models are locally structurally identifiable.  An evolutionary optimisation algorithm for parameter inference in the proposed method is developed using a loss function derived from Chen-Fliess expansions of stochastic differential equations.  The method is demonstrated by fitting the proposed modular network to grouped epileptic EEG data in both sensor- and source- levels. We identify significant changes in the parameters that characterise dynamic networks within and between brain regions. We reveal network disruptions, due to imbalance of excitatory-inhibitory interneurons and other altered epileptic brain connectivities, before and during seizure periods, and contrast them between case and control groups. The performance of the proposed method is also evaluated on synthetic data.

\paragraph {Key words:}{ Differential network of dynamic causal models, M/EEG scans, epileptic brain network, dynamic causal nets, and mixed-effects models.}


\section {Introduction}
 Non-invasive imaging techniques electro- and magneto-encephalography (EEG and MEG) are often used for dynamic causal discovery and causality quantification in group comparison with neurological disorders such as stroke, motor neuron disease, Alzheimer's disease, Parkinson's disease, and epilepsy \citep{Choi2023, Aljalal2022, Ghassemkhani2025, Pyrzowski2015, Ogrim2012, Rosch2018} among others.  M/EEG are promising tools for studying brain dynamics  and enable researchers and clinicians to measure and characterise brain activity in fine temporal detail thus improving the understanding  of the etiology of neurological conditions as well as allowing the development of brain-based technologies.

 M/EEG imaging revealed a rich catalogue of behaviour in both resting and active brains..  For example, \cite{Papadopoulou2015} used spectral EEG data to profile epileptic seizure activity in the brain. \cite{Ozbek2021} demonstrated that the frontal alpha/theta power ratio derived from the resting-state EEG data  is a promising biomarker for distinguishing early-onset Alzheimer's disease from healthy controls. \cite{Duffy2012} found that EEG spectral coherence discriminates children with autism from healthy controls.

 Using MEG spectral data, \cite{vanEs2025} showed that, although brain network dynamics are stochastic, their activities have a robust cyclical pattern. \cite{Zhang2025} proposed a novel method for detecting mild traumatic brain injury by using spectral likelihood ratio statistics of MEG scan data when controls are not homogenous.
\cite{CoronelOliveros2026} characterised a previously introduced multiple frequency Jansen-Rit neural mass model with inhibitory synaptic plasticity.

Complementary to the aforementioned spectral analysis, dynamic causal modelling (DCM) of M/EEG data offers a powerful way to infer effective connectivity within the brain through coupled neuronal circuits with certain biophysics-informed parameters that are responsible for what are observed in the brain \citep{Friston2003, Kiebel2008, Grefkes2008a, Grefkes2010, Herz2014, Rehme2011, Rosch2018, Forrester2020, Singh2025}. Effective connectivity, referring to the causal influences that different parts of the brain exert over each other, provides a direct understanding of the neural mechanisms underlying various cognitive processes in the brain.  Instead of modelling individual neurons, DCM describes the average behaviour of neural populations. In particular,  allowing for dynamic causal interactions between distinct brain regions, DCM can reveal biophysics-informed links between microscopic neural activity and macroscopic brain phenomena.  By adjusting model parameters, researchers can simulate the transition from normal brain rhythms to pathological states, providing valuable insights into disease mechanisms and potential treatment strategies \citep{Wendling2024, Friston2019}. \cite{Sotero2007} used neural circuits to simulate EEG signals, helping to elucidate how the macroscopic electrical activity recorded from the scalp emerges from underlying neural dynamics.  Using the DCM approach,  \cite{Papadopoulou2015} tracked slow modulations in synaptic gain for a cohort of epileptic patients. \cite{Chen2009} employed spectral DCM on MEG scans to infer functional asymmetries in inter- and cross-frequency-coupling between brain regions during the period that subjects visually processes faces. This approach can be used to model brain activity on a whole brain scale.\cite{Schmidt2014} developed a modular network model that accounts for both node-driven and network-driven synchrony.  With the Human Connectome Project dataset, \cite{Singh2025} developed a large scale, data-driven mean field cortical model for activities in the brain, involving hundreds of interacting neural populations.

Following \cite{Friston2019}, a multiple cortical column DCM can be described by the following set of  neural mass stochastic differential equations:
\begin{eqnarray}\label {dcm1}
d{\x}(t) &=& f(\x(t), \u(t); \bthe,\K)dt +\Gamma(\bthe)d\w(t),~~~~\x(0)=\x_0; ~~~~y(t) = h(\x(t)),
\end{eqnarray}
where $\w(t)$ is a multivariate Gaussian white noise process, $\x(t)$ is a vector of $p$ hidden states, $\x_0$ contains initial values of states, $y(t)$ is the partial observation of hidden states, $\u(t)$ is an external input, $f$ is a nonlinear vector field, $h$ is a function from $\mathbb{R}^{p}\rightarrow\mathbb{R}$ with $\bthe $ being a vector of local node parameters, $\K$ a matrix of connection parameters between nodes, $\Gamma(\bthe)$  is a degenerate diffusion coefficient matrix. A specific choice of the vector field $f$ in the model (\ref{dcm1}) will lead to a six first-order stochastic Jansen-Rit model \citep{Ableidinger2017}.
 In the past two decades, considerable progress has been made in the development of dynamic causal modelling \citep{Friston2019, Park2021, Forrester2020, Singh2025}, and references therein. Despite this, a brain-wide search for dynamic causal abnormalities across spatial scales remains difficult, partly because of the challenges in modelling neural connectivity at a large scale.

The first challenge is about how to solve and sample from high-dimensional nonlinear neural mass differential equations when the parameters involved are only partially identifiable.
 \cite{Kulik2023} used a fourth order stochastic Runge-Kutta (RK) method to numerically solve the coupled  differential equations involved in the Jansen-Rit model. In these situations, recomputing the Jacobian-updating integrator continuously is required to solve the equations, making the computation infeasible when fitting many pairwise coupled DCMs to the data.  On the other hand, a truncated Chen-Fliess expansion \citep{Chen1954, Fliess1981} can solve the above differential equations. However, it still remains difficult to perform a Bayesian data analysis when the vector fields are highly nonlinear.

The second challenge is that some states $\x(t)$ are only partially observed at discrete time points. Then the exact likelihoods for the transition densities are difficult to trace. To estimate these likelihoods, \cite{Pokern2009} proposed  a data-augmentation algorithm to impute the hidden states.  The computation was intensive and it did not always  converge to the true values when many local hypoelliptic diffusions were included.

The third challenge is about how to implement large scale dynamic causal models, subject to a limited resource of computation  \citep{Honey2007, Li2020, Rosch2018, Singh2025}. To reduce the model dimension, \cite{Forrester2020} made a very restrictive assumption that the cortical columns are homogeneous in the sense that the column parameters are kept the same across the columns. \cite{Ziaeemehr2025} presented a valuable software package, named ``Virtual Brain Inference'' (VBI), that enables faster and more efficient parameter inference in dynamical system models of whole-brain activity. Their simulation-based inference approach can handle an 89-dimensional modular network of simplified Jansen--Rit models, with one global parameter and one local connectivity parameter for each Jansen--Rit model. However, scaling this approach to a modular network comprising a set of Jansen--Rit models over $68$ regions, with a parameter space of $4828$ dimensions, remains challenging, as discussed later.

The fourth challenge is about heterogeneity of human brains even when case and control subjects are matched by ages and genders \citep{Zhang2025, Li2025}. Some existing studies constrained model parameters to be the same across all modelled regions \citep{Hauke2026}. It is unclear how to take individual brain differences into account in a dynamic causal model with M/EEG data.  Finally, although the excitation-inhibition imbalance across columns has been shown to underlie an epileptic form of brain activity \cite[e.g., ][]{Friston2016},  it is unclear how such a spatial distribution pattern is associated with epilepsy.

Here, to address these challenges, we first develop an EEG data-driven network of conditionally coupled dynamic causal models (NccDCM) across multiple cortical regions for cases and controls respectively, where each node corresponds to a functional unit that can be represented by a neural mass model, and with edges informed by effective connectivity. For simplicity, neural mass activities are described by stochastic Jansen-Rit models \citep{Jansen1995, Ableidinger2017, Forrester2020}.

We then calculate  differential causal nets between cases and controls to quantify abnormal causality in patients. By these nets, we are able to identify intrinsic and extrinsic dynamic connectivity patterns that are associated with brain diseases. In the proposed network, each node describes a set of neuron populations which may receive input signals from other nodes. The strengths of directed connectivity between nodes are regularised by biophysics-informed parameters.

The above network structure allows sub-networks to be nested within a composite network, thereby providing a basis for the subsequent differential analysis of brain systems at different levels of detail.  The following assumptions are made in the NccDCM: Within each node, neuronal populations are represented by three types of neuronal populations: pyramidal cells (P), excitatory interneurons (E), and inhibitory interneurons (I). These populations form two feedback loops: a fast excitatory feedback loop involving pyramidal cells and excitatory interneurons, and a slower inhibitory feedback loop involving pyramidal cells and inhibitory interneurons. The model incorporates external inputs, typically representing background neuronal activity, which drive the excitatory interneurons. In our setting, these inputs include signals from other brain regions and rough-path noise representing background fluctuations. The node output is the average relative membrane potential of the pyramidal cell population, which serves as a proxy for the EEG signal.
The coupling within a node and between nodes are regularised by structural parameters. However, it should be noted that the precise nature of the within-node model is not restrictive and more complex models can be incorporated.

A conditional inference of the proposed model is implemented as follows: We first estimate NccDCMs by ccDCM screening, followed by inferring disease-associated causal nets from estimated contrasts between case and control subjects.  The conditional estimation is carried out through minimising a profiled loss function based on what is called Chen-Fliess expansions of stochastic differential equations \citep{Ableidinger2017, Chen1954, Fliess1981, Lyons1998}. In the Chen-Fliess expansions, we solve the stochastic differential equations by linear combinations of iterative integrals of the vector of field functions. These estimated ccDCMs are then assembled via a post analysis of variance \citep{Anderson2017}. Unlike the existing approaches, which assume parameters to be known from diffusion-imaging data and previous literature \citep[e.g., ][]{Forrester2020}, we directly estimate local parameters in each node conditionally on inputs from other nodes by using subject's EEG data. This enables the inference of the large number of node parameters feasible even when nodes are heterogeneous. The estimated NccDCM allows us to investigate how neural activities in these channels/nodes influence each other, particularly, when defected channels/nodes exhibit pathological dynamics compared to healthy controls. Changes in these parameters reflect alterations in the balance between excitation and inhibition, compensatory adaption to abnormal inputs.  We evaluate the proposed methodology on both simulated data and a recently published benchmark dataset of scalp EEG recordings for a group of pediatric epilepsy patients and their controls \citep{OK2021}. See Figure S4, 
 the Online Supplementary Materials for an example of  interictal epileptiform discharges (IEDs)  in the data and the bifurcation behaviour of the fitted ccDCM.

 A healthy, balanced brain connectivity is critical for maintaining brain functions. In contrast, an abnormal brain connectivity can change subject's behavior by affecting certain areas involved in seizure propagation.
Epilepsy is a group of neurological disorders characterised by abnormal spontaneous brain activity, involving multiscale changes in brain functional organizations.
 The system epilepsy hypothesis posits that the enduring susceptibility to generate seizures in epilepsy is due to the specific vulnerability of the brain system \citep{Fisher2005}. Abnormal firing of neurons causes defected local circuit function, leading to alterations of macroscale functional activities during seizure propagation \citep{Burman2018}. The virtual brain model has demonstrated that a combination of a global shift in the brain's dynamic equilibrium and locally hyperexcitable network nodes provides a mechanistic explanation for the epileptic brain during interictal resting state \citep{Courtiol2020, Yang2024}.  IED in routine EEG recordings is crucial evidence of epilepsy in patients. \cite{OK2021} published a benchmark dataset for detecting IED.  However, it is not clear to what extent the IEDs affect macroscale intrinsic dynamics and microcircuit organizations, that supports their pathological relevance.
 In the real epileptic data analysis, we treat IEDs as electroencephalographic markers for epilepsy. IEDs reflect transient cortical hyperexcitability in neuron populations and are useful biomarkers in diagnosis and treatment of epilepsy. We are in particular interested in characterising the IED associated epileptic networks via differential dynamic causal nets, which shed light on the mechanisms of seizure propagation in brain regions.

The contributions of this paper are two-fold. First, we develop a novel network architecture that leverage the region-wise coupling across a whole brain via hierarchical mixed-effects modelling. We show that all ccDCMs are locally structurally identifiable. We provide a conditional inference scheme to estimate NccDCM. By the post analysis of variance, we demonstrate that the proposed modelling is promising in discovering novel differential dynamic causal nets and in providing a deep insight into abnormal connectivity in case subjects compared to healthy controls. Secondly, fitting the proposed NccDCM to an epileptic case-control dataset in sensor and source levels, we identify comprehensive neuronal circuits for a whole brain and reveal biophysics-informed dynamic causal nets that underlie defected brain connectivity in epileptic patients.

The remaining paper is organised as follows.  The details of the proposed methodology and optimisation algorithm are provided in Section 2. The applications of the proposed methods to the IED dataset and synthetic data are presented in Section 3.   Discussion and conclusions are made in Section 4.

\section{Methods and materials}
In this section, we model subject's EEG data using modular networks, where each node is represented by a conditionally coupled dynamic causal model (ccDCM). By conditional coupling, we mean that the output of each node is treated, in turn, as the response variable, while the neural populations within the output-node serve as internal regressors and the other nodes serve as external regressors. We further assume that the model parameters for subjects in the groups follow a mixed-effects model, thereby accounting for both group-level and subject-specific variations. The ccDCM for each node is then estimated by minimising an estimated loss function, defined as minus the estimated log-likelihood, using an evolutionary optimisation algorithm. The loss function is derived from the Chen–Fliess solution to the stochastic Jansen–Rit differential equations. Finally, edges in differential dynamic causal nets are identified by testing for significant differences in the model parameters between the groups using a post hoc permutational multivariate analysis of variance (PERMANOVA) \citep{Anderson2017}.

\begin{figure}[htbp]
\centering

\begin{tikzpicture}[
    >=Stealth,
    every node/.style={inner sep=0pt}
]

\node (a) {
\begin{minipage}{0.48\textwidth}
    \centering
    \includegraphics[height=5.5cm,width=\linewidth]{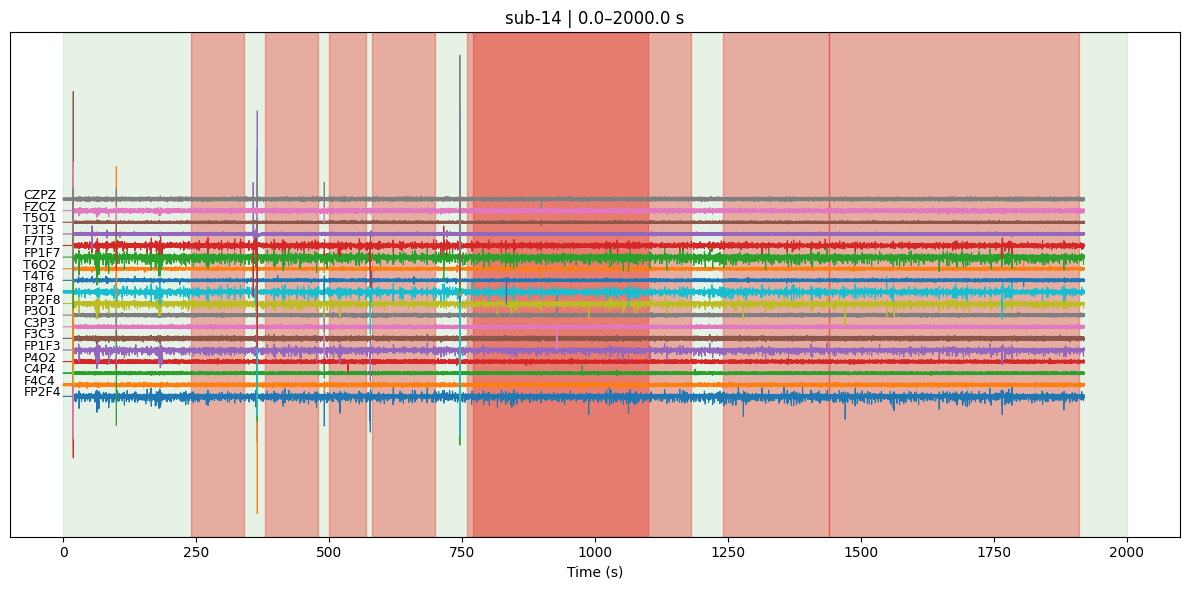}
    \tiny (a) Fitting ccDCM to the standardised control
\end{minipage}
};

\node[right=1cm of a] (b) {
\begin{minipage}{0.48\textwidth}
    \centering
    \includegraphics[height=5.5cm,width=\linewidth]{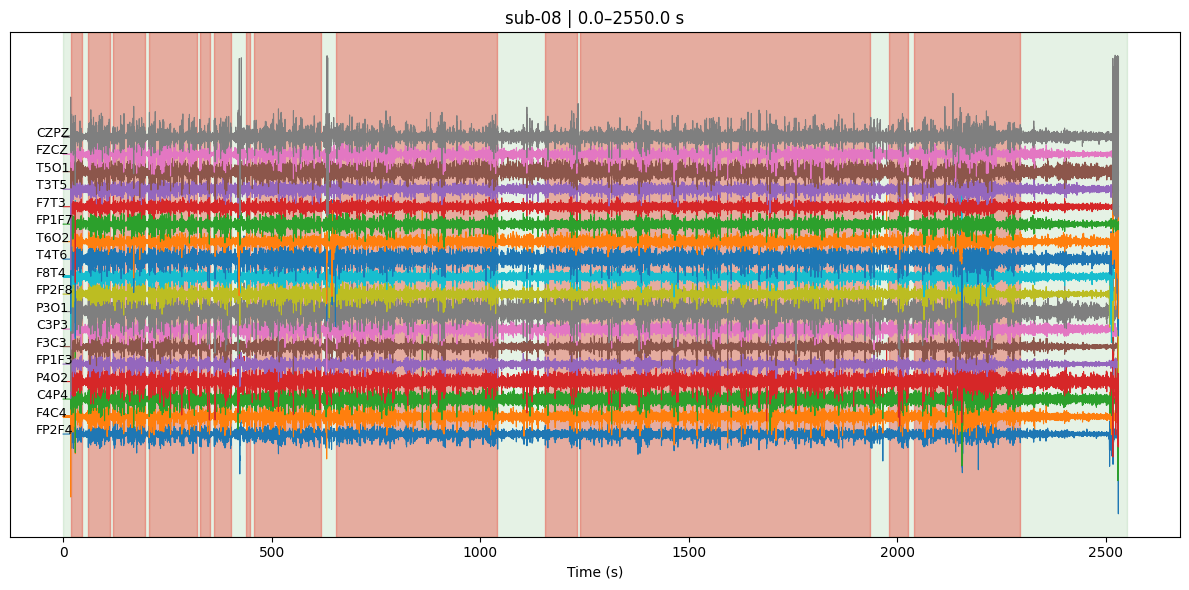}
    \tiny (b) Fitting ccDCM to the standardised case
\end{minipage}
};

\node[below=1cm of a] (c) {
\begin{minipage}{0.48\textwidth}
    \centering
    \includegraphics[height=4.8cm,width=\linewidth]{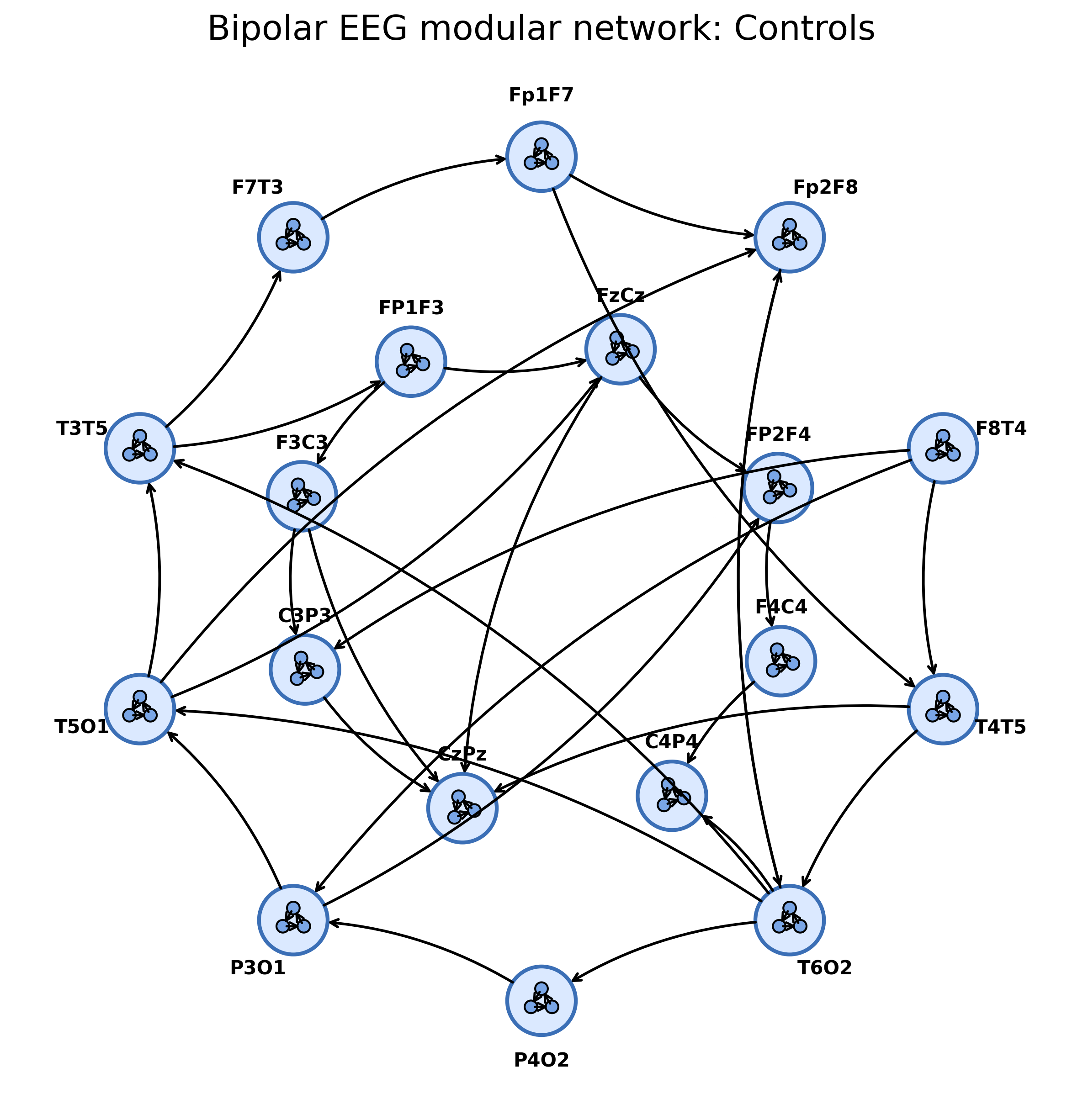}
    \tiny (c) Control network
\end{minipage}
};

\node[right=1cm of c] (d) {
\begin{minipage}{0.48\textwidth}
    \centering
    \includegraphics[height=4.8cm,width=\linewidth]{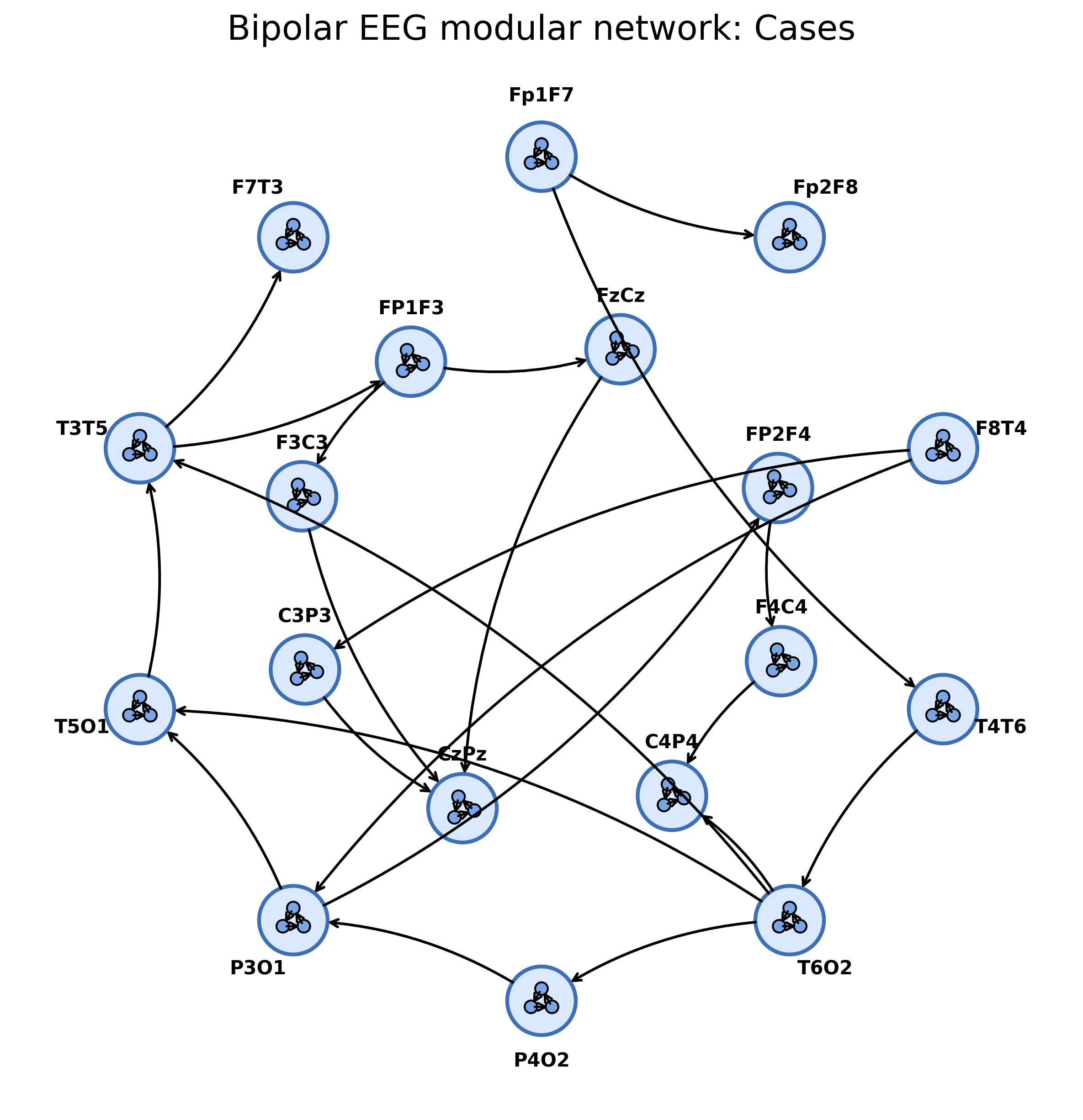}
    \tiny (d) Case network
\end{minipage}
};

\path ($(c.south)!0.5!(d.south)$) ++(0,-3.0cm)
      node (e) {
\begin{minipage}{0.45\textwidth}
    \centering
    \includegraphics[height=4.8cm,width=\linewidth]{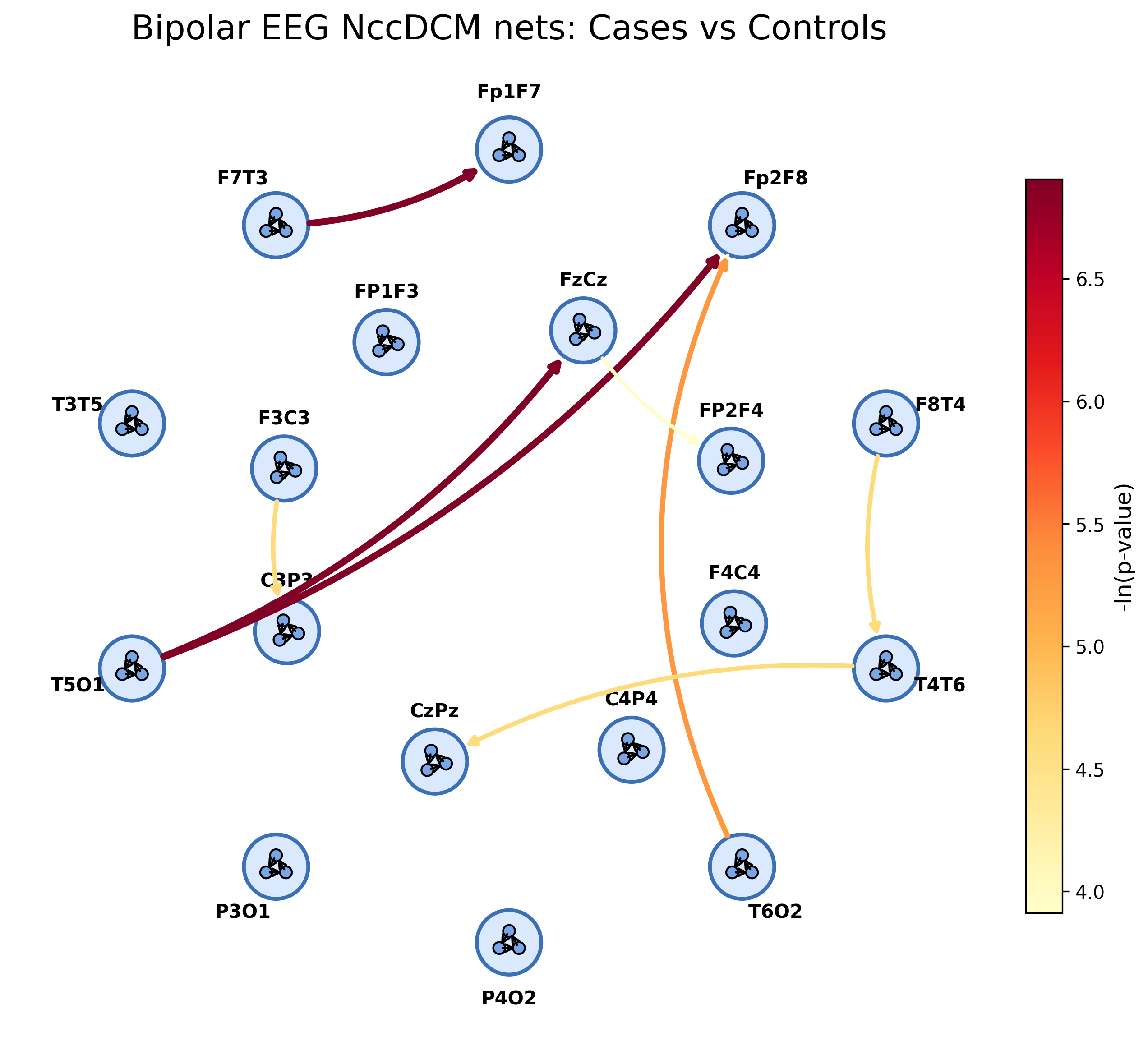}
    \tiny (e) Differential net
\end{minipage}
};

\node at ($(c.east)!0.5!(d.west)+(0,0.3cm)$) {\Huge $\mathbf{-}$};

\draw[->,thick] (a.south) -- (c.north);
\draw[->,thick] (b.south) -- (d.north);

\draw[->,thick] (c.south) -- ([xshift=-8mm]e.north);
\draw[->,thick] (d.south) -- ([xshift=8mm]e.north);

\end{tikzpicture}

\caption{Flowchart for generating differential NccDCM nets from sensor data. Step 1: Case and control time series are standardised and segmented (first row), and the corresponding modular networks are derived. Red shading indicates ictal periods. Each network node comprises three interconnected subcircles representing the three-state Jansen–Rit model. Step 2: Group of control modular networks are subtracted from the corresponding group of case networks via their parameters (second row). Step 3: Statistical significance testing of the network differences yields differential NccDCM nets. Step 4: Adjustment for multiple input effects. The colour bars are in the scale of $-\ln(\mbox{p-value})$, where $\mbox{p-value}$ is BH-adjusted.}
\label{Fig:scheme}

\end{figure}

\subsection{Models}

To account for multiple sources of variability in the data, the network of coupled canonical dynamic causal models (NccDCM) is constructed hierarchically. The observable outputs of each node are modelled conditionally on a set of parameters, which are themselves assigned probabilistic distributions. Nodes are connected by directed edges, each characterised by a transmission rate. The node and edge parameters are modelled using a mixed-effects framework to capture both group-level and subject-specific variations.
To identify NccDCM differences between the groups, for each pair of nodes, treating one as an output node and the other as an input node, we test for differences between two NccDCMs per parameter. We assign a directed edge from the input node to the output node if the testing result is significant after multiple testing corrections. This gives rise to a differential DCM net per parameter. Their performance can deteriorate by confounding effects of indirect and direct inputs. For example, suppose that the data are produced from a three-node ccDCM model where node $1$ links to node $2$ while node $2$ links node $3$. Given the data, a false direct link from node $1$ to node $3$ can be generated from the pairwise comparison.  This problem can be addressed by inputting both nodes $1$ and $2$ into the ccDCM. Under the multiple input framework, the influence of node $1$ on node $3$ is often weaker than the influence of node $2$ on node $3$ in terms of p-values. Therefore, the edge importances, quantified in terms of p-values, can be used to distinguish between direct and indirect links between nodes.
 In practice, for each node, we first select marginally significant inputs as multiple inputs in the model, refit the model to the data and keep those who can survive in the Benjamini-Hochberg  multiple testing adjustment.  Such an adjustment is called multiple input adjustment.  The architectures of NccDCM and differential DCM nets are illustrated by the schematic diagrams in  Figure \ref{Fig:scheme}. The further mathematical details are given as follows.

\vspace {0.5cm}
{\bf NccDCM and its  identifiability.}  Each node is represented by the Jansen-Rit model, a biologically inspired model of the three interconnected neural populations in a cortical column: pyramidal projection neurons, excitatory interneurons, and inhibitory interneurons.  These populations form feedback loops, enabling the model to capture the complex dynamics of cortical activity.
At dendritic spines, incoming spikes are converted into postsynaptic potentials (PSPs) through convolution with an alpha-type response function.
Let hidden states  $x_0, x_1$ and $x_2$ denote the hidden states of these populations in terms of total membrane potential alterations in the synapses respectively.
 To simplify the notations,  we also let $L_a$ and $L_b$ denote the differential operators,
 \begin{eqnarray*}
L_a&=&\partial^2/\partial t^2+2a\partial/\partial t+a^2,\\
L_b&=&\partial^2/\partial t^2+2b\partial/\partial t+b^2,
\end{eqnarray*}
and $S(\cdot)$ denote the sigmoid function of neural firing,
\begin{eqnarray}\label{SF}
  \mathrm{S}(\nu)=\frac{\nu_{\max}}{1+\exp(r(\nu_0-\nu))},
\end{eqnarray}
where  $a^{-1}$ and  $b^{-1}$  are known time constants, $r$ is a known scale constant  and $\nu_{\max}$ is a known maximum firing rate.    Let $z(t)$ denote external inputs. Then, the Jansen-Rit model can be described by differential equations below:
\begin{eqnarray*}
L_ax_0(t)&=&AaS(x_1(t)-x_2(t)),\\
L_ax_1(t)&=&Aa(Kz(t)+0.8CS(Cx_0(t))).\\
L_bx_2(t)&=&0.25BbCS(0.25Cx_0(t))
\end{eqnarray*}
with output $y(t)=x_1(t)-x_2(t).$ Here, the parameters $A$ and $B$ stand for the average excitatory and inhibitory synaptic gains respectively, the parameter $C$ governs the strengths of the synaptic connectivity between the three neural populations, $C_1=C, C_2=0.8C,$ $ C_3=0.25C$ and $C_4=0.25C$ \citep{Cabrera2023}. The first equation 
 describes the dynamics of the pyramidal-cell population. Its input depends on the net pyramidal-cell membrane potential, represented by the difference between the excitatory and inhibitory postsynaptic potential.  The second equation describes the excitatory postsynaptic potential acting on the pyramidal cells. It receives contributions from  some external inputs, often representing background neuronal activity, coupling signals from other regions and recurrent excitatory input from the pyramidal-cell population, scaled by the connectivity parameters $C_1$ and $C_2$. The third equation describes the inhibitory postsynaptic potential acting on the pyramidal cells. The inhibitory input is driven by pyramidal-cell activity through the sigmoid function, with connectivity parameters $C_3$ and $C_4$. The excitatory and inhibitory pathways have different time constants, represented by \(1/a\) and \(1/b\), respectively.

 Although the Jansen-Rit equations are derived from biological principles, it is not completely  determined by these principles. So, the model contains unknown quantities $(A,B,C, K, \nu_0,r)$ that must be inferred from partially observed data before it can be used for prediction or decision making. In this paper, for simplicity, we fix $r$ at a typical value and adjust it via scaling output $y(t).$ So,  unknown parameters  $(A,B,C, K, \nu_0)$.   Estimating the Jansen-Rit model is highly dependent on its structural identifiability. Non-identifiable parameters can create a flat likelihood landscape, leading to biased, inaccurate estimates of the parameters. Moreover, estimates of identifiable parameters can be biased if other parameters are non-identifiable \citep{Heinrich2025}. 
Here, a parametric system  is said to be structurally identifiable if  any change
in the parameter results in a change in the output. Similarly, we define structural observability for hidden states. 

We identify non-identifiable parameters in the model via symmetries of differential equations.
 Note that under the following scaling transformations
\begin{eqnarray*}
A&\longrightarrow& \lambda A, B\longrightarrow \lambda B, C\longrightarrow \lambda^{-1} C, K \longrightarrow \lambda^{-1} K, \nu_0 \longrightarrow \nu_0, \\
x_0&\longrightarrow& \lambda x_0, x_1\longrightarrow x_1, x_2\longrightarrow x_2,
\end{eqnarray*}
 the model preserves its output. The non-identifiability arises because the unknown $x_0$, $B$, $C$, $K$ can absorb arbitrary changes in $A$, making simultaneous recovery of these parameters impossible. The symmetry also suggests a reparametrisation scheme for the Jansen-Rit model: $B_A=A/B,$ $C_A=AC,$ $K_A=AK$, $\nu_0=\nu_0$, $x_{A0}=x_0/A$, $x_1=x_1$ and $x_2=x_2.$ After the reparametrisation, we have the form:
\begin{eqnarray}\label{JR1}
L_ax_{A0}(t)&=&aS(x_1(t)-x_2(t)),\nonumber\\
L_ax_1(t)&=&a[K_Az(t)+0.8C_AS(C_Ax_{A0}(t))],\nonumber\\
L_bx_2(t)&=&0.25B_A^{-1}bC_AS(0.25C_Ax_{A0}(t))
\end{eqnarray}
with the same output $y(t)=x_1(t)-x_2(t)$ as before.  The model (\ref{JR1})  is called canonical form of the Jansen-Rit model. In this form,  fixing $A=1$ for the synaptic gain of the excitatory neuron population,  $B_A^{-1}$ is the synaptic gain of inhibitory neuron populations, $C_A$ measures the strengths of local connectivity between the populations produced by pyramidals, and $K_A$ measures the synaptic effect of the external inputs on the excitatory  neuron population.

Following \cite{Yates2009}, parameter values $(B_A,C_A,K_A,\nu_0)$ and $(B^*_A,C^*_A,K^*_A,\nu^*_0)$ are called equivalent, written $(B_A,C_A,K_A,\nu_0)\sim$ $(B^*_A,C^*_A,K^*_A,\nu^*_0)$   if there exists a positive constant $\tau$ such that the associated outputs $y$ and $y^*$ have the same values over time interval $[0,\tau].$
We
call a Jansen-Rit model locally structurally identifiable if there exists a small parametric ball in which any two  different parameter values are non-equivalent. In the Appendix A, by using the symbolic computing software SIAN \citep{Hong2019}, we show that the canonical Jansen-Rit model is locally structurally identifiable under Condition $\bf{(C0)}$.
By a Lie symmetry analysis in  S1, 
the  Online Supplementary Material, we also show that the canonical Jansen-Rit model has no non-trival Lie symmetry. In particular, we reveal that there is no parameter $A$ anywhere in the reparameterised system. This is consistent with the recent experimental results in \cite{Hauke2026}.  In practice, fixing $A$ at a typical value has no effects on estimating $(B_A,C_A,K_A,\nu_0)$. 

\begin{figure}[htbp]
    \centering

    \begin{minipage}[c]{0.35\textwidth}
        \centering

\begin{tikzpicture}[
    node_style/.style={
        circle,
        draw,
        thick,
        minimum size=0.9cm,
        inner sep=0pt,
        font=\large
    },
    arrow/.style={
        ->,
        >=stealth,
        thick,
        shorten >=4pt,
        shorten <=4pt
    }
]

\node[node_style] (P) at (0,2.5) {$P$};
\node[node_style] (E) at (-2,0) {$E$};
\node[node_style] (I) at (2,0) {$I$};
\node[node_style] (z) at (-2.5,1.8) {$z$};

\draw[arrow] (z) to [bend right=25]
node[left, above=2pt, pos=0.88] {$K_A$}
 (E);

\draw[arrow]
    (P) to[bend right=25]
    node[left, pos=0.48] {$C_A$}
    (E);

\draw[arrow]
    (E) to[bend right=25]
    node[right, above=3pt, pos=0.38] {$0.8C_A$}
    (P);

\draw[arrow]
    (P) to[bend left=25]
    node[right, pos=0.48] {$0.25C_A$}
    (I);

\draw[arrow]
    (I) to[bend left=25]
    node[left, above=3pt, pos=0.38] {$0.25C_A$}
    (P);

\draw[dotted, thick]
(E) node[midway] { $\mbox{E/I ratio } B_A$ } (I);

\end{tikzpicture}

    \end{minipage}
    \hfill
    \begin{minipage}[c]{0.63\textwidth}
        \centering

        \includegraphics[
            width=\linewidth,
            height=6cm,
            keepaspectratio,
            clip,
            trim=20 0 320 10
        ]{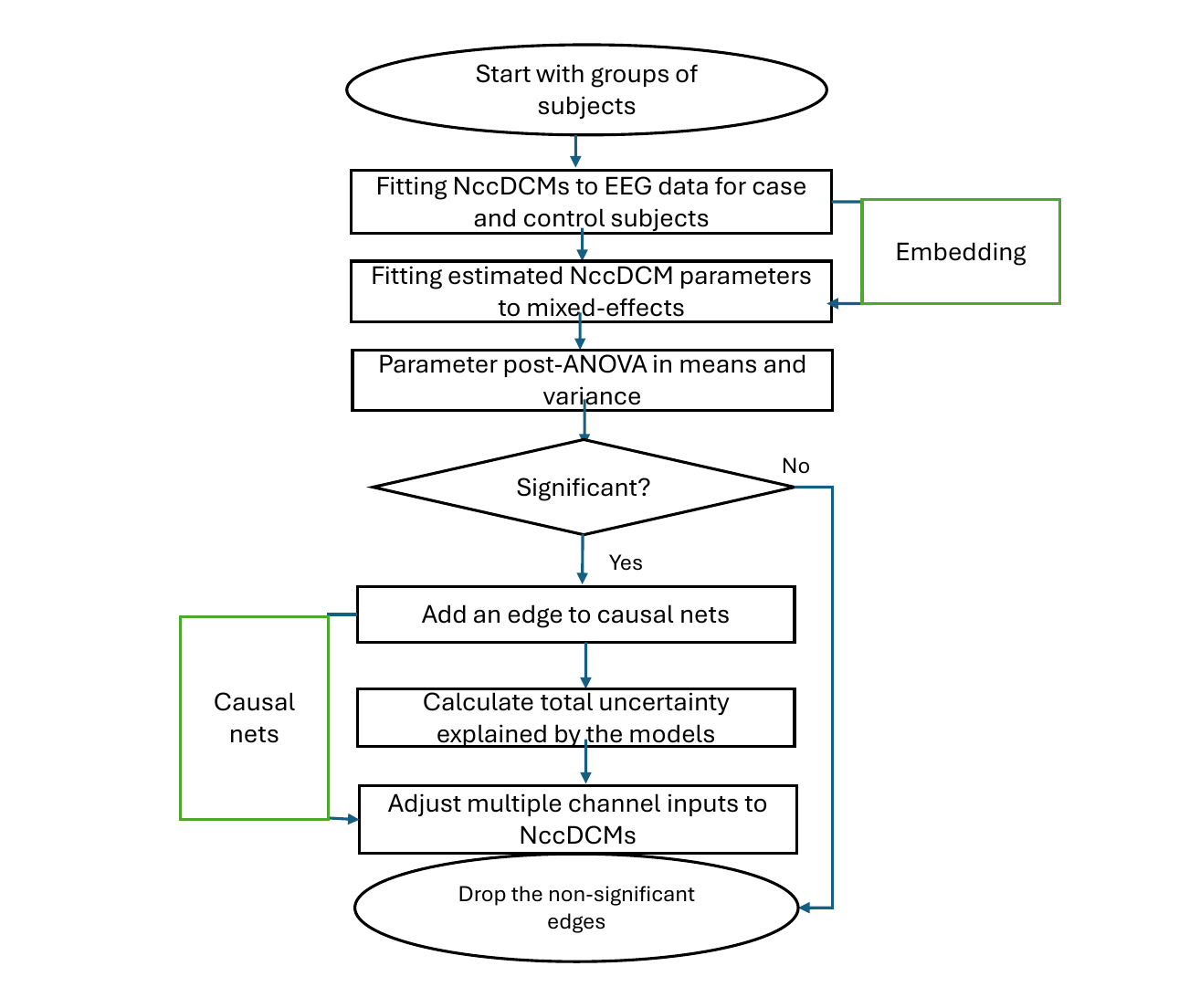}

    \end{minipage}

    \caption{Overview of the NccDCM analysis. Left: Jansen-Rit node model, showing the interactions
    between local neuron populations  Pyramidal (P), Excitatory (E), and  Inhibitory (I) under external input (z). Right: The flowchart of the NccDCM analysis.}
    \label{scalp}

\end{figure}

\vspace {0.5cm}
To model grouped EEG data, more notations are needed.  Let $k=0, 1$ denote group indices and $N_k$ the size of group $k$. We collected $J$ time series over the time interval $[0,T]$ from each subject using an EEG system. These time series are either sensor time series or source time series obtained from source localisation. For each subject, the time series are partitioned into $Q$ segments, corresponding to disjoint time intervals. In the epilepsy study, these intervals correspond to seizure or pre-seizure periods as shaded in the time series plots of Figure \ref{Fig:scheme}.
In the following, these time series are treated as the emissions of a $J$-node network, each node emits a time series.

\vspace {0.5cm}
Let $x^{(i)k}_{A0nq}(t), x^{(i)k}_{1nq}(t)$, and $x^{(i)k}_{2nq}(t)$ denote the hidden states of the pyramidal cell, excitatory interneuron, and inhibitory interneuron populations, respectively, associated with the observed time series $y^{(i)k}_{nq}(t)$ at node $i$, for subject $n$ in group $k$ during time interval $q$. For notational simplicity, the indices $k$, $n$, and $q$ are suppressed when unambiguous. The hidden states evolve according to three coupled second-order ordinary differential equations describing the mean membrane-potential dynamics driven by incoming firing rates within the neural mass. These rates are transformed by the sigmoid activation function in equation (\ref{SF}). The second-order system can equivalently be written as a coupled system of first-order differential equations. After suppressing the indices $(k,n,q)$ from the model parameters, the dynamics of the three neuronal populations at node $i$, subject to inputs from other nodes and background noise, can be expressed as
\begin{eqnarray}\label{sc}
  dx^{(i)}_{A0}(t) &=& x^{(i)}_{A3}(t)\,dt, ~~~ dx^{(i)}_{1}(t) = x^{(i)}_{4}(t)\,dt, ~~~ dx^{(i)}_{2}(t) = x^{(i)}_{5}(t)\,dt, \nonumber\\
  dx^{(i)}_{A3}(t) &=& \left[a\, \mathrm{S}^{(i|j)}\left(y^{(i)}(t)\right) - 2a\,x^{(i)}_{A3}(t) - a^2\,x^{(i)}_{A0}(t)\right]dt, \\
  dx^{(i)}_{4}(t) &=&  \left\{a\left[K^{(i|j)}_A\,\mathrm{S}^{(i|j)}\left(y^{(j)}(t)\right) + 0.8C^{(i|j)}_{A}\,\mathrm{S}^{(i|j)}\left(C^{(i|j)}_{A}\,x^{(i)}_{A0}(t)\right)\right]\right.\nonumber\\
  &&\left. - 2a\,x^{(i)}_{4}(t) - a^2\,x^{(i)}_{1}(t)\right\}dt +a d\tilde{\w}^{(i|j)}(t), \nonumber\\
  dx^{(i)}_{5}(t) &=& \left[(1/B^{(i|j)}_A)b0.25\,C^{(i|j)}_{A}\,\mathrm{S}^{(i|j)}\left(0.25C^{(i|j)}_{A}\,x^{(i)}_{A0}(t)\right) - 2b\,x^{(i)}_{5}(t) - b^2\,x^{(i)}_{2}(t)\right]dt. \nonumber
\end{eqnarray}
The output of node $i$ is given by
$
y^{(i)}(t)=x^{(i)}_1(t)-x^{(i)}_2(t).
$
 For node $i$, subject $n$ in group $k$, and time interval $q$, after suppressing the indices $(k,n,q)$ from the model parameters, we denote $\sigma^{(i|j)}_A= A^{(i|j)}\sigma^{(i)} $ and define the multiple stochastic input $\tilde{ \w}^{(i)}(t)$ in the above model  by
\begin{eqnarray*}
d\tilde{ \w}^{(i|j)}(t)&=&\left[\sum_{j_1 \not=i, j} K^{(i|j_1)}_{A}\,\mathrm{S}^{(i|j_1)}\left(y^{(j_1)}(t)\right)\right]dt +\sigma^{(i|j)}_{A}d\w^{(i)}(t),\\
  \mathrm{S}^{(i|j)}(x(t))&=&\frac{\nu_{\max}}{1+\exp(r(\nu^{(i|j)}_{0}-x(t)))},
\end{eqnarray*}
where  \(  \w^{(i)}(t) \) is a Gaussian noise process, $K^{(i|j)}_{A}$ is a signal transmission rate from node $j$ to node $i$, $\sigma^{(i|j)}_{A}$ is the background noise level,  and $\nu^{(i|j)}_{0}$ is the value for which a $50\%$  of
the maximum firing rate is attained  by neurons in node $i$.  Let  the parameters  $B^{(i|j)}_{A}$ and $ C^{(i|j)}_{A}$ be the values of the parameters $B_A,C_A$ of the neural populations within node $i$ in the presence of input from node $j.$

Under the assumption that the model parameters are identical across all nodes and all subjects, \cite{Forrester2020} introduced a similar model to describe the modular evolution of the neural populations in each node in the presence of external inputs from the remaining nodes.
However, their homogeneity assumption is found not valid later in our EEG data analysis. Our work helps bridge the gap between theory and practice by extending \cite{Forrester2020} to the setting where the model parameters are allowed to vary across nodes and subjects. The equations are solved sequentially for node $i=1,…,J,$ fixing the parameters associated with the remaining nodes.
In light of  \cite{Cabrera2023}, \cite{Grimbert2006}, \cite{Ditlevsen2025} and  \cite{Ableidinger2017}, we set the ranges of these parameters in Table \ref{paramer}.

\begin{table}[ht]
    \centering
    \caption{Ranges of biophysical parameters in ccDCM. 
}

    \scalebox{0.8}{\begin{tabular}{lll}
        \toprule
        \textbf{Parameter} & \textbf{Description} & \textbf{Value} \\
         $B^{(i|j)}_{A}$ & Ratio of average excitatory-inhibitory synaptic gains & $ [0.1, 1]$ \\
        $a^{-1}$ & Time constant of excitatory postsynaptic potential & 10 ms \\
        $b^{-1}$ & Time constant of inhibitory postsynaptic potential & 20 ms \\
        $C^{(i|j)}_A$ & Scaled effective synaptic gain  & $[0,2500] mV $\\
        $K^{(i|j)}_A$ & Scaled edge from node $j$ to node $i$  & $[0, 2500]$  \\
        $v_{\text{max}}$ & Maximum firing rate of the neural populations & 5$ s^{-1}$ \\
        $v^{(i|j)}_{0}$ & Value for which 50\% of the maximum firing rate is attained & $[0,20]$ mV \\
        $r$ & Slope of the act. function at $v^{(i|j)}_{0}$  & 0.56 mV$^{-1}$ \\
        $\sigma_A$ &    Noise level in the region                                          & $[5, 400]$\\
        \bottomrule
    \end{tabular}}
\label{paramer}
\end{table}

Let $I_k$ denote the number of active inputs to  node $k$ with the output time series $\by_k$. Assume that

({\bf C0}): the data matrix $(\by_i)_{i\not=k, i\in I_k}$ has the full column rank.

Then, under the condition {\bf (C0)}, the resulting ccDCMs are shown to be locally structurally identifiable in the Appendix A.  A Lie symmetry analysis  in S1, 
the Online Supplementary Material proves that there are no non-trival Lie symmetry on the states $x_{A0}(t), x_1(t), x_2(t)$.

To summarise, in the ccDCM of each node,  the dynamic behavior of three neural populations are determined by the above stochastic differential equations.   These populations are  interconnected through the coupling parameters in the equations, which  determine the average evolution rates of the postsynaptic potential.
The extrinsic parameter \( K^{(i|j)} \) regulates the influence of neural populations in node $j$ on the excitation of neural populations in node $i$ and thus determine the edge between nodes $i$ and $j$.
 A large \( K^{(i|j)} \) increases the strength of neural synchronization, changing oscillatory activity or pathological dynamics such as epileptic discharges, while decreasing \( K^{(i|j)} \) reduces inter-regional coupling, giving rise to more localised dynamics. Neural gain refers to how strongly a neuron circuit changes its response when its input changes.This input–output relationship is modulated by  E/I balance $B^{(i|j)}_A$ and mediated by external inputs with tuning parameter \( K^{(i|j)} \) .
In simplified terms, changing excitation alters the drive to a neural population, while inhibition can regulate the magnitude and selectivity of the resulting response. Inhibitory interneurons can therefore help prevent excessive amplification and allow circuits to adjust their sensitivity to incoming information.
This is important for the stability of cognitive functions such as attention, sensory processing, and adaptive responses to changing environmental demands\citep{Ferguson2020}.

Depending on the model parameters, the ccDCM model can reproduce a range of oscillatory dynamics corresponding to canonical EEG frequency bands, including alpha, beta, and gamma rhythms \cite{David2003}. Furthermore, depending on the parameter regime and external inputs, the model can exhibit irregular, transient, and epileptiform activity \cite{Ableidinger2017}.
 It also allows us to investigate how neural populations interact under both deterministic and stochastic settings, and how coupling between nodes modulates large-scale brain rhythms observed in E/MEG recordings \citep{Forrester2020}. As the nonlinear sigmoid function \( \mathrm{S}^{(i|j)} \) is globally Lipschitz continuous, it follows from  Theorem 6.2.2 of \cite{Arnold1974} that  the Eq.~\eqref{sc} has a path-wise unique and \( \mathcal{F}_t \)-adapted solution. In the real data analysis, we fix $a$, $b$, $v_{\max}$ and $r$ at their typical values.  It is a common practice to rescale the standardised time series data by a constant before fitting a NccDCM (see the  software MNE 1.12). This can be further refined by letting $r$ be estimated from the data. Furthermore,  it follows from the Appendix A that under the condition ({\bf C0}), the ccDCMs are locally structurally identifiable. This implies that the output time series in each node keeps the same only possible for a finite number of parameter values. In particular, the initial values of these hidden states are locally structurally identifiable. As the hidden state trajectories can be uniquely determined by the initial values via the Jansen-Rit differential equations. Therefore, all hidden states are identifiable/observerable. However, this holds for noiseless data.  EEG data are very noisy. So, in practice,  due to the small number of observed sensor time series, the so-called practical identifiability of these parameters can arise. This may lead to some confoundings in estimating the coupling parameter $K_A$ and the local connection parameters 
$C_A.$

\vspace {0.5cm}
{\bf Mixed-effects models for deriving differential DCMs.}
To facilitate the description of the nested parameters in different groups, we write them in vectors:
\[
\bthe^{(i|j)k}_{nq}=(\theta^{(i|j)k}_{nq1},...,\theta^{(i|j)k}_{nq4},\theta^{(i)k}_{nq5})^T=( B^{(i|j)k}_{Anq},  C^{(i|j)k}_{Anq},  K^{(i|j)k}_{Anq},  \nu^{(i|j)k}_{0nq},\sigma^{(i)k} _{Anq})^T.
\]
   We assume that these scaled parameters follow the mixed-effects model
\begin{eqnarray}\label{mixed-effects}
\bthe^{(i|j)k}_{nqc}=\bdel_{qc}+\balp^{(k)}_{qc}+\bbet^{(i|j)}_{qc}+c_{01}\balp^{(k)}_{qc}\odot\bbet^{(i|j)}_{qc}+\r^{(i|j)k}_{nqc}, \quad k=0, 1; \quad q=1,...,Q,
\end{eqnarray}
where $\odot$ denotes the component-wise multiplication, $\balp^{(1)}_{qc}+\balp^{(2)}_{qc}=0$, $\sum_{i\not=j}\bbet^{(i|j)}_{qc}=0 $, and $c_{01}$ is the coefficient satisfying  $c_{01}\sum_q|\beta^{(i|j)}_{qc}+1|\not=0$ and $c_{01}\sum_q|\alpha^{(i|j)}_{qc}+1|\not=0$.

\subsection{Algorithms }
Simultaneously estimating parameters for all nodes in the model (\ref {sc}) is difficult as the model is ill-posed. Instead, we decompose the problem into a set of pairwise conditional models  ccDCMs. For each target node, we fit a set of ccDCMs, in which each remaining node is treated in turn as an external input to the target node.    However, brain networks are inherently multivariate, where node-outputs are strongly dependent. In particular, EEG sensor data may be spatially degenerate, such that the mapping from bipolar data to reference monopolar data is not one-to-one. As a result, a strong extrinsic coupling between nodes via parameter $K_A$ can be confounded with local parameters ($B_A, C_A, \nu_0$). A screening step to identify marginally significant edges in differential DCM nets at a pre-selected level can help reduce the confounding. Taking these significant edges and nodes as multiple inputs to the target node, we refit Fitting a ccDCM to the data. The resulted p-values will be adjusted due to multiple input effects. The details are as follows.

\vspace {0.5cm}
{\bf Estimating ccDCMs.}
We fit the neural mass differential equations in Eq. (\ref{sc}) to the time series data for each pair of nodes in three steps below.

\vspace {0.5cm}
 {\bf Step 1:  Solve the neural mass differential equations for node pair $(i,j)$.} For a pairwise screen, we use a misspecified model by ignoring the bias term in $d\tilde{ \w}^{(i|j)k}_{nq}(t)$. We compute the solutions of the neural mass differential equations iteratively over a sequence of time-instances with the help of
the Chen-Fliess expansion in terms of iterative Stratonovich integrals \citep{Fliess1981}. In Section \ref{joint_es} and  S13, 
the Online Supplementary Materials,  we show the advantage of the Chen-Fliess expansion method over the commonly used Euler-Maruyama (EM) and the stochastic RK in terms of ARE.

For subject $n$ in the group $k$, the time series emitted by node $i$ during the $q$-th period are partitioned into $D$ blocks, each containing $b_q$ time-instants $t^d_{0:b_q-1}=(t^d_0, ...,t^d_{b_q-1} )$. The time series for the $d$-th block is given by
$\Y^{(i)k}_{nq}(t^d_{0:b_q-1}) =\left(y^{(i)k}_{nq}(t^d_0 ), y^{(i)}_{nq}(t^d_1),...,y^{(i)}_{nq}(t^d_{b_q-1})\right)$, $d=1,2,...,D.$ Let $x^{(i)k}_{snq}(t),  s=0,1,2$ denote the corresponding three hidden neural states. Let $-\{A3\}=\{A0,1,2,4,5\}$, $-\{1,4\}=\{A0,2,3,5\}$ and $x^{(i)k}_{Inq}(t)=(x^{(i)k}_{snq}(t))_{s\in I}.$ Denote the truncated Chen-Fliess expansions of the conditional expectations of $y^{(i)k}_{nq}(t)$ and $x^{(i)k}_{snq}(t)$, $ s=A0, 2, 3, 5$, given  by the forms
\begin{eqnarray}\label{expansion}
y^{(i)k}_{nq}(t_m^d)&\approx&g\left(t_m^d|x^{(i)k}_{-\{A3\}nq}(t_0^d), y^{(i)k}_{nq}(t_0^d),\dot{y}^{(i)k}_{nq}(t^d_0), y^{(j)k}_{nq}(t^d_0),\bthe^{(i|j)k}_{nq}\right)\non\\
x^{(i)k}_{A0nq}(t^d_0)&\approx&g_0\left(t^{d-1}_{b_q-1}|x^{(i)k}_{\{A0,A3\}nq}(t^{d-1}_0), y^{(i)k}_{nq}(t^{d-1}_0), \dot{y}^{(i)k}_{nq}(t^{d-1}_0),\bthe^{(i|j)k}_{nq} \right),\non\\
x^{(i)k}_{2nq}(t^d_0)&\approx & g_2\left(t^{d-1}_{b_q-1}|x^{(i)k}_{-\{1,4\}nq}(t^{d-1}_0),\bthe^{(i|j)k}_{nq}\right),\non\\
x^{(i)k}_{A3nq}(t^d_0)&\approx & g_3\left(t^{d-1}_{b_q-1}|x^{(i)k}_{\{A0,A3\}nq}(t^{d-1}_0),y^{(i)k}_{nq}(t^{d-1}_0), \dot{y}^{(i)k}_{nq}(t^{d-1}_0),\bthe^{(i|j)k}_{nq}\right),\non\\
x^{(i)k}_{5nq}(t^d_0)&\approx & g_5\left(t^{d-1}_{b_q-1}|x^{(i)k}_{-\{1,4\}nq}(t^{d-1}_0),\bthe^{(i|j)k}_{nq}\right),
\end{eqnarray}
where $g(t|\cdot), g_0(t|\cdot), g_2(t|\cdot), g_3(t|\cdot)$ and $g_5(t|\cdot)$ are defined in the Appendix B. Then the truncated Chen-Fliess expansions of the above Jansen-Rit differential system can be written as a conditionally deterministic function, say $g$ or $g_m$, plus a stochastic error term. Moreover, in the expansions of $ x^{(i)k}_{A0nq},x^{(i)k}_{2nq}(t)  $, the stochastic error terms, only appearing in the terms of orders larger than $3$, are  bounded by
$|o_p((t-t_0)^4)|$, while in the expansion of $y^{(i)k}_{nq}(t)$, the stochastic terms, not appearing in the term of order $1$,  is of the form $\sigma^{(i|j)k}_{Anq}a\int_{t_0}^t(t-s)d\tilde{\w}^{(i)k}_{nq}(s)+o_p((t-t_0)^3)$. Therefore, these approximate functions of the conditional expectations  in (\ref{expansion}) provide one-step predictions of  $y^{(i)k}_{nq}(t)$ and $x^{(i)k}_{snq}(t)$. Note that in the misspecified model, the ignored bias is absorbed in the error terms.

\vspace {0.5cm}
{\bf Step 2: Construct loss functions.}
Let $t^d_0=t_{b_q-1}^{d-1}$. We randomly initiate $x^{(i)k}_{-\{1,4 \}nq}(t^1_0)$ by the standard multivariate norm with mean value $(0,...,0).$ Given the observed $D$ block data, we can iteratively make one-step forward predictions of  both the hidden states and the node outputs over time-grids by using the truncated Chen-Fliess expansions and  the difference operator as follows. For the first block,
\begin{eqnarray*}
\dot{y}^{(i)k}_{nq}(t^1_0)&\approx& (y^{(i)k}_{nq}(t^1_1)-y^{(i)k}_{nq}(t^1_0))/(t^1_1-t^1_0),\\
\hat{y}^{(i)k}_{nq}(t^1_m)&\approx&g\left(t^1_m|x^{(i)k}_{-\{A3 \}nq}(t^1_0), y^{(i)k}_{nq}(t^1_0), \dot{y}^{(i)k}_{nq}(t^1_0), y^{(j)k}_{nq}(t^1_0),\bthe^{(i|j)k}_{nq}  \right),
\end{eqnarray*}
where $x^{(i)k}_{\{A0:5\}nq}(t_0^1)=(0, y^{(i)k}_{nq}(t^1_0),0,0,\dot{y}^{(i)k}_{nq}(t^1_0),0)^T.$
 In general, for $m=1,...,b_q-1$ and $d=2,..., D$,
\begin{eqnarray*}
\dot{y}^{(i)k}_{nq}(t^d_0)&\approx& (y^{(i)k}_{nq}(t^d_1)-y^{(i)k}_{nq}(t^d_0))/(t^d_1-t^d_0),\\
\hat{y}^{(i)k}_{nq}(t^d_m)&\approx&g\left(t^{d}_m|x^{(i)k}_{-\{A3 \}nq}(t^d_0), y^{(i)k}_{nq}(t^d_0), \dot{y}^{(i)k}_{nq}(t^d_0), y^{(j)k}_{nq}(t^d_0),\bthe^{(i|j)k}_{nq}  \right),\\
\hat{x}^{(i)k}_{A0nq}(t^d_0)&\approx&g_0\left(t^{d-1}_{b_q-1}|x^{(i)k}_{\{A0,A3 \}nq}(t^{d-1}_0), y^{(i)k}_{nq}(t^{d-1}_0),\dot{y}^{(i)}_{nq}(t^{d-1}_0),\bthe^{(i|j)k}_{nq} \right),\\
\hat{x}^{(i)k}_{2nq}(t^d_0)&\approx& g_2\left(t^{d-1}_{b_q-1}|x^{(i)k}_{-\{1,4\}nq}(t^{d-1}_0), \bthe^{(i|j)k}_{nq} \right),\\
\hat{x}^{(i)k}_{1nq}(t^d_0)&=&y^{(i)k}_{nq}(t^d_0)+\hat{x}^{(i)k}_{2nq}(t^d_0),\\
\hat{x}^{(i)k}_{A3nq}(t^d_0)&\approx&g_3\left(t^{d-1}_{b_q-1}|x^{(i)k}_{\{A0,A3 \}nq}(t^{d-1}_0), y^{(i)k}_{nq}(t^{d-1}_0),\dot{y}^{(i)}_{nq}(t^{d-1}_0),\bthe^{(i|j)k}_{nq} \right),\\
\hat{x}^{(i)k}_{5nq}(t^d_0)&\approx& g_5\left(t^{d-1}_{b_q-1}|x^{(i)k}_{-\{1,4\}nq}(t^{d-1}_0), \bthe^{(i|j)k}_{nq} \right),\\
\hat{x}^{(i)k}_{4nq}(t^d_0)&=&\dot{y}^{(i)k}_{nq}(t^d_0)+\hat{x}^{(i)k}_{5nq}(t^d_0).
\end{eqnarray*}

For the subject $n$ in group $k$ and node pair $(i,j)$  the loss function/negative log-likelihood function in node $i$, with a regressive input from  node $j$, can be written as
\begin{eqnarray*}
L^{(i|j)k}_{nq}( \bthe^{(i|j)k}_{nq})=\frac{D}{2}\log 2\pi+\frac 12
\sum\limits_{d=1}^{D}\log G^{(i|j)k}_{nqd}
+\frac{1}{2}\sum\limits_{d=1}^D\frac{\left(y^{(i)k}_{nq}(t_{b_q-1}^d)-\hat{y}^{(i)k}_{nq}(t_{b_q-1}^d)\right)^2}{G^{(i|j)k}_{nqd}},
\end{eqnarray*}
where $G^{(i|j)k}_{nqd} = (\sigma^{(i)k}_{Anq}a)^2(t^{d}_{b_q-1}-t^d_0)^3/6$.

\vspace {0.5cm}
{\bf Step 3: Minimise the loss function by genetic evolutionary algorithms.}  Minimising the above loss function with respect to the model parameters, we obtain the estimate $\hat{\bthe}^{(i|j)k}_{nq}.$ Note that  S2, 
 the Online Supplementary Material demonstrates that the above non-convex loss function contains multiple local minima as only partial observations on hidden states are available in dynamic causal modelling. We employ a genetic evolutionary algorithm of \cite{Zhang2009}, called adaptive differential evolution with optional external archive
(JADE), to search for global minima. JADE evolves a population of candidate solutions through mutation, crossover, and selection, aiming to explore the landscape of the loss function around a global minimum. See S2, 
 the Online Supplementary Material for the further details of the JADE.

 To mitigate the systematic bias present in the above estimation procedure, we further employ a bootstrap bias correction method. See  S4, 
the Online Supplementary Material.

\vspace {0.5cm}
{\bf Identify differential causal nets for case and control groups. }
Having obtained estimated  networks of ccDCMs for cases and controls respectively, we want to find differential causal nets by contrasting cases and controls in terms of the estimated within-node parameters $(\hat{B}^{(i|j)k}_{Anq}, \hat{C}^{(i|j)k}_{Anq}$, $\hat{\nu}^{(i|j)k}_{0nq})  $ and the estimated between-node parameters $\hat{ K}^{(i|j)k}_{Anq}$ respectively. The procedure is implemented into two steps: testing and screening followed by the adjustment for multiple-input effects.
In the differential causal net for the $c$-th component of the estimated parameter vector,  we assign a directed link from node $j$ to node $i$  if the estimated values $(\hat{\bthe}^{(i|j)1}_{nqc})_{1\le q\le Q}, 1\le n\le N_1$, in the cases are significantly different from their counterparts in the controls.

\vspace {0.5cm}
{\bf Step 1:  Hypothesis testing}. In order to test for the significant differences between cases and controls, we treat the case-control memberships and node memberships as factors in the following two-factor mixed-effects model
\begin{eqnarray}\label{anova}
\hat{\bthe}^{(i|j)k}_{nq}=\bdel_q+\balp^{(k)}_q+\bbet^{(i|j)}_q+c_{01}\balp^{(k)}_q\odot\bbet^{(i|j)}_q+\r^{(i|j)k}_{nq}, \quad k=0, 1; \quad q=1,...,Q,
\end{eqnarray}
where $\odot$ denotes the component-wise multiplication and we write each estimated parameter as the summation of a baseline, fixed-factors related to case-control memberships and nodes, and subject-specific random factors.
Here, the significance of the case-control factor or the interaction factor between the case-control and nodes implies that case-control memberships make a contribution to the group differences while the significance of the node-factor shows that the group difference should be accounted in study. The significance of random-effects (e.g., the difference of the variations across groups) implies that subject-specific variations play a role in differencing the groups from each other.

To make the above model identifiable, we assume that $\balp^{(1)}_q+\balp^{(2)}_q=0$, $\sum_{i\not=j}\bbet^{(i|j)}_q=0 $, and $c_{01}$ is the coefficient satisfying  $c_{01}\sum_q|\beta^{(i|j)}_{qc}+1|\not=0$ and $c_{01}\sum_q|\alpha^{(i|j)}_{qc}+1|\not=0$ . Under the assumption that the parameters for different subjects are independent of each other,  the expected Manhattan distances between the groups and within the groups admit
\begin{eqnarray}\label{eanova}
E\sum_q|\theta^{(i|j)1}_{nqc}-\theta^{(i|j)0}_{mqc}|&=&\sum_q E|r^{(i|j)0}_{1qc}-r^{(i|j)1}_{1qc}|, \mbox{ under } H_0\non\\
E\sum_q|\theta^{(i|j)k}_{nqc}-\theta^{(i|j)k}_{mqc}|&=&\sum_qE|\theta^{(i|j)k}_{nqc}-\theta^{(i|j)k}_{mqc}|,\quad k=0, 1.
\end{eqnarray}
 Approximating the expectations in equation (\ref{eanova}) by the U-statistics \cite{Lee1990}, we have the following statistics: For $1\le c\le 5$, $k=0,1$, and $1\le i\not=j\le J$,
\begin{eqnarray*}
\sigma^{(i|j)}_{ckk}&=&\frac 2{N_k(N_k-1)Q}\sum_{1\le m<n\le N_k}\sum_{1\le q\le Q}|\theta^{(i|j)k}_{mqc} -\theta^{(i|j)k}_{nqc}|,\\
\sigma^{(i|j)}_{c01}&=&\frac 1{N_0N_1Q}\sum_{1\le m\le N_0, 1\le n\le N_1}\sum_{1\le q\le Q}|\theta^{(i|j)0}_{mqc} -\theta^{(i|j)1}_{nqc}|
\end{eqnarray*}
which are the empirical versions of the within-group and between group average distances.
For $1\le c\le 5$ (estimable parameters),  to test the above null hypothesis $H_0$,  we consider the test statistics  $\bF^{(i|j)}=(F^{(i|j)}_{1},...,F^{(i|j)}_{5})^T$ with
\begin{eqnarray*}
F^{(i|j)}_{c}=\frac{(N_0+N_1)\sigma^{(i|j)}_{c01}}{N_0\sigma^{(i|j)}_{c00}+N_1\sigma^{(i|j)}_{c11}}.
\end{eqnarray*}
We permuted the case-control memberships $M$ times with default value $M=5999$.  For each permutation, we recomputed the value of $F^{(i|j)}_{c}$, obtaining $M$ permuted $F$-values.
 The permuted p-values $p_{c}$ is then defined by counting the proportion of the permuted F-values that are less than the observed value of $F^{(i|j)}_{c}$. We also run a permuted $F$-test for differences in the variances of  the random factors of patients and controls. Similarly, we test the null hypothesis that $\beta^{(i|j)}_{qc}=0$  for all $(i,j)$.

\vspace {0.5cm}
{\bf Pairwise screening.} We select node pairs that have p-values at the level $0.01$ or at a more stringent level $0.005$ such that for each node, the multiple inputs meet the identifiability condition $({\bf C0})$.

\vspace {0.5cm}
{\bf Step 2: Adjustment for multiple input effects.} In the above pairwise screening, we only account for the marginal effects of multiple input nodes on a target node. This may result in confounding between direct and indirect effects. To mitigate the problem, we adjust the multiple-input effects for the node pairs that survived in the previous screening step below. For each target node, we identify input nodes that are linked to it and survived in the previous screening step.  Taking these nodes as multiple inputs,  we refit a ccDCM
to the time series emitted from the target node for each subject in the two groups. We then perform the mixed-effects model-based tests for differences in the estimated model parameters between the groups. We replace the pairwise p-values by these new p-values. Finally, the Benjamini–Hochberg (BH) correction is
applied to the resulting p-values across all tested node pairs to control the false discovery rate. We retain the node edges with BH-adjusted p-values less than or equal to $0.05.$

\vspace {0.5cm}
{\bf Identifying differential causal nets for subjects paired groups.}
Understanding differences of epileptic patients between the preictal period and ictal period is crucial for gaining deeper insights into the mechanisms underlying epileptic seizures. While most studies focus on the ictal states, examining the transition from the preictal phase to the ictal phase can reveal early neural signatures that precede seizure onset. Identifying such signatures are important not only for elucidating pathophysiological dynamics of epilepsy but also for improving the characterisation of the transition from  the preictal to the ictal and thus developing early intervention strategies.  We performed multiple permutation-based Wilcoxon signed-rank tests for significant NccDCM changes from the preictal period to the ictal period.  See  S3,
 the Online Supplementary Material for the details.

\subsection{Epilepsy dataset}
{\bf Data description}.
This publicly available dataset contains EEG recordings  in $18$ EEG channels for $21$ subjects who underwent EEG tests for epilepsy.
  Longitudinal bipolar montage is used for channel configuration for the EEG recording, and the considered channels are FP2F4, F4C4, C4P4, P4O2, FP1F3, F3C3, C3P3, P3O1, FP2F8, F8T4, T4T6, T6O2, FP1F7, F7T3, T3T5, T5O1, FZCZ, CZPZ, and BP3REF ($18$ EEG channels and 1 ECG channel). See Table \ref{tab:eeg_positions} in \ref{Section:S6},
 the Online Supplementary Material for the brain regions which the electrode-pairs cover. EEG signals were band-pass filtered between 1–40 Hz and re-referenced implicitly through the bipolar montage. No additional re-referencing was applied to preserve the original spatial derivations. Segmentation information provided by the owner in an accompanying metadata identified ictal intervals via start and end times; all remaining intervals were treated as interictal periods.  Ethical permission for research use was also in
place with informed consent for data usage for research
in the anonymised format and approved by the ethics
committee of the organisation which the data owner belongs to.

We project these channels onto two concentric rings- outer and inner- based on their spatial positions on the scalp. A sensor-level differential causal net per parameter is built by assigning significant edges to these channels.
 To test for significant edges, sensor  interictal epileptiform discharge (IED) time series from $11$ subjects were identified and confirmed by neurologists while the sensor time series from the remaining $10$ subjects were free from IEDs. Furthermore,  for each of $8$ patients, we were able to define $6$ segments of IED time series, each of length $20$ seconds. These segments were respectively matched by the segments of sensor time series of $10$ controls at the same time-periods. The dataset is preprocessed using independent component analysis with six chosen components, embedded in the software dSPM. We divide each sensor time series into epochs, each of size $2$ seconds.
  Here, we aim to analyse this dataset to identify abnormal dynamic causal nets associated with the IED in the sensor and source levels.

\vspace {0.5cm}
{\bf Source localisation with bipolar EEG sensor data.}
EEG sensors record linear mixtures of signals originating from multiple sources in the brain. Although the forward model from source signals to sensor measurements is linear, the nonlinear dynamics of the Jansen–Rit model do not generally commute with this mixing operation. See the Appendix B for the details.  Consequently, Jansen–Rit parameters estimated from sensor-level mixtures cannot, in general, be inversed to obtain the corresponding source-level parameters through a linear transformation.
So, to construct differential causal nets in the source level, we need to localise the intracranial sources that generated observed EEG measurements on the scalp. This is an ill-posed inverse problem because a very large number of different sources can produce the same potential field on the scalp.
For this purpose, we first convert the bipolar sensor measurements to a dataset of the monopolar format, to which the distributed source localisation package dSPM
\citep{Michel2019} is applied. Specifically, let $B_2(t)$ and $B_1(t)$ denote the bipolar and monopolar sensor measurements, respectively, and let M denote the transformation matrix from the monopolar to the bipolar montage: $B_2(t)=MB_1(t).$ Let $L$ denote the lead-field matrix derived from the average surface model in the software FreeSurfer. The monopolar measurements can then be represented by the forward model $B_1(t)=L\beta(t)+E(t), $ where $\beta(t)$ denotes the underlying source time series and $E(t)$  represents error terms. We obtain  the following monopolar representation of the measurements from the bipolar data using the Moore–Penrose pseudoinverse of $M$
\[
B_2(t)=M L\beta(t)+ME(t).
\]
In practice, applying dSPM to $M^{+}B_2(t)$, we obtain the estimate of source time series, $\hat{\beta}$,  at grid points in the source-space. See S9, 
 the Online Supplementary Material for the theory. We then group the grid points into the $68$ cortical regions defined by the Desikan–Killiany (D-K) atlas and using sign-flipped means to average the estimated source time series over the grid points within each region. This results in $68$ estimated regional source time series, one for each D-K cortical region. We project these regions/nodes onto three concentric rings— outer, middle, and inner—based on their spatial locations on the cortical surface.
See the abbreviations of these region names in S10, 
the Online Supplementary Material.
 A source-level differential causal net per parameter is constructed by assigning significant edges to these nodes. Note that the $K_A$ differential causal nets measure the significant difference between the input-output relationships in the case-control groups while the $B_A$, $C_A$ and $\nu_0$ differential causal nets show impacts of these input-output relationships on the local node differences between the case-control groups.

\section{Numerical results}
In this section, we aimed to investigate the behavior of the proposed models and to evaluate the performance of the proposed estimating method using simulated data as well as the real epileptic dataset of \cite{OK2021}. We first conducted sensor- and source-level analyses of the case-control data, followed by analyses of the preictal–ictal data and, finally, simulation studies.

\subsection{Sensor differential dynamic causal nets}

\begin{figure}[htbp]
  \centering
  \begin{minipage}{0.4\textwidth}
    \centering
    \includegraphics[width=\linewidth, height=4cm]{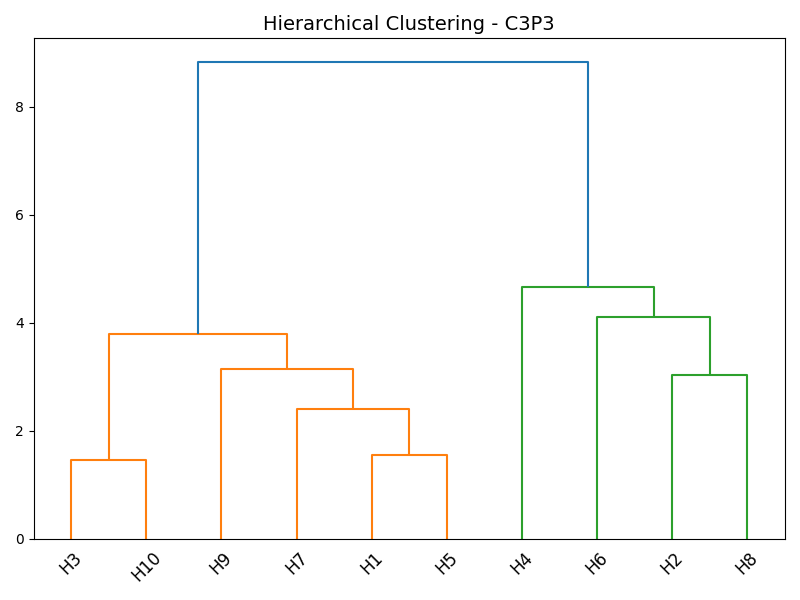}
    \caption*{Controls}
  \end{minipage}
  \hspace{0.01\textwidth}
  \begin{minipage}{0.4\textwidth}
    \centering
    \includegraphics[width=\linewidth, height=4cm]{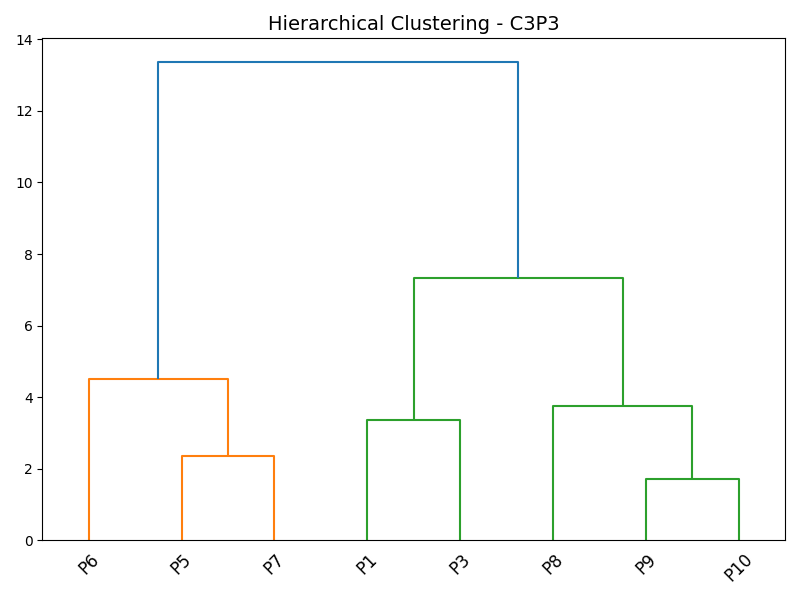}
    \caption*{Cases}
  \end{minipage}
  \caption{Dendrograms for clustering subjects based on the time series data derived from the channel C3P3 for the control and case groups.}
  \label{Heterogeneity:control}
\end{figure}

We began with a subject-heterogeneity analysis for control-case sensor time series data using hierarchical clustering. We took the C3P3 channel as an example. The subject's EEG time series in the C3P3 and other channels were extracted from the same time period across all subjects. We calculated the feature vector of marginal mean, variance and the autocorrelations at the first $50$ lags for each subject's EEG time series. Subject-heterogeneity in the C3P3 and other channels was assessed through hierarchical clustering of these feature vectors, where the Euclidean distance based average linkage method was adopted to generate dendrograms. The results for the C3P3 and other channels, displayed in Figure~\ref{Heterogeneity:control} and Figure~ \ref{Heterogeneity:controlb} in Section \ref{Section:S8},
 the Online Supplementary Material, showed evidence of subject-heterogeneity in the case and control groups.  These findings gave us a rationale to fit the NccDCM individually for each subject to capture subject-specific dynamics. This approach allowed us to make group-level inferences while accounting for individual variability through a mixed-effects model.

In the following analysis, we focused on these $8$ patients against $10$ controls. See a flowchart of the data analysis in Figure~\ref{scalp}.

\begin{figure}[htbp]
    \centering
    \begin{minipage}[b]{0.45\textwidth}
        \centering
        \includegraphics[height = 5cm,width=\linewidth]{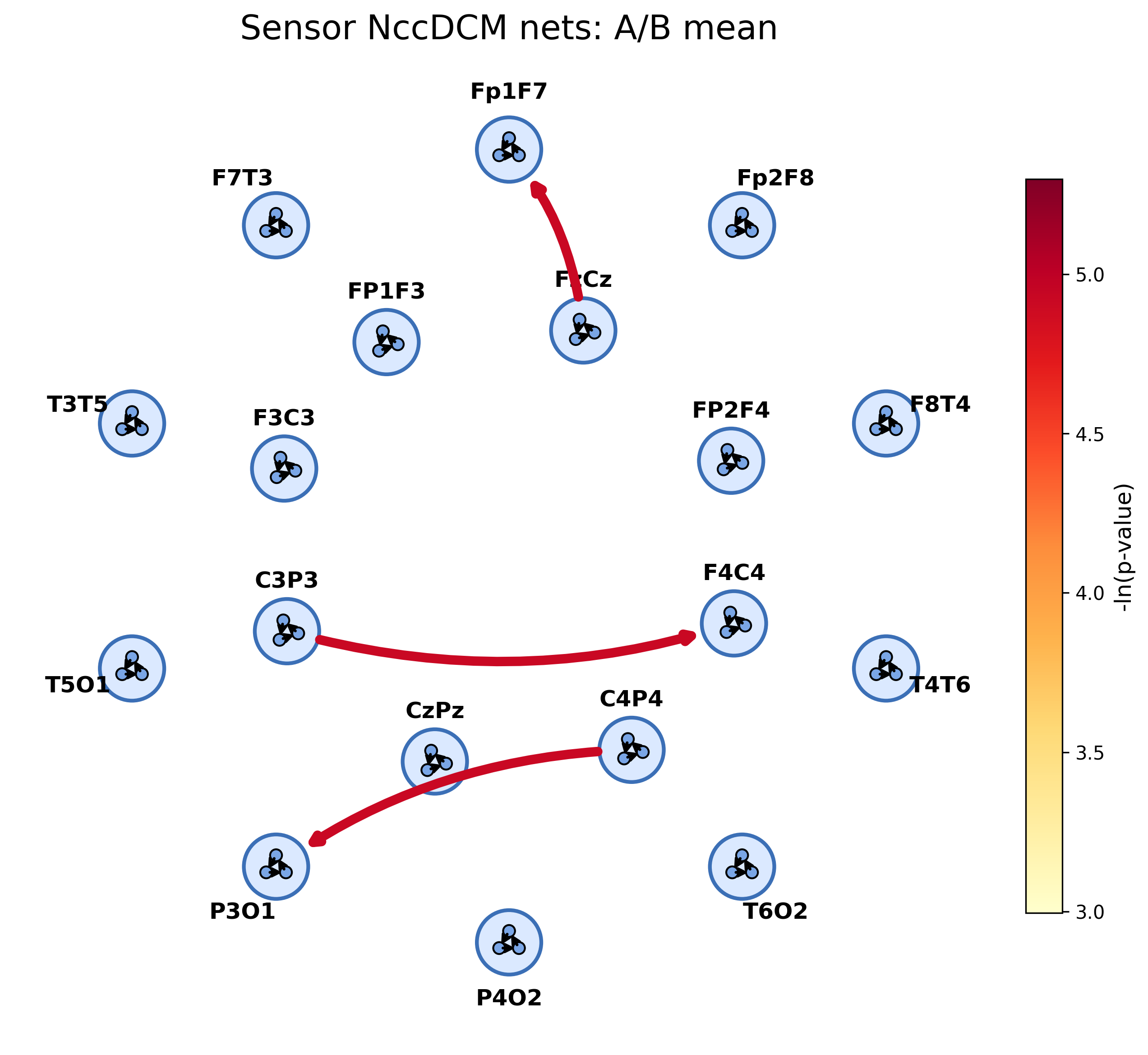}
        \tiny (a)
    \end{minipage}
    \hfill
    \begin{minipage}[b]{0.45\textwidth}
        \centering
        \includegraphics[height = 5cm,width=\linewidth]{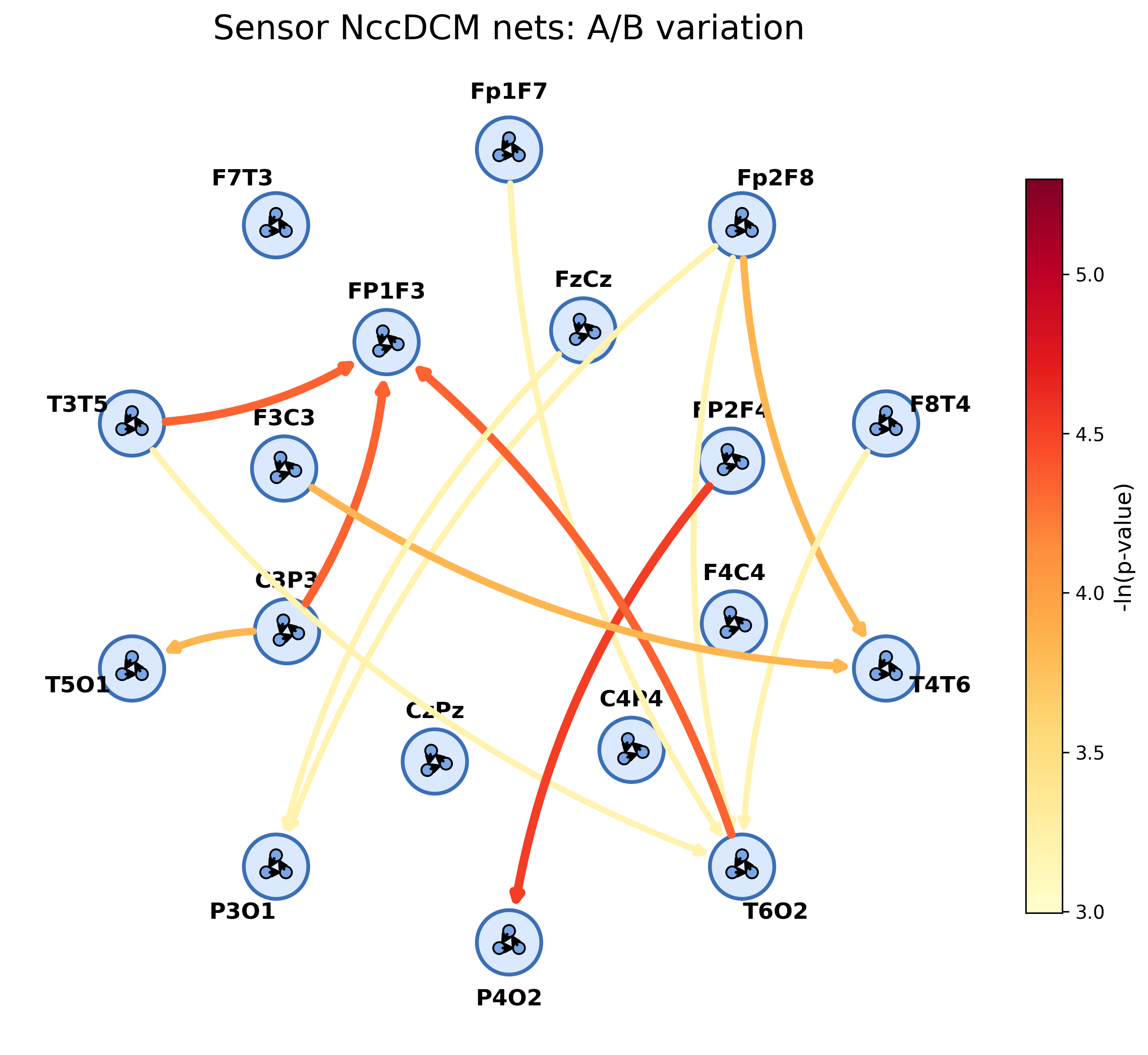}
        \tiny (b)
    \end{minipage}
    \vspace{0.4cm}

    \begin{minipage}[b]{0.45\textwidth}
        \centering
        \includegraphics[height = 5cm,width=\linewidth]{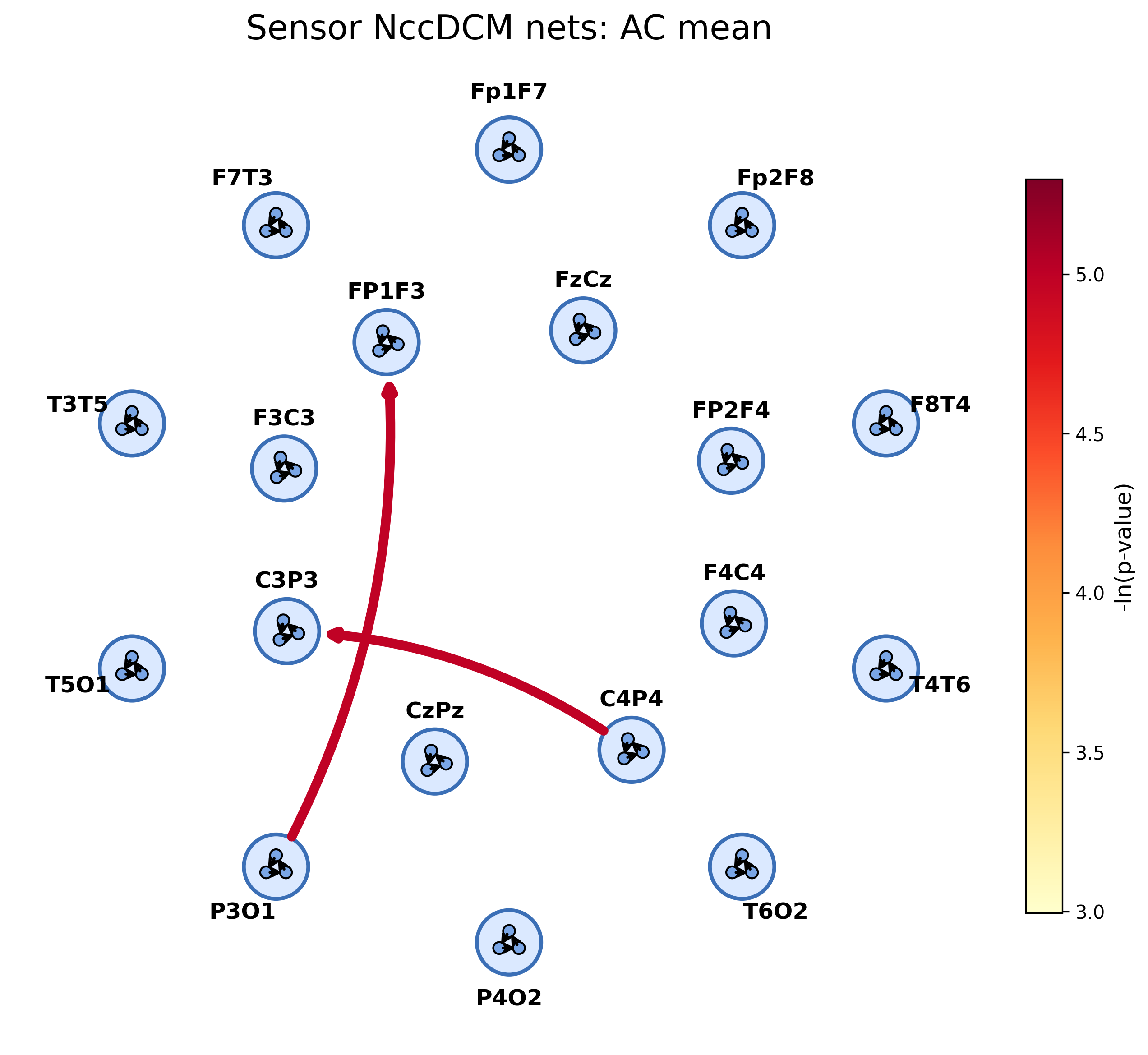}
        \tiny (c)
    \end{minipage}
    \hfill
    \begin{minipage}[b]{0.45\textwidth}
        \centering
        \includegraphics[height = 5cm,width=\linewidth]{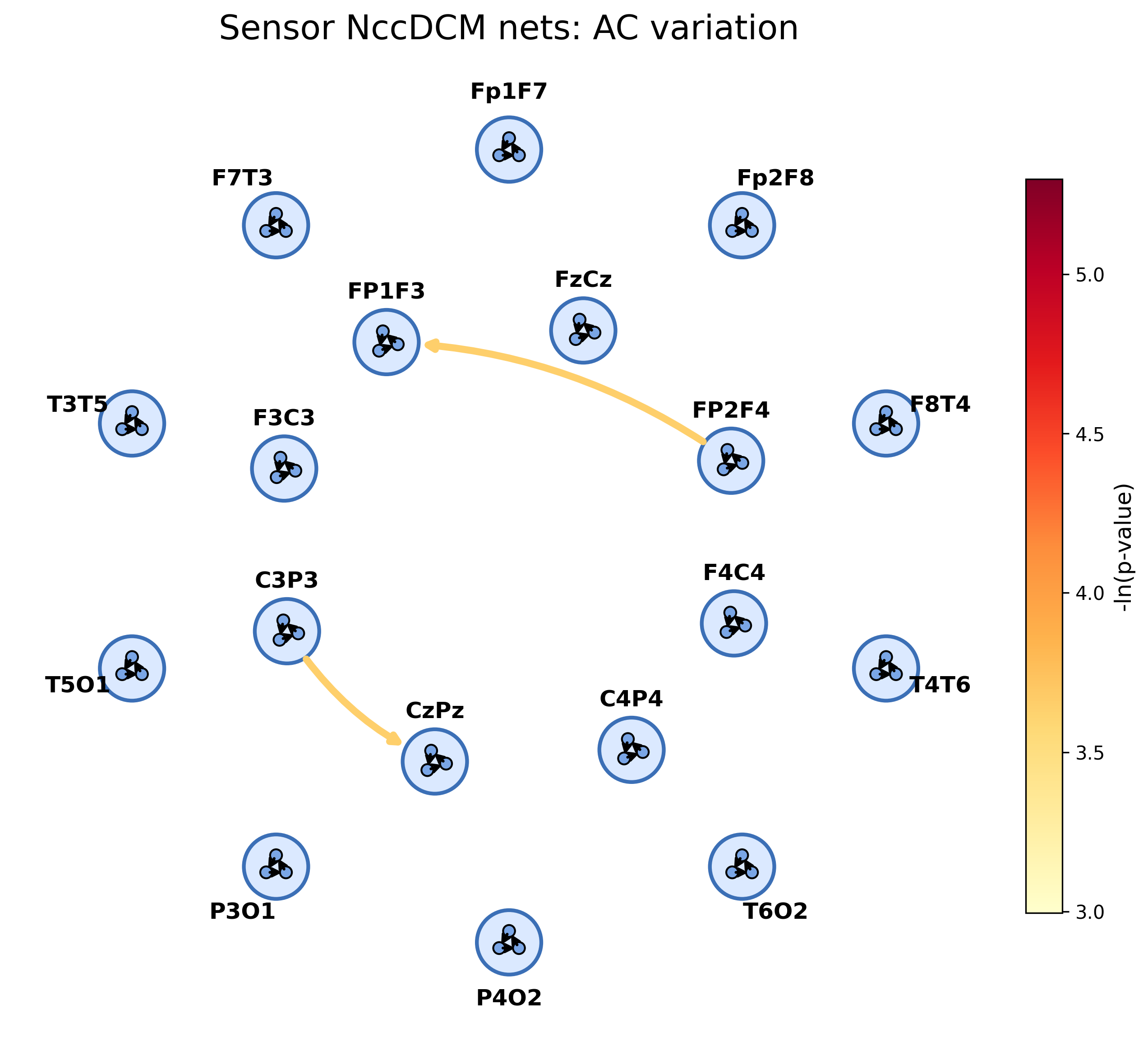}
        \tiny (d)
    \end{minipage}

    \begin{minipage}[b]{0.45\textwidth}
        \centering
        \includegraphics[height = 5cm,width=\linewidth]{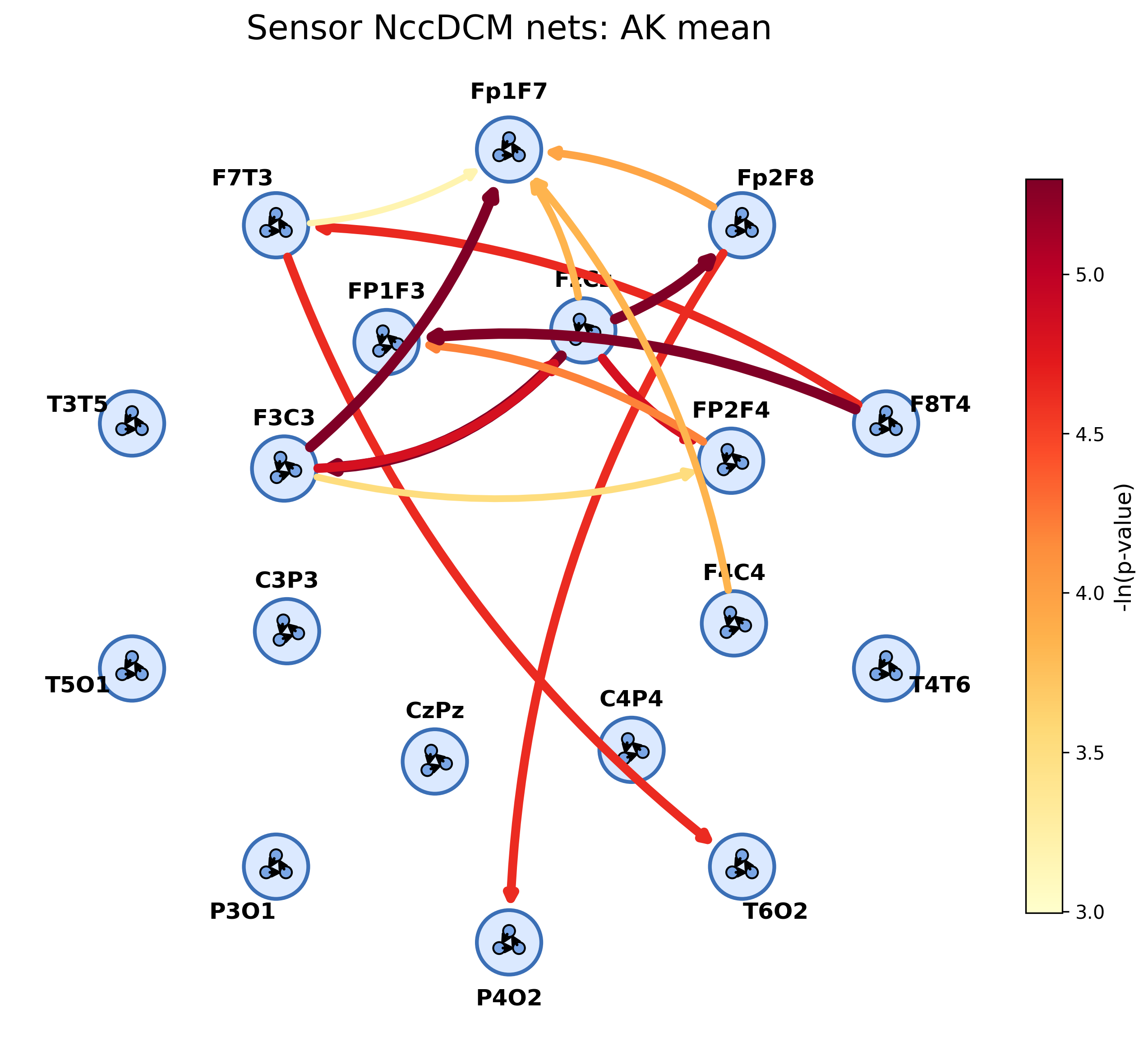}
        \tiny (e)
    \end{minipage}
    \hfill
    \begin{minipage}[b]{0.45\textwidth}
        \centering
        \includegraphics[height = 5cm,width=\linewidth]{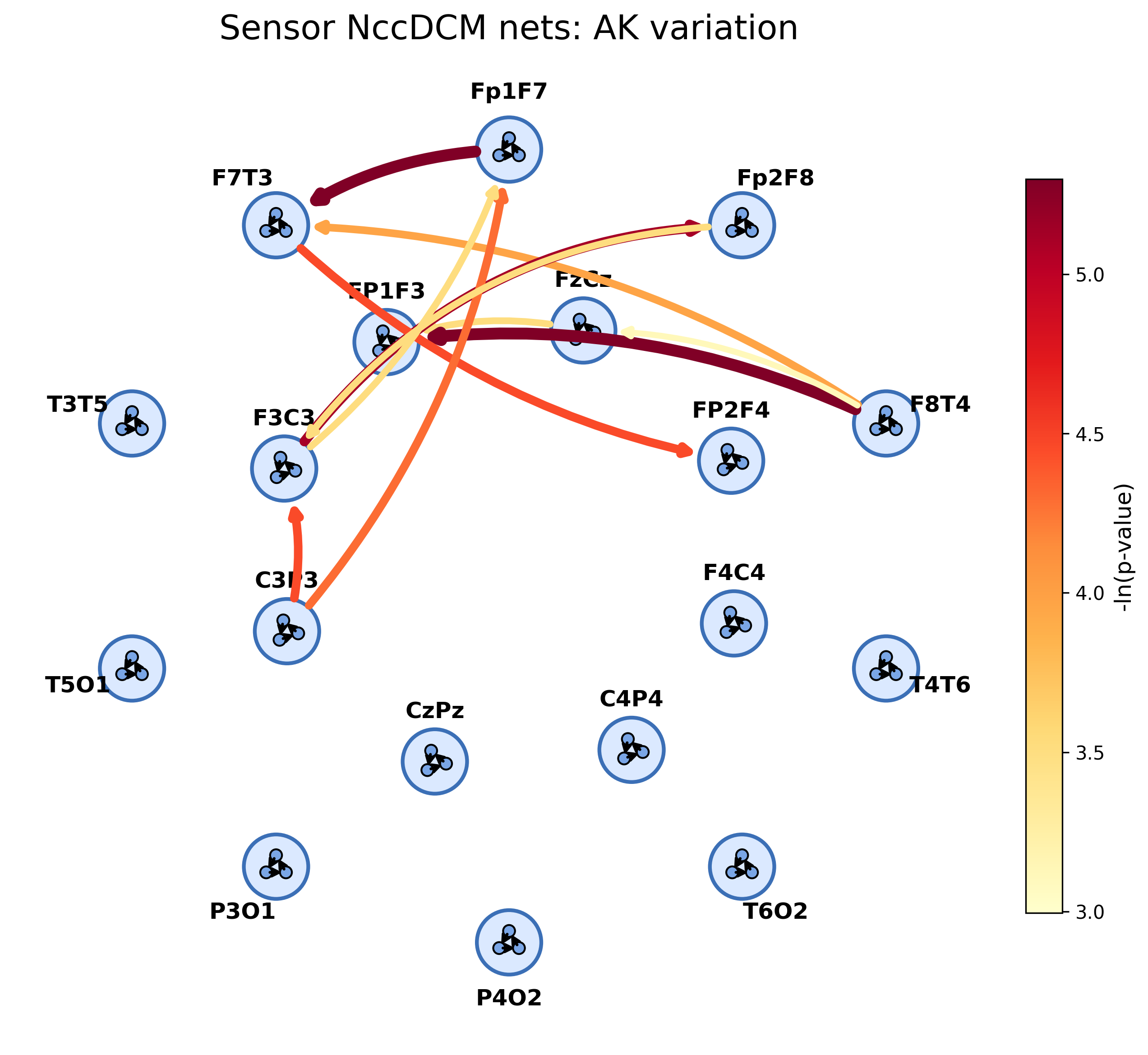}
        \tiny (f)
    \end{minipage}
    \caption{Sensor differential dynamic causal nets in mean and variation for parameters after the adjustments for multiple testing. (a) and (b): $B_{A}=A/B.$  (c) and (d): $ C_{A}=AC$. (e) and (f): $ K_{A}=AK$. The colour bars are in the scale of $-\ln(\mbox{p-value})$, where $\mbox{p-value}$ is  BH-adjusted. }
    \label{Fig:effetivenetwork}
\end{figure}

\begin{figure}[htbp]
    \centering
    \begin{minipage}[b]{0.45\textwidth}
        \centering
        \includegraphics[height = 5cm,width=\linewidth]{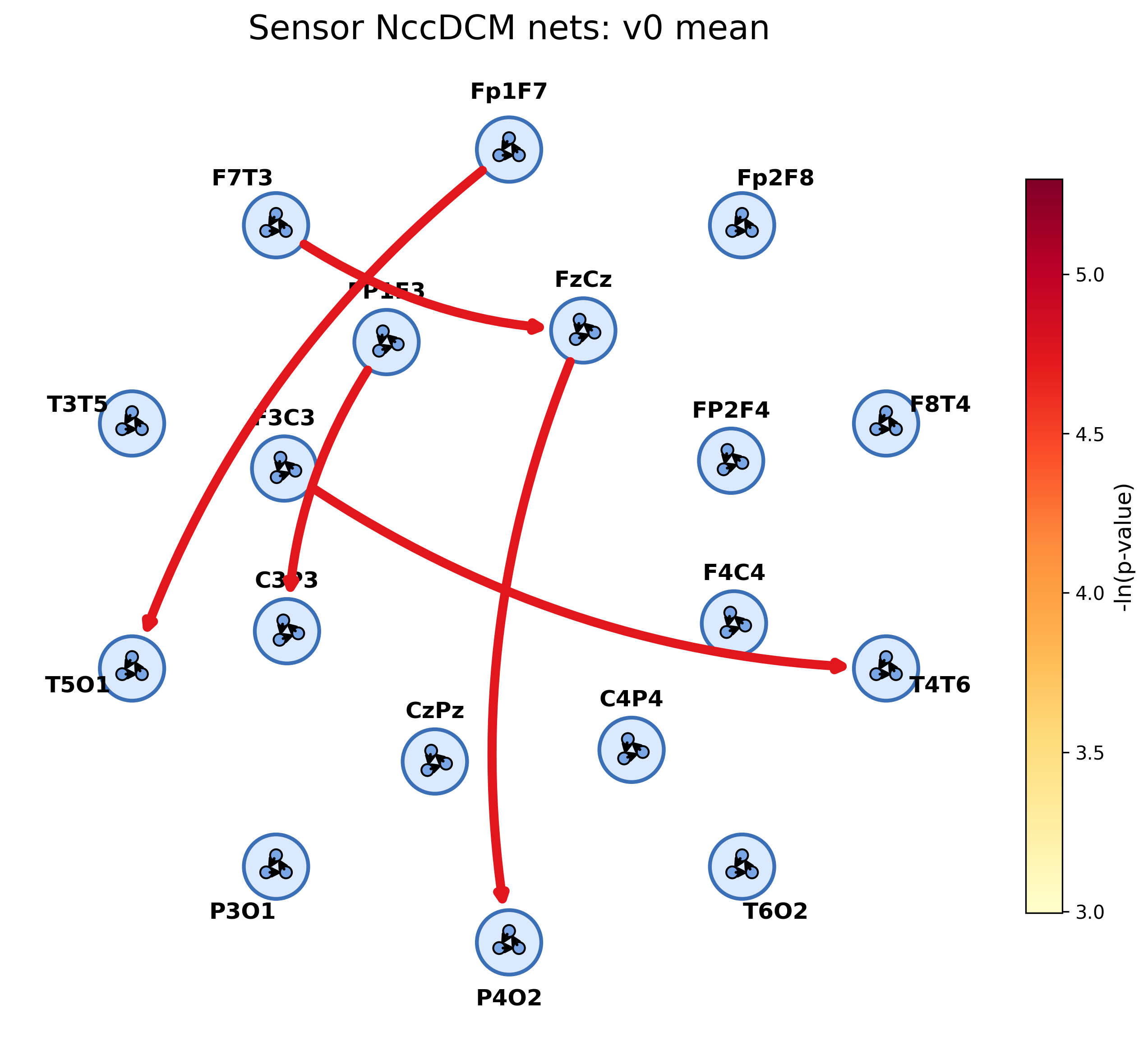}
        \tiny (a)
    \end{minipage}
\vspace{0.5cm}
    \begin{minipage}[b]{0.45\textwidth}
        \centering
        \includegraphics[height = 5cm,width=\linewidth]{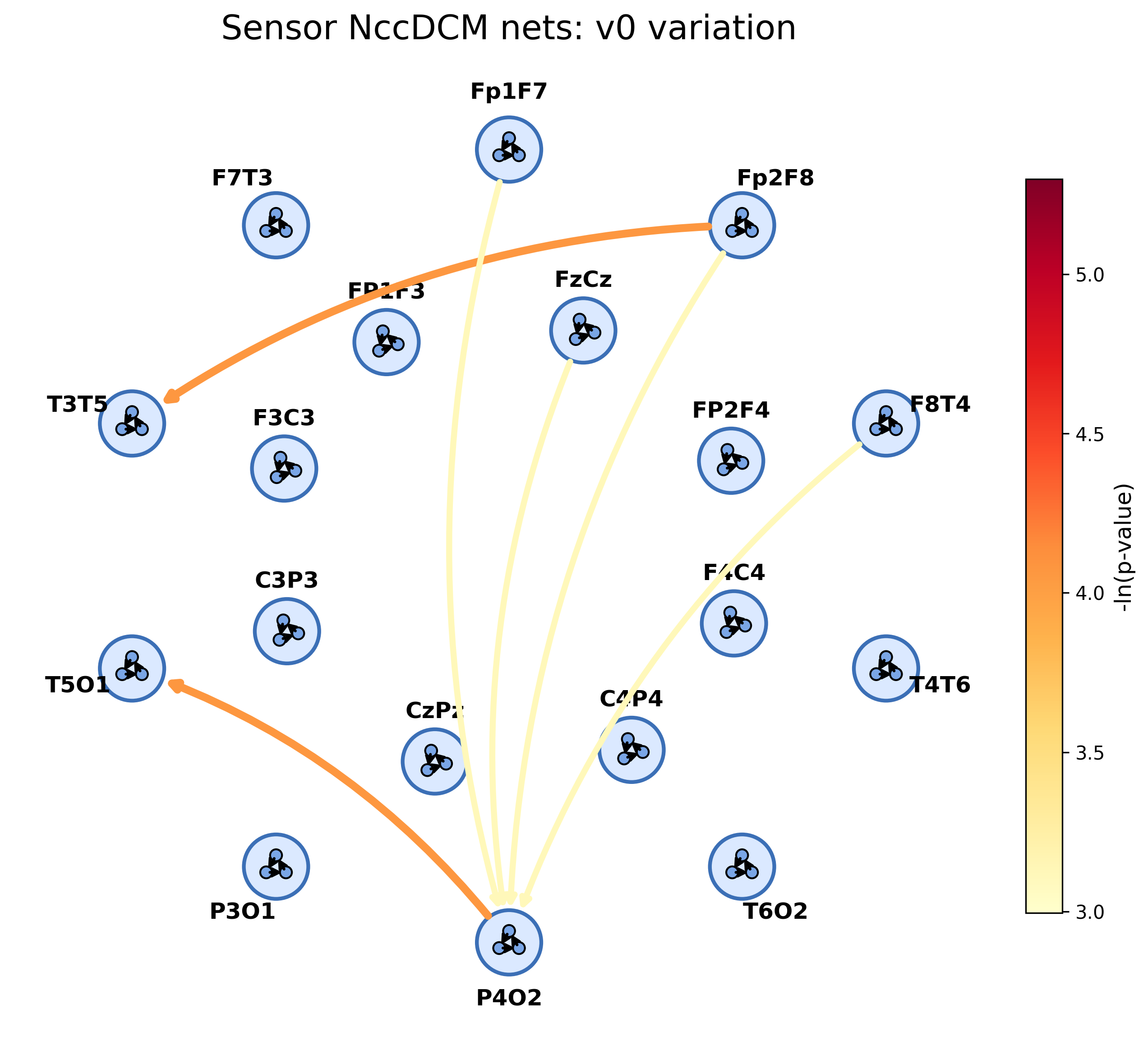}
        \tiny (b)
    \end{minipage}
    \hfill

    \begin{minipage}[b]{0.45\textwidth}
        \centering
        \includegraphics[height = 5cm,width=\linewidth]{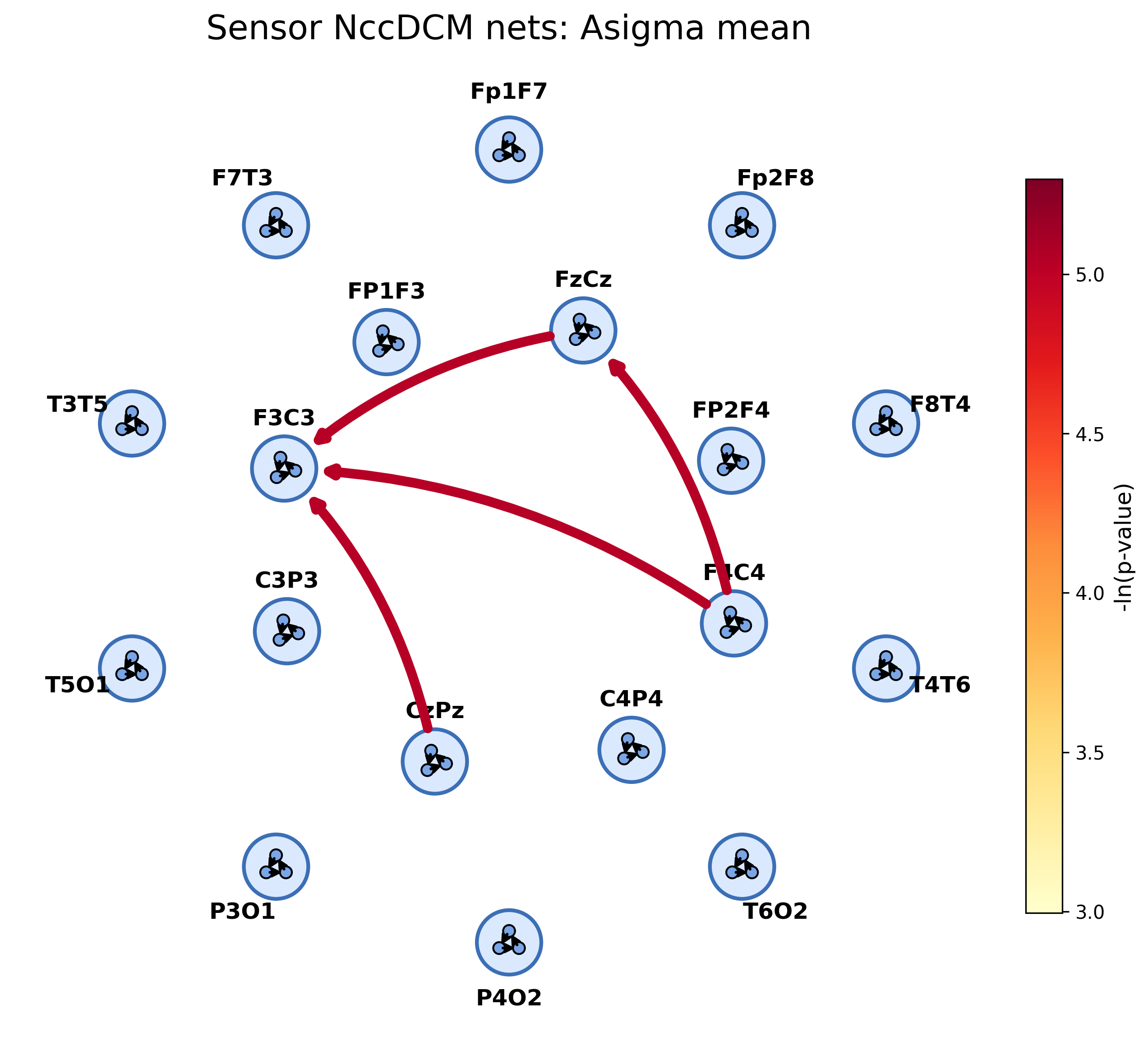}
        \tiny (c)
    \end{minipage}
\vspace{0.5cm}
    \begin{minipage}[b]{0.45\textwidth}
        \centering
  \includegraphics[width=5cm,height=5cm]{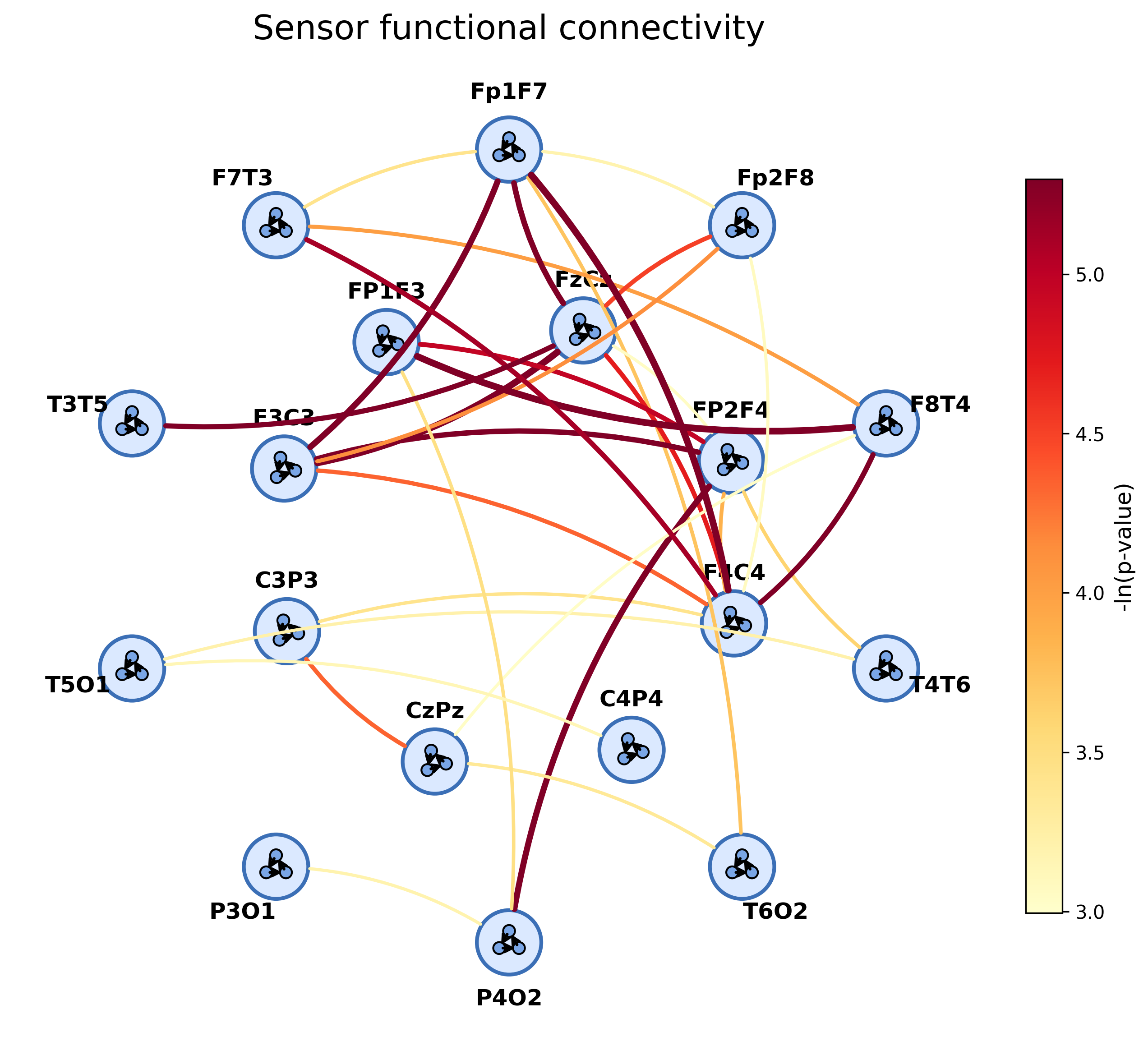}\\
       \tiny (d) 
    \end{minipage}
  \vspace{0.4cm}
    \caption{Sensor differential dynamic causal nets in mean and variation for parameters after the adjustments for multiple testing.   (a) and (b):  $\nu_0$.  (c): $\sigma_{A}=A\sigma$.
(d): Sensor differential causal nets using the maximum lagged correlations-based functional connectivities between case and control groups at the level $5\%$. However, these edges in the nets were not significant at the level of $5\%$ after the adjustments for multiple testing. The colour bars are in the scale of $-\ln(\mbox{p-value})$, where $\mbox{p-value}$ is  BH-adjusted. }
    \label{Fig:effetivenetworkvariation}
\end{figure}

\vspace {0.5cm}
{\bf Differential dynamic causal nets in mean and variation.}
For each pair of channels, we first fit ccDCMs to recorded patients' IED time series and their counterparts in controls respectively, obtaining the estimated values of parameters in the ccDCMs after the corrections of multiple testing and multiple input-effects. We then perform a post-analysis of variance on these estimated parameters to reveal the IED-associated causal patterns in terms of channel excitation-inhibition ratio of synaptic gain, the strength of within-channel connectivity, the strength of between-channel connectivity and the contrasts of synaptic transmission between cases and controls during seizure periods.
 The manhattan distance was used in these post-analyses of variance.  Based on the adjusted $(1-p)$-values, we constructed the adjacency matrix to reflect the degree of difference between the case and control groups. To facilitate the presentation, we retained only the p-values that are less than $95\%$. The resulting differential dynamic causal nets are displayed in Figures \ref{Fig:effetivenetwork} and \ref{Fig:effetivenetworkvariation}. In the nets, the directionality was indicated by arrows from an input channel to an output channel. The value of $-\ln(p)$ increased from $3$ to $5.5$ as colors are ranging from yellow to dark red to yellow. From  Figure \ref{Fig:effetivenetwork}, we revealed the following causal nets:

\begin{itemize}
  \item $B_{A}$: In Figure \ref{Fig:effetivenetwork}(a)(b), there were $3$ and $6$ channel pairs to be claimed to differ cases from controls significantly in mean and variance respectively, where the front channels FP1F7, FP1F3, FP2F8 and the back channels C3P3, P3O1 were involved. 
  \item $C_{A}$: In Figure \ref{Fig:effetivenetwork} (c)(d), there were $2$ and $2$  channel pairs to be found to have significant differences between cases and controls in mean and variance respectively. Notably, these differences were most pronounced in channel pairs where prefrontal regions served as output channels, including the front channels FP1F3 and FP2F4 and the back channel P3O1.
  \item $K_{A}$: In Figure \ref{Fig:effetivenetwork}(e)(f), $16$ and $10$ channel pairs exhibited significant differences between cases and controls in mean and variance respectively. These differences were predominantly observed in the frontal regions,  including FP1F7, FP1F3, FP2F8, FZCZ and FP2F4.
  \item $\nu_0 $:  In Figure \ref{Fig:effetivenetworkvariation}(a)(b), there were $11$ channel pairs which showed significant differences between cases and controls. The most pronounced differences were in the prefrontal and central-temporal-parietal region, where F7T3, FZCZ, FP1F7, FP1F3, C3P3, T5O1, P4O2 and T4T6 were involved.

\item $\sigma_A$: In Figure \ref{Fig:effetivenetworkvariation}(c), there were $4$ channel pairs to be significant, where the front channels F4C4, FZCZ and F3C3 were involved.
\end{itemize}

In summary, for the parameters  $B_{A}, C_{A} , K_{A}$, $\nu_0$ and $\sigma_A$, the  most pronounced differences between the case and control groups were in channel pairs covering frontal, temporal and back brain regions.
 In particular,  FP1F7, FP1F3, FP2F8, FZCZ and FP2F4 are related to the left and right dorsolateral and ventrolateral prefrontal cortices. This finding highlighted the prominent role of these prefrontal areas in driving network-level alterations.  An altered excitation/inhibition (E/I) balance, often involving decreased inhibition, has been identified as a robust biomarker in various neurodegenerative diseases such as Alzheimer's disease (AD) and frontotemporal dementia (FTD). Our analysis suggested a mechanism that epilepsy shared with other neurological diseases that changes in the inhibitory connections in the frontal cortex disrupted normal circuit operations and brain network dynamics.  The above findings centred on FP1F7, FP1F3, FP2F8 and FP2F4. Unfortunately, these derivations could also carry most eye-movement and blink artefacts, more so in children. These artefacts might not be removed completely by independent component analysis in preprocessing EEG recordings. EEG channels might not map onto individual cortical areas very well due to volume conduction effectsl \citep{vandenBroek1998}.  Note also that FP1F3 and F3C3 share F3, whose outputs are coupled by construction with opposite signs. Therefore,  edges between adjacent channels could be partially caused by the montage design. In general, the bipolar montage cancelled signals with low spatial-frequency \citep{Zaveri2006}. This could make sign-dependent claims on $B_A$ difficult in the sensor space. This explained why the $B_A$ diifferential dynamic causal net in Figure~\ref{Fig:effetivenetwork}(a) is very sparse. In contrast,
the source-space connectivity analysis in the following subsection could significantly improve the analysis \citep{Nguyen2023}, where we revealed not very sparse differential dynamic causal nets in the prefrontal, temporal and posterior cortical regions.

\begin{figure}
\centering
  \includegraphics[width=10cm,height=7cm]{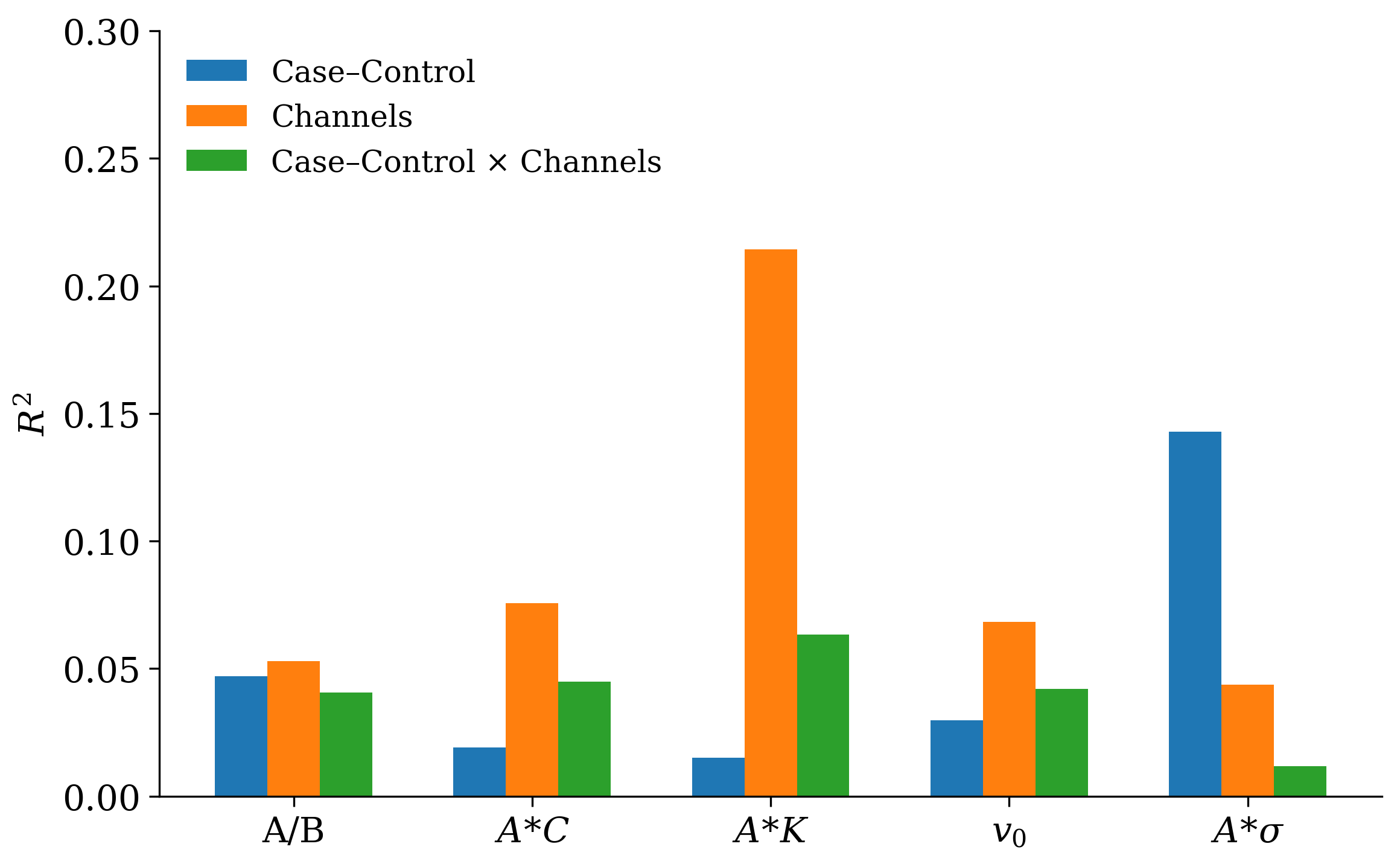}\\
  \caption{The percentages of uncertainty in the estimated sensor parameters which was explained by the NccDCM-based mixed-effects models, $R^2$. See \cite{Anderson2017} for the definition of $R^2$.
 Here, $B_A=A/B, C_A=AC, K_A=AK$ and $\sigma_A=A\sigma$.}
 \label{R2ANOVA}
\end{figure}

{\bf Post ANOVA tables for the NccDCM.}
We treated the fitting of the NccDCM to each subject as a transformation of its EEG measurements into the values of the estimated network parameters, which inherited the uncertainty in the original data.
We used the \texttt{adonis2} function in the R software to perform post multivariate ANOVA on these values and to test the significance of case-control and channels as two factors.   The numerical results are displayed in Figure \ref{R2ANOVA} and in Tables \ref{tab:permanova_AB_SENSOR_STANDARD} to \ref{tab:permanova_Asigma_SENSOR_STANDARD} in the Section \ref{Section:S7},
 the Online Supplementary Material. The results indicated that the identified dynamic causal nets were highly significant for each parameter. The percentages of variations in the estimates  explained by the estimated dynamic causal nets, due to Case-Control factor plus its interaction with Channels, for $B_{A}$,  $C_{A}$ ,  $K_{A}$, $\nu_0$
and $\sigma_A$ were respectively $14\%$, $14\%$, $29\%$, $14\%$ and $19.9\%$  in means and $14\%$, $25.9\%$, $44\%$, $17.6\%$ and $0\%$ in variations. The percentages of variations in the estimated parameters, $R^2$ values confirmed the significant heterogeneity of the network parameters across the channels. For example, around 6.3\% (p-value 0.000167)
 of the variations in the connection rate $K_A$ between regions was due to the channel difference.

{\bf Lagged correlation-based functional connectivity analysis.}
To compare with lagged correlation analysis, we calculated lagged correlation coefficients using the method in S5, 
 the Online Supplementary Material.
The analysis revealed $31$ channel pairs with
$p$-values less than $0.05$ in testing the difference of lagged correlations between the control and case groups as displayed in Figure \ref{Fig:effetivenetworkvariation}(d). The channel pairs (FP2F4, P4O2), (FP2F4, F3C3), (F4C4, FP1F7), (FP1F3, F8T4), (F3C3, FP1F7), (F3C3, FZCZ), and (FP1F7, FZCZ) with $p$-values were smaller than $0.005$. Among them,  channel pairs (F4C4, FP1F7), (F3C3, FP1F7) and (F3C3, FZCZ) were also identified in the $K_{A}$ based NccDCM analysis. However, unlike the previous NccDCM analysis, there were no significant pairs left at $5\%$ after the BH correction for multiple testing. This showed an advantage of the NccDCM analysis over the conventional functional connectivity analysis. The similar advantage was also found in source level but omitted.

\subsection{Source differential dynamic causal nets}

We first applied the bipolar source localisation to the Epilepsy EEG dataset to obtain an estimated source time series for each of $68$ D-K regions. We fit a NccDCM with $68$ nodes to the estimated source time series data. The results, displayed in Figures~\ref{Fig:ACv0}$\sim$\ref{R2ANOVASource} below. 
\begin{itemize}
\item
The $B_A$ differential causal nets in mean and variation contain a mixture of temporal, parietal, frontal, occipital regions, on both sides of the brain. The E/I ratio maintains the stability, precise neuronal coding and information processing in neural circuits. The changes of the E/I ratios, due to abnormal input-to-output relationships between these regions, are linked to neurological disruptions generated by epilepsy.

In \cite{Yang2023}, significant increases were observed in frequency-specific correlations between the anterior cingulate cortex and the anterior insular cortex, as well as in the excitation/inhibition ratio during seizures.
The cortical regions identified in the present causal nets were previously reported as epilepsy-related structural abnormalities in the following brain network systems \citep{Whelan2018,Lariviere2022}:
 Default Mode Network (DMN) (Posterior cingulate, middle temporal, inferior parietal, entorhinal/medial temporal regions and temporal pole);
 Frontoparietal/executive control network (Precuneus, superior parietal, inferior parietal, caudal middle frontal, parts of supramarginal and inferior frontal cortex);
Dorsal attention network (Superior parietal, inferior parietal and frontal control regions); Sensorimotor network (Precentral and postcentral);
 Visual network (Pericalcarine, cuneus,  lingual, fusiform and lateral occipital); Auditory network (Transverse temporal, banks of superior temporal sulcus, middle temporal, supramarginal and pars orbitalis); Cognitive-control systems (Anterior cingulate, orbitofrontal cortex and frontal regions).
The fact that these nets involve regions belonging to the DMN, attention/executive,  sensorimotor, visual and auditory systems could indicate alterations in a distributed network-level, rather than simply abnormalities in a single functional system.
For example,  the insula is anatomically concealed, so as physiologically, with a cryptic role in epilepsy while
entorhinal cortex is particularly relevant to temporal-lobe epilepsy. But the other regions can reflect propagation networks, secondary effects, or group-level cortical changes rather than the seizure-onset zone.

\item $C_A$ differential causal nets show some fairly coherent epilepsy-network alternation patterns  related to the following network systems:
 Medial temporal (Entorhinal, parahippocampal, temporal pole and fusiform); Temporo-parietal (Superior temporal, supramarginal , superior parietal);
Cingulate/salience (Rostral \& caudal anterior cingulate, isthmus cingulate and insula);
Frontal/executive (Superior frontal,  frontal pole, medial/lateral orbitofrontal and pars opercularis/orbitalis);
 Visual (Lingual, lateral occipital and fusiform).

Compared to healthy controls, in a few structural MRI analyses,  thickness differences of the ipsilateral temporopolar, parahippocampal, entorhinal, and fusiform gyri, contralateral pars triangularis, and bilateral precuneus, superior frontal and caudal middle frontal gyri were observed in left, but not right of the brains with mesial temporal lobe epilepsies and hippocampal sclerosis \citep{Whelan2018}. Note that the parameter $C_A$ scales inter-regional coupling, governing how strongly these regions excite/inhibit one another  as demonstrated in Figure~\ref{scalp}. Compared to the healthy subjects, the changes of input-to-output relationships between these regions significantly modified regional coupling structures in the brain.

\item The $K_A$ differential dynamic causal nets in mean and variation define direct measurements of alterations in a distributed network in terms of input-output coefficients between regions. They were consistent with the findings in a strucure analysis of focal and generalised epilepsy \citep{Lariviere2022}. The regions involved were actually reported in a number of epilepsy studies. For example, \cite{Ogren2018} reported abnormalities involving the insula, frontal pole, lateral orbitofrontal, cingulate, paracentral, postcentral, cuneus and pericalcarine cortices in generalized tonic-clonic seizures. \cite{Sammarra2024} reported cortical-thickness abnormalities in inferior parietal, inferior temporal, middle temporal, temporal pole and isthmus-cingulate regions in mesial temporal lobe epilepsy (MTLE). \cite{Wu2018} also identified several of these regions as altered network features in  temporal lobe epilepsy (TLE).

\item The $\nu_0$ differential causal nets in mean and variation involve the regions which have also well-documented associations with epilepsy-related structural brain abnormalities, particularly in large ENIGMA-Epilepsy studies  \citep{Whelan2018}.

\item The $\sigma_A$  differential causal net in variation implies that  regions lh-caudalmiddlefrontal and rh-caudalanteriorcingulate both  have a  high degree of inputs from other regions. This was consistent with emerging evidences that the left caudal middle frontal gyrus played a role in front networks of the brains with epilepsy \citep{Luo2014}.

\end{itemize}

{\bf Post ANOVA tables for the NccDCM.} In the source space, we performed post multivariate ANOVA on the estimated parameter values  and to test the significance of case-control and regions as two factors. The results, displayed in Figure~\ref{R2ANOVASource} and Tables~\ref{tab:permanova_AB_SOURCE_STANDARD}--\ref{tab:permanova_Asigma_SOURCE_STANDARD}, show the region heterogeneity indicated by the significant interactions of two factors.

\begin{figure}[htbp]
    \centering
    \begin{minipage}[b]{0.45\textwidth}
        \centering
        \includegraphics[height = 5cm,width=\linewidth]{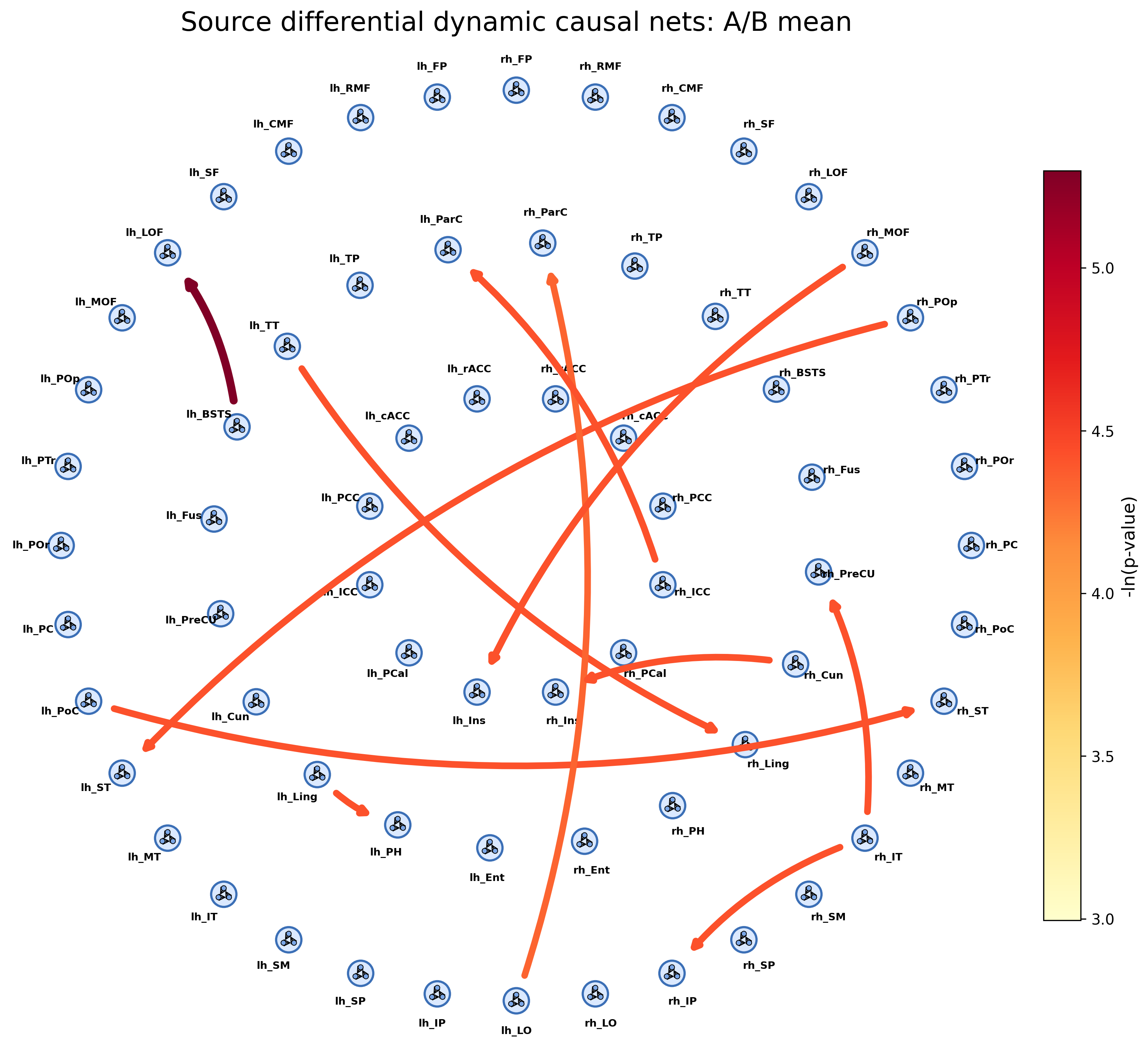}
        \tiny (a)
    \end{minipage}
    \hfill
    \begin{minipage}[b]{0.45\textwidth}
        \centering
        \includegraphics[height = 5cm,width=\linewidth]{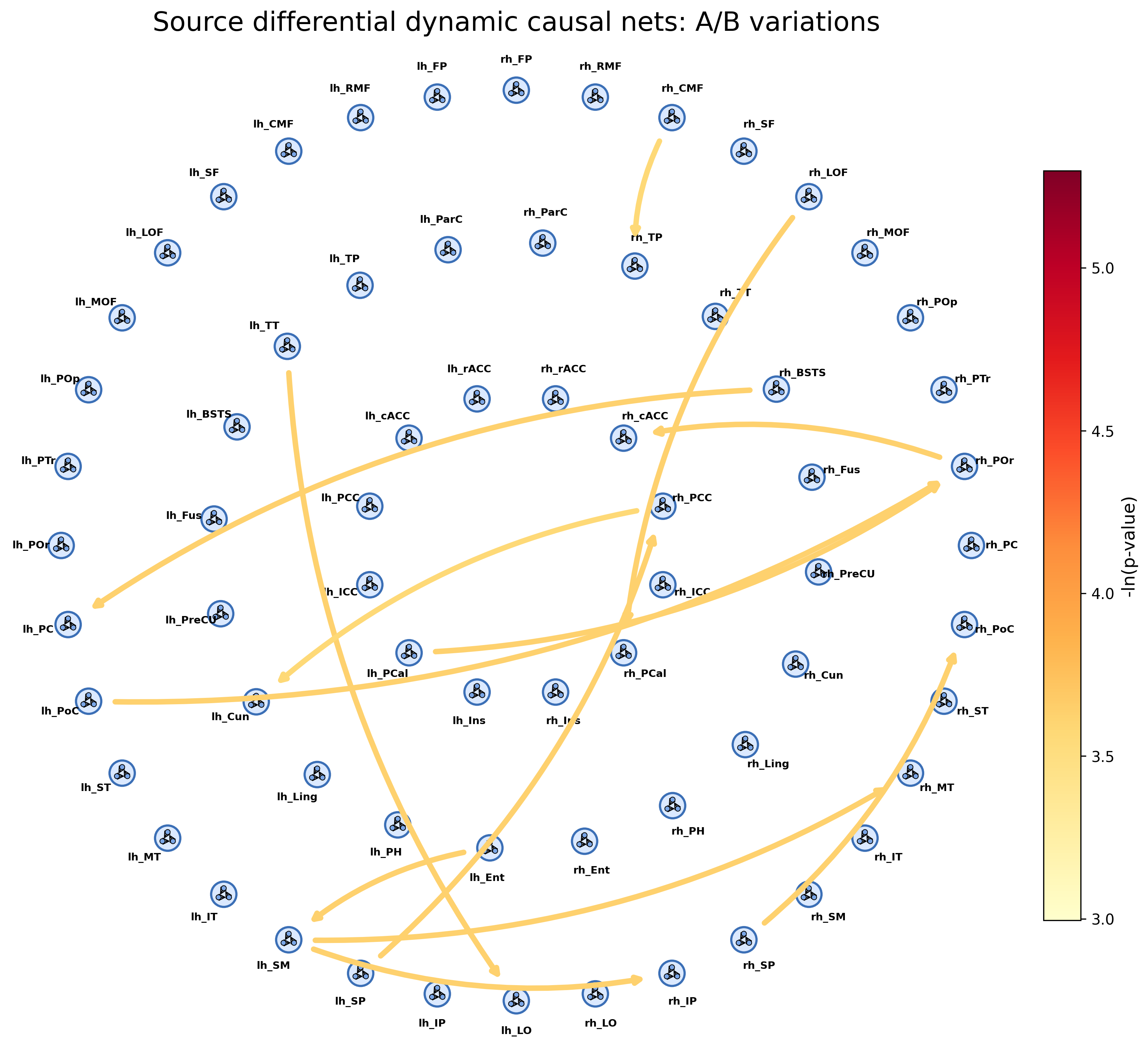}
        \tiny (b)
    \end{minipage}
    \vspace{0.4cm}

    \begin{minipage}[b]{0.45\textwidth}
        \centering
        \includegraphics[height = 6cm,width=\linewidth]{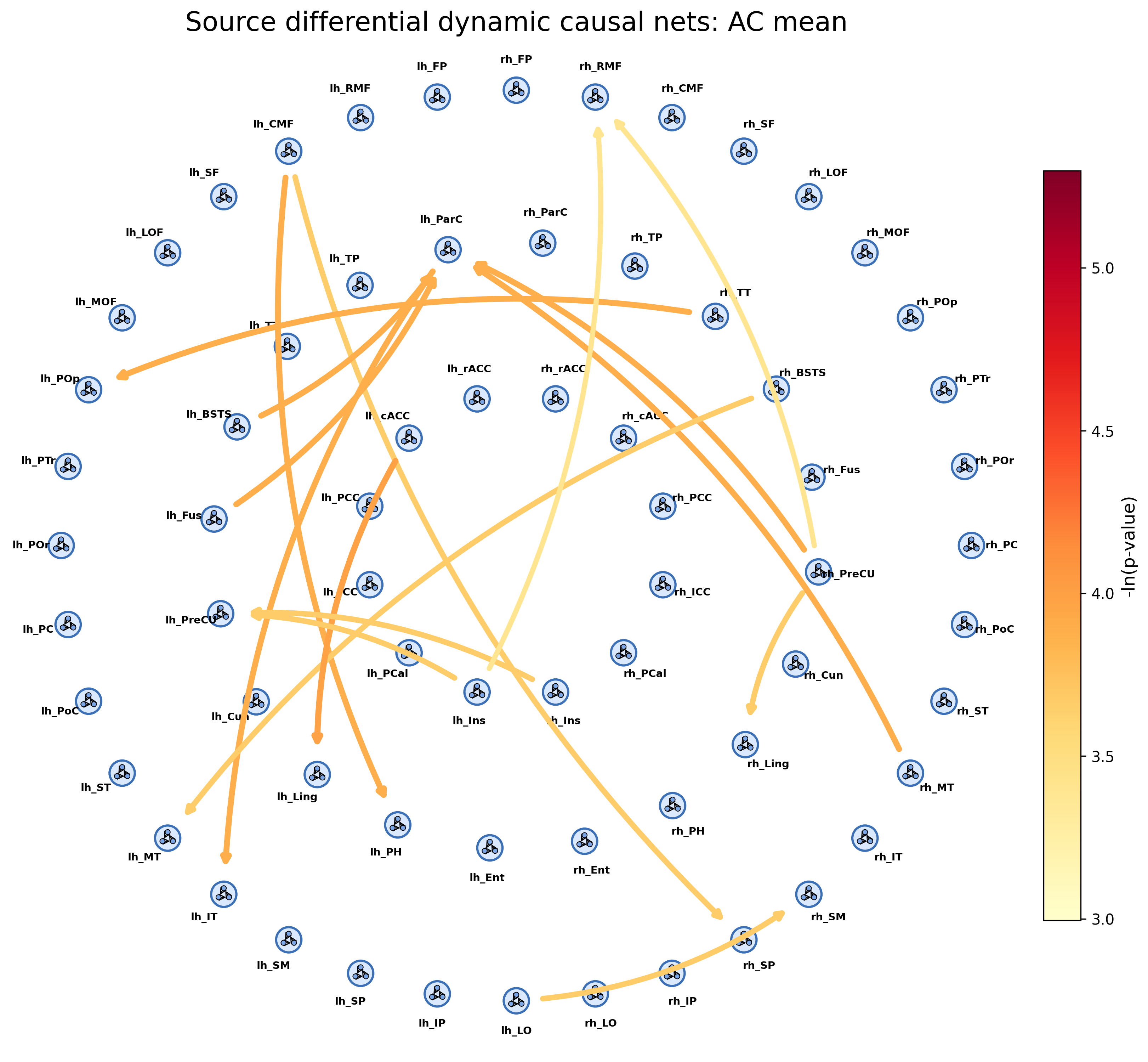}
        \tiny (c)
    \end{minipage}
    \hfill
    \begin{minipage}[b]{0.45\textwidth}
        \centering
        \includegraphics[height = 6cm,width=\linewidth]{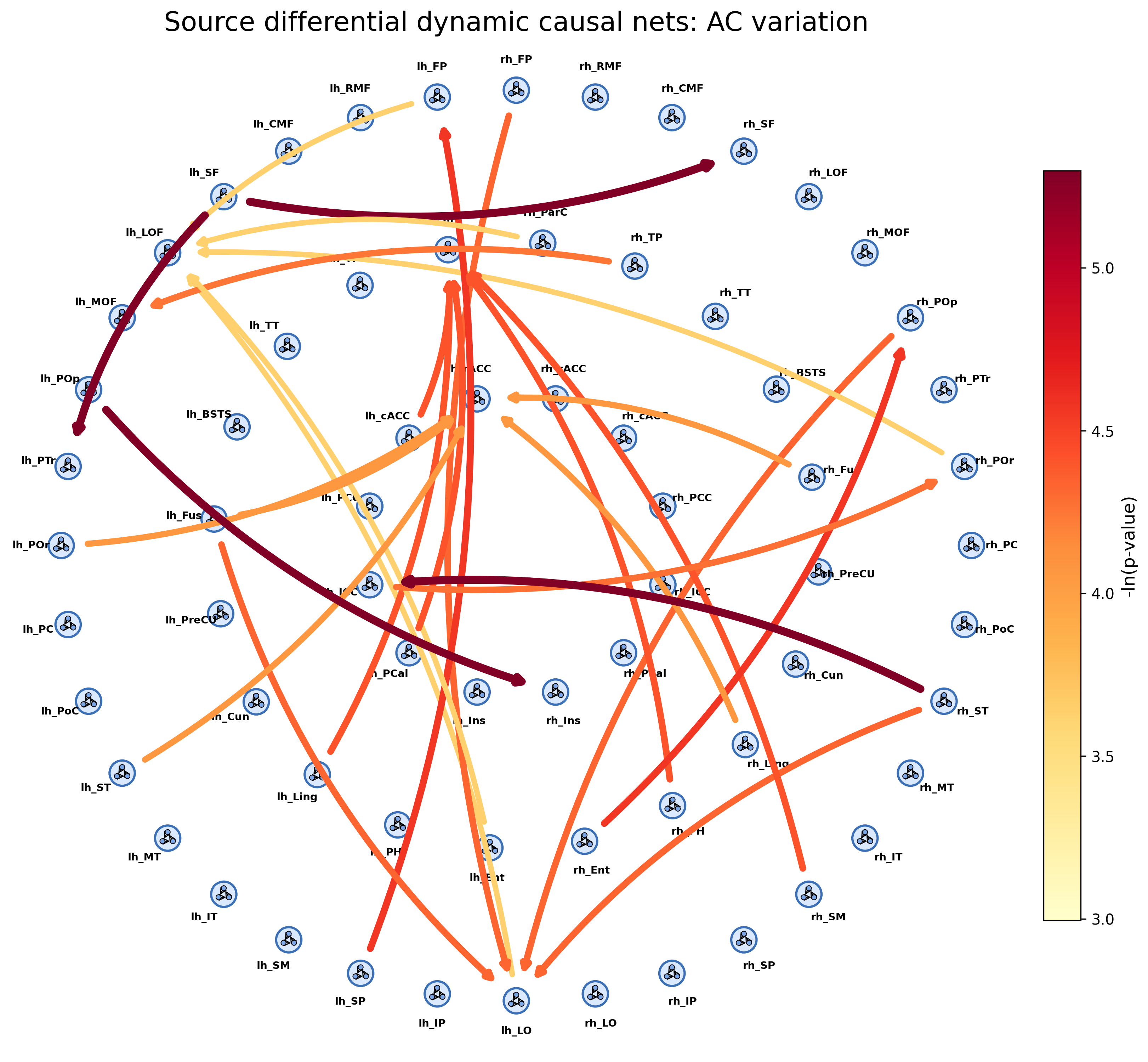}
        \tiny (d)
    \end{minipage}
    \vspace{0.4cm}

    \caption{Source differential dynamic causal nets for parameters after the corrections for multiple testing and for multiple-input effects. (a) and (b): $B_{A}$  mean and variation. (c) and (d):  $C_{A}$  mean and variation. The colour bars are in the scale of $-\ln(\mbox{p-value})$, where $\mbox{p-value}$ is BH-adjusted.}
    \label{Fig:effetivenetworksource}
\end{figure}

\begin{figure}[htbp]
    \centering
    \begin{minipage}[b]{0.45\textwidth}
        \centering
        \includegraphics[height = 5cm,width=\linewidth]{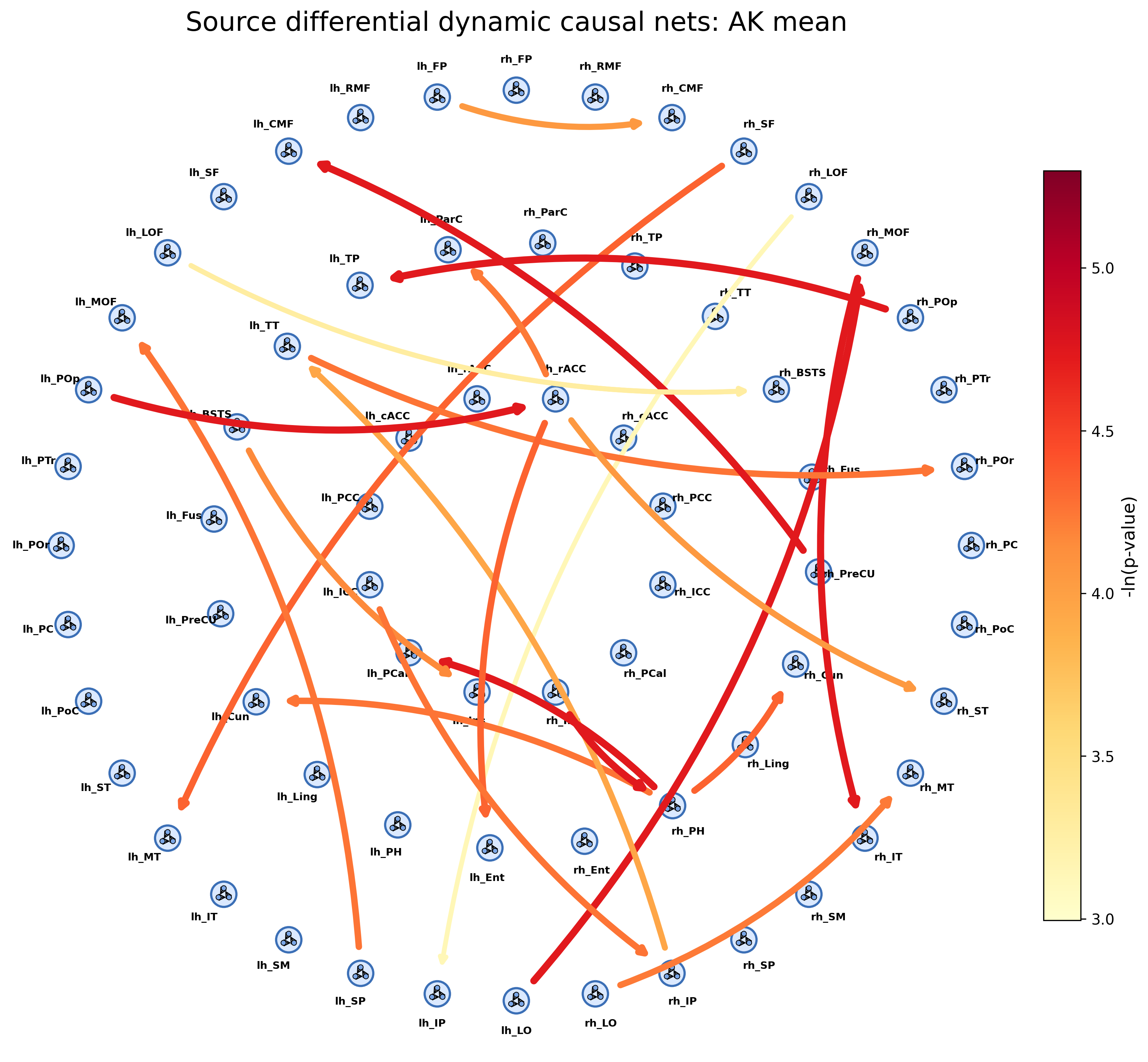}
        \tiny (a)
    \end{minipage}
    \hfill
    \begin{minipage}[b]{0.45\textwidth}
        \centering
        \includegraphics[height = 5cm,width=\linewidth]{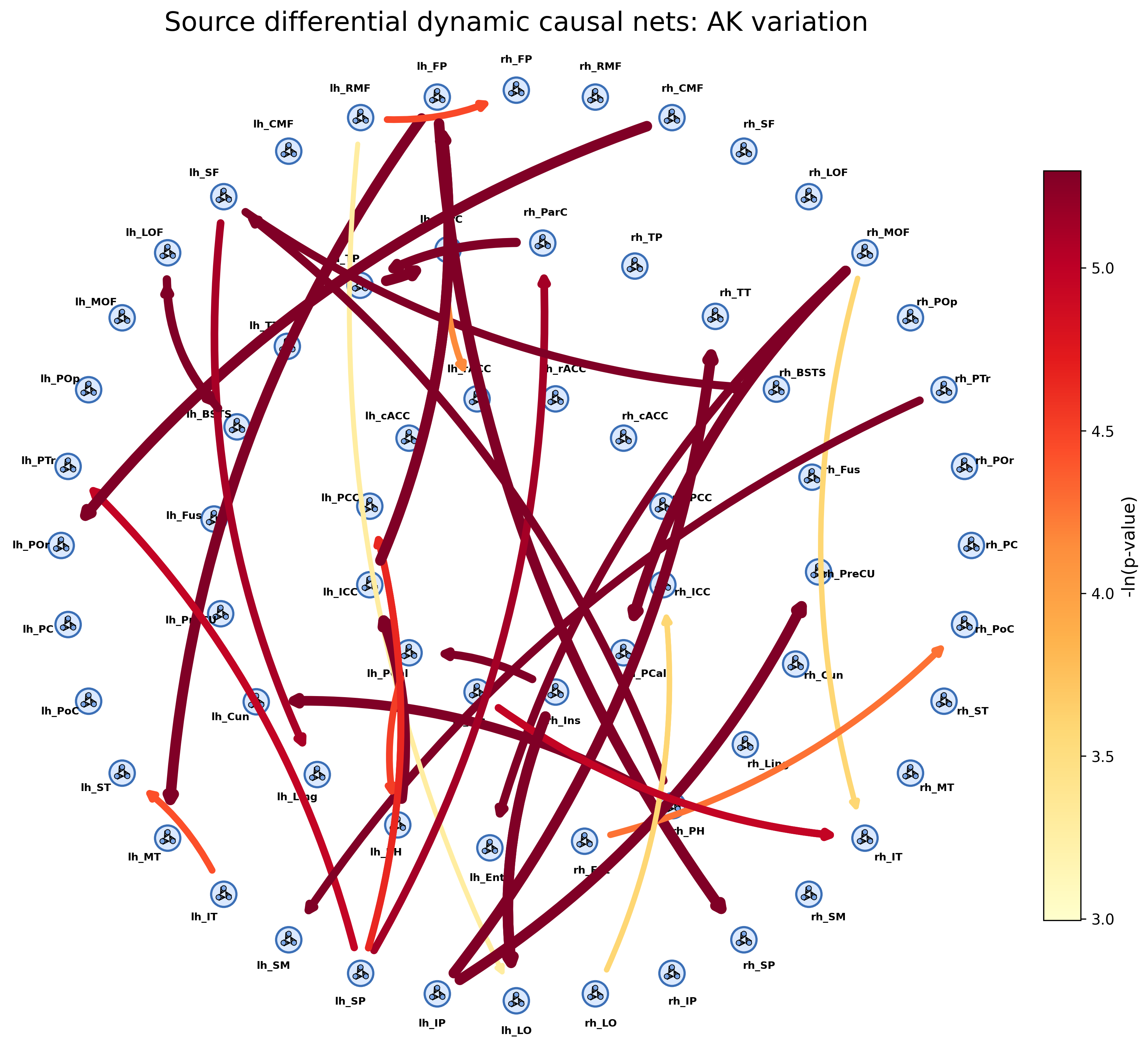}
        \tiny (b)
    \end{minipage}
    \vspace{0.4cm}

    \begin{minipage}[b]{0.45\textwidth}
        \centering
        \includegraphics[height = 6cm,width=\linewidth]{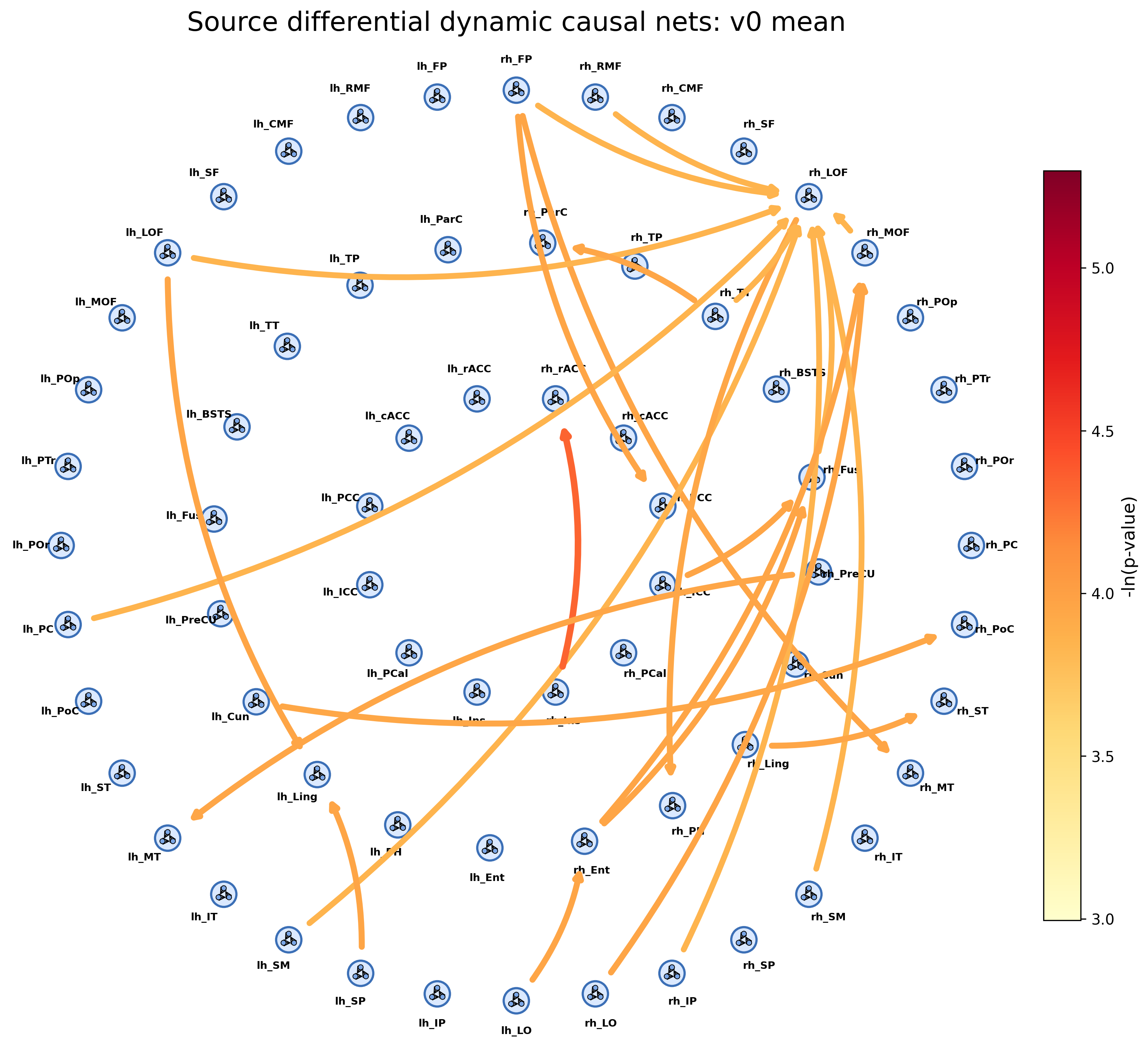}
        \tiny (c)
    \end{minipage}
    \hfill
    \begin{minipage}[b]{0.45\textwidth}
        \centering
        \includegraphics[height = 6cm,width=\linewidth]{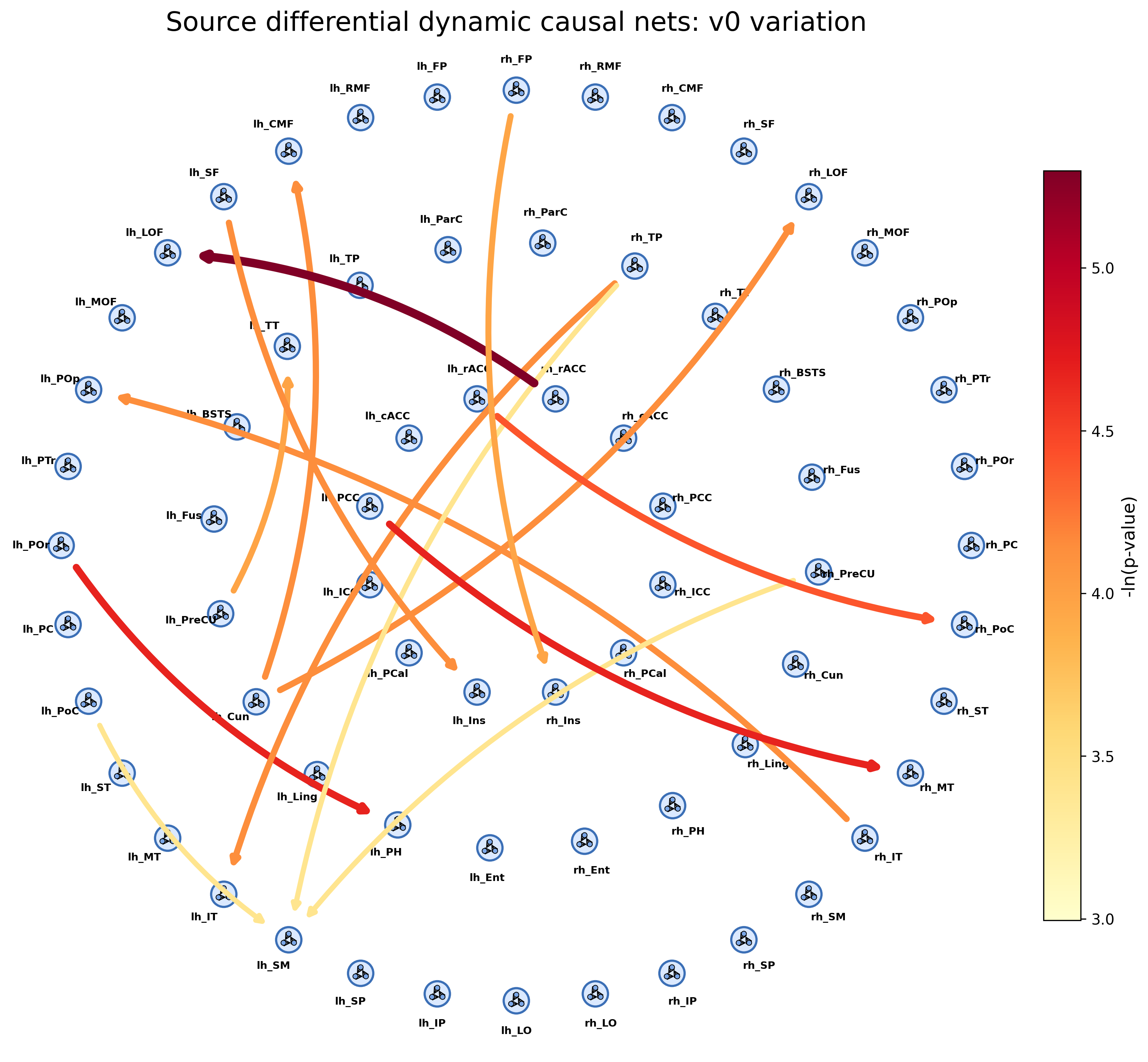}
        \tiny (d)
    \end{minipage}
    \vspace{0.4cm}

   \begin{minipage}[b]{0.45\textwidth}
        \centering
        \includegraphics[height = 5cm,width=\linewidth]{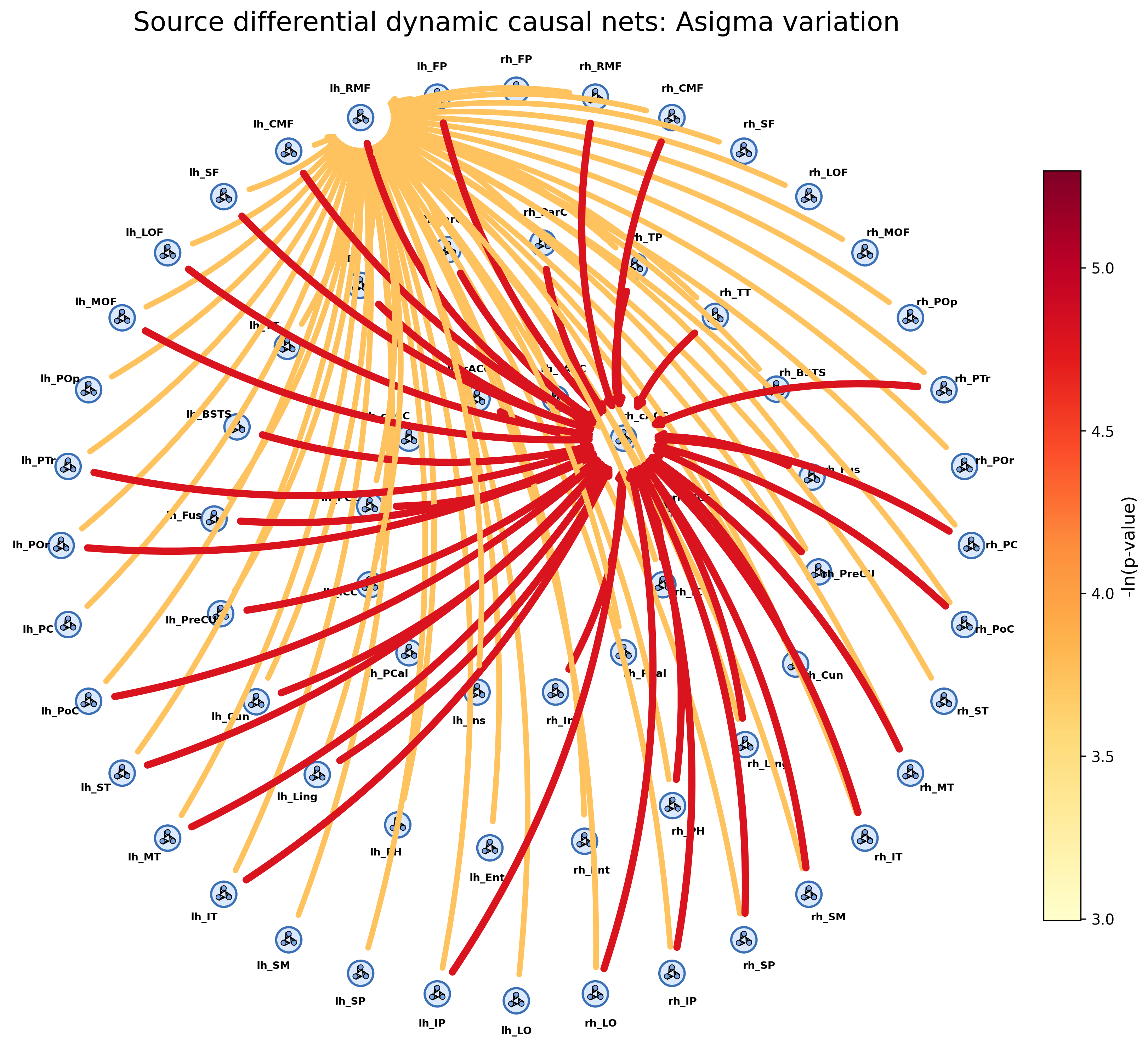}
        \tiny (e)
    \end{minipage}
\hfill
   \begin{minipage}[b]{0.45\textwidth}
        \centering
  \includegraphics[width=5cm,height=5cm]{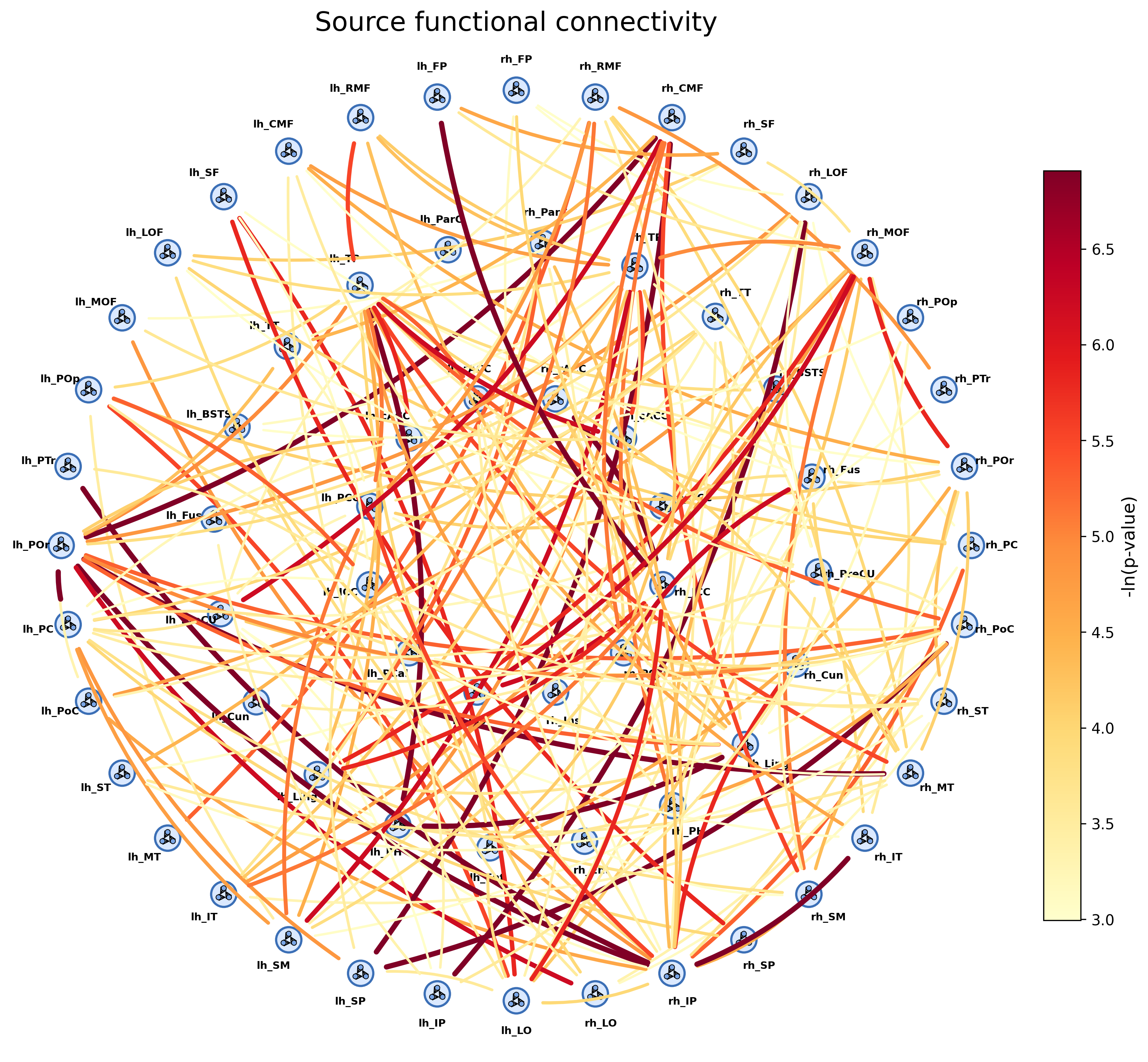}\\
       \tiny (f)
    \end{minipage}

    \caption{Source differential dynamic causal nets for parameters after the corrections for multiple testing and for multiple-input effects. (a) and (b): $K_{A}$ mean and variation. (c) and (d):  $\nu_0 $  mean and variation. (e): $\sigma_A$ variation. (f): Source differential causal nets using the maximum lagged-correlation-based functional connectivity between the case and control groups at the level $5\%$. However, these edges in the nets were not significant at the level of $5\%$ after the adjustments for multiple testing. The colour bars are in the scale of $-\ln(\mbox{p-value})$, where $\mbox{p-value}$ is BH-adjusted. }
    \label{Fig:ACv0}
\end{figure}

\begin{figure}
\centering
    \begin{minipage}[b]{1.\textwidth}
        \centering
        \includegraphics[height = 6cm,width=\linewidth]{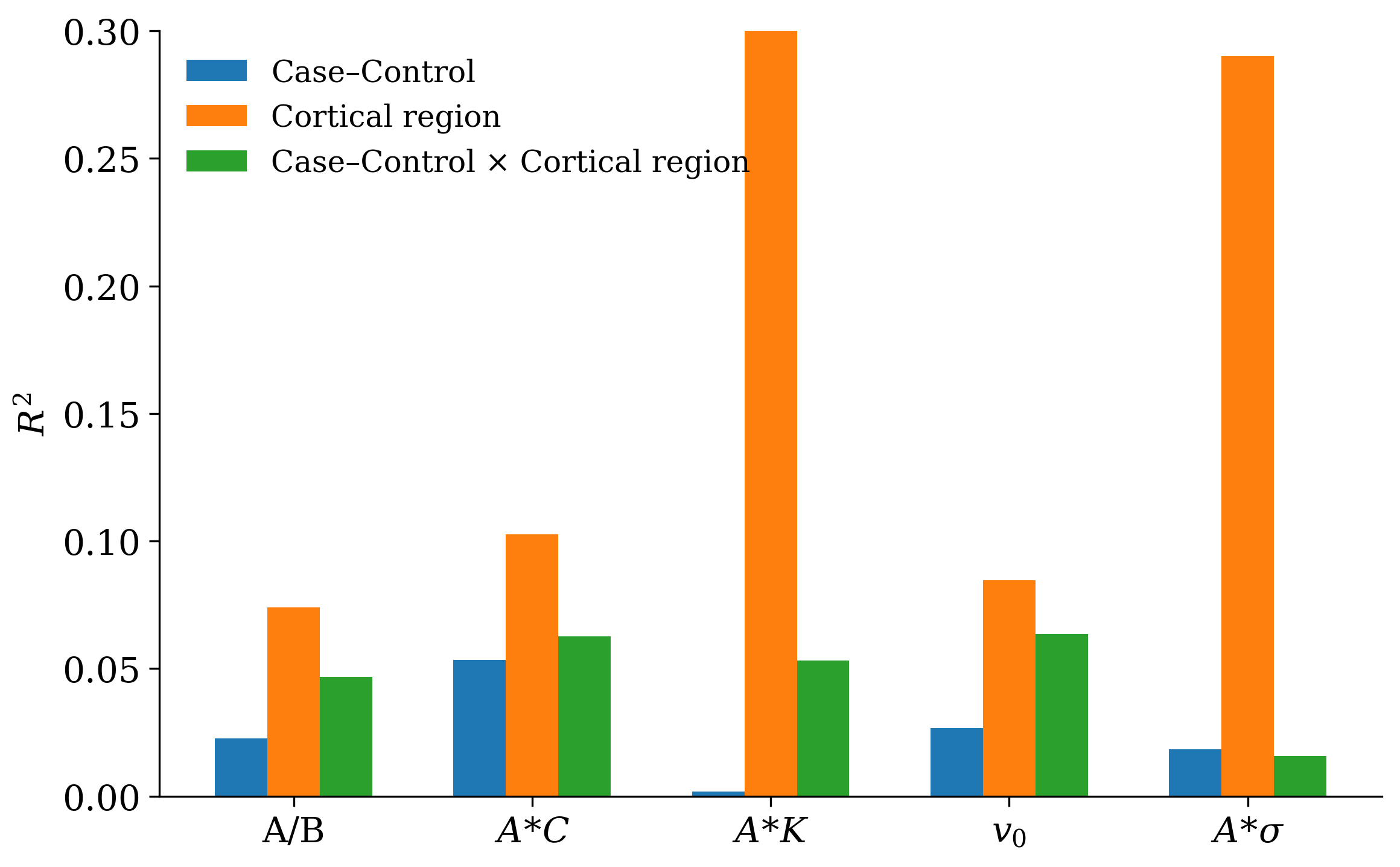}
    \end{minipage}
  \caption{ The percentages of uncertainty in the estimated source parameters which was explained by the NccDCM-based mixed-effects models, $R^2$. See \cite{Anderson2017} for the definition of $R^2$. Here, $B_A=A/B, C_A=AC, K_A=AK$ and $\sigma_A=A\sigma$. }
  \vspace{0.4cm}
 \label{R2ANOVASource}
\end{figure}

\subsection{Differential dynamic causal nets between preictal and ictal periods}

\paragraph{Description of pair-matched data}
According to the annotations provided by \cite{OK2021}, in the preictal periods, the epileptic subjects P1, P3 and P5$\sim$P9 were selected with six $20$ seconds long preictal segments of time series identified. P2, P4 and P10 were excluded from further analysis as there were no six  $20$ seconds long preictal segments identified. Note that the preictal-ictal period dataset was period-matched for each subject while the case-control dataset was not pair-matched.

\begin{figure}[htbp]
    \centering
    \begin{minipage}[b]{0.45\textwidth}
        \centering
        \includegraphics[height = 6cm,width=\linewidth]{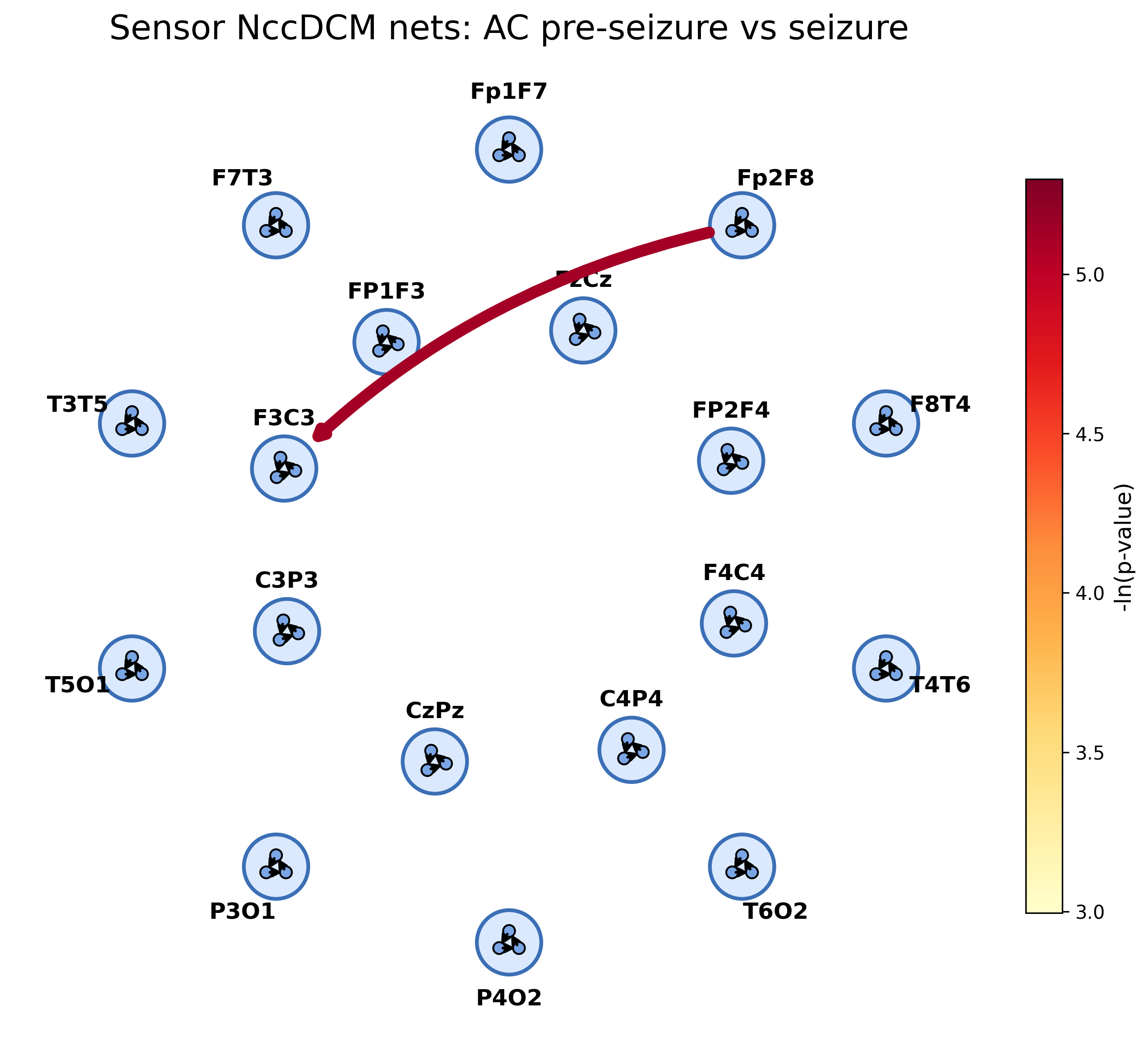}
        \tiny (a)
    \end{minipage}
  \hfill
    \begin{minipage}[b]{0.45\textwidth}
        \centering
        \includegraphics[height = 6cm,width=\linewidth]{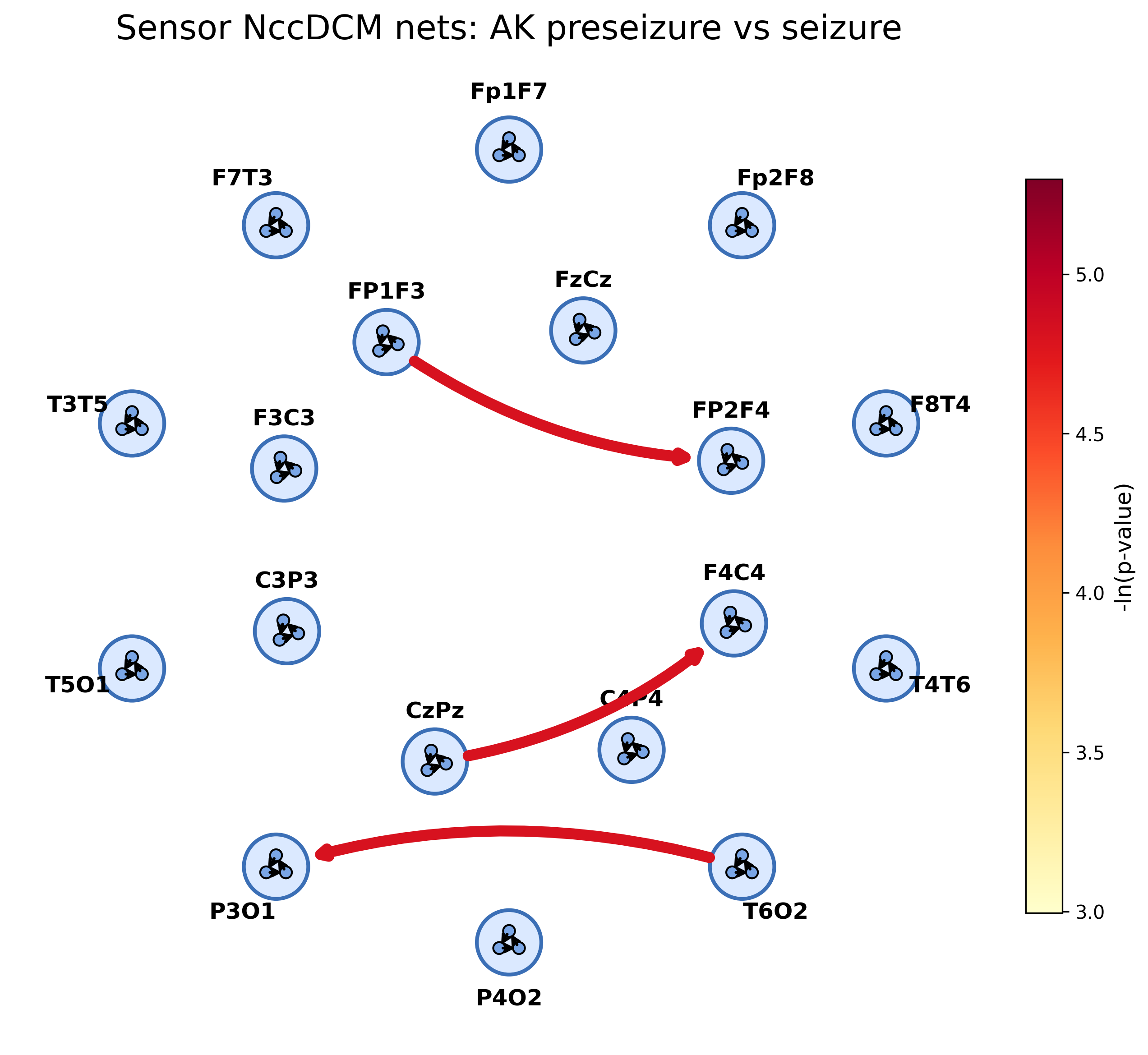}
        \tiny (b)
    \end{minipage}
       \vspace{0.4cm}

    \begin{minipage}[b]{0.45\textwidth}
        \centering
        \includegraphics[height = 6cm,width=\linewidth]{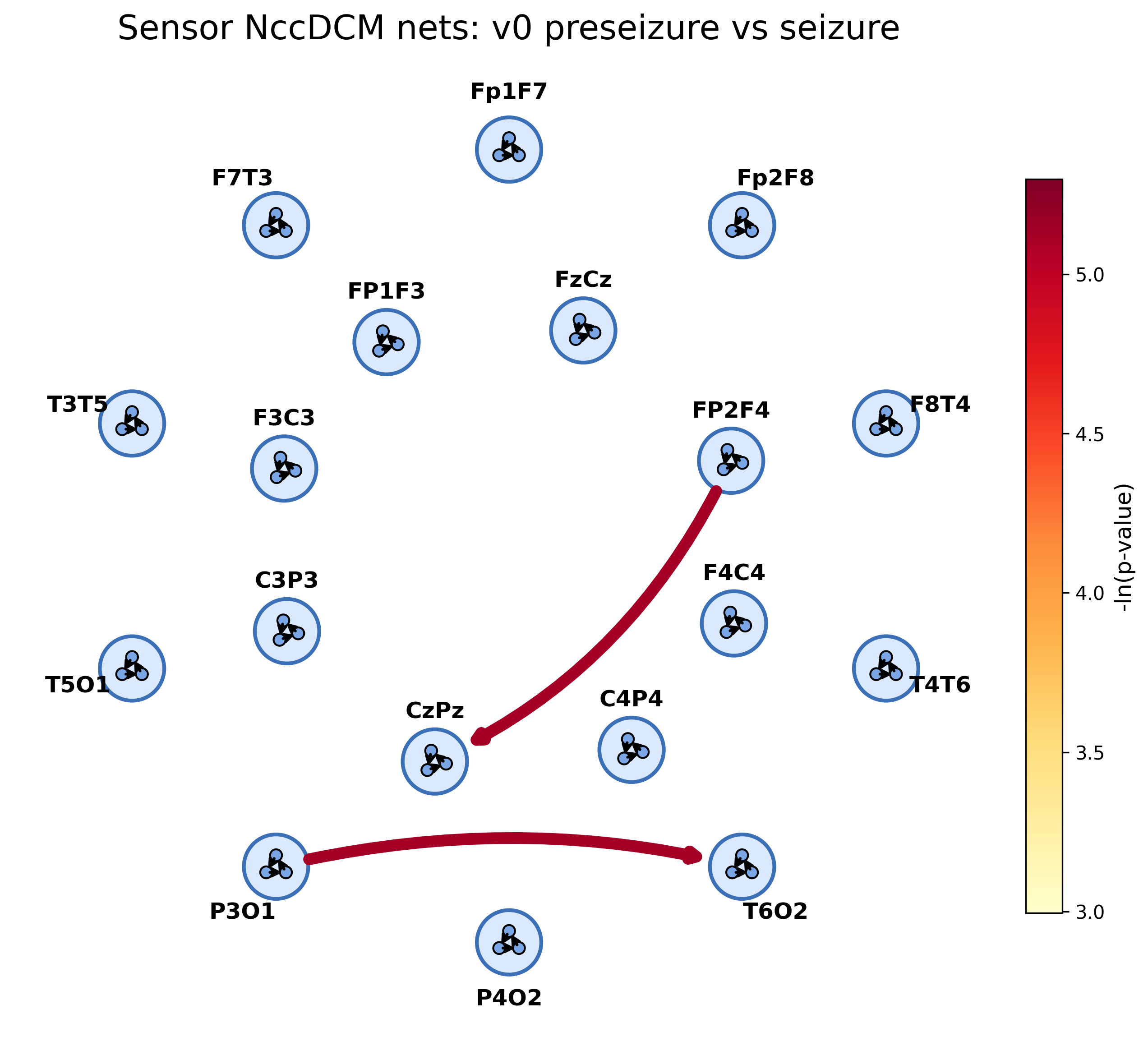}
        \tiny (c)
    \end{minipage}
 \hfill
    \begin{minipage}[b]{0.45\textwidth}
        \centering
        \includegraphics[height = 6cm,width=\linewidth]{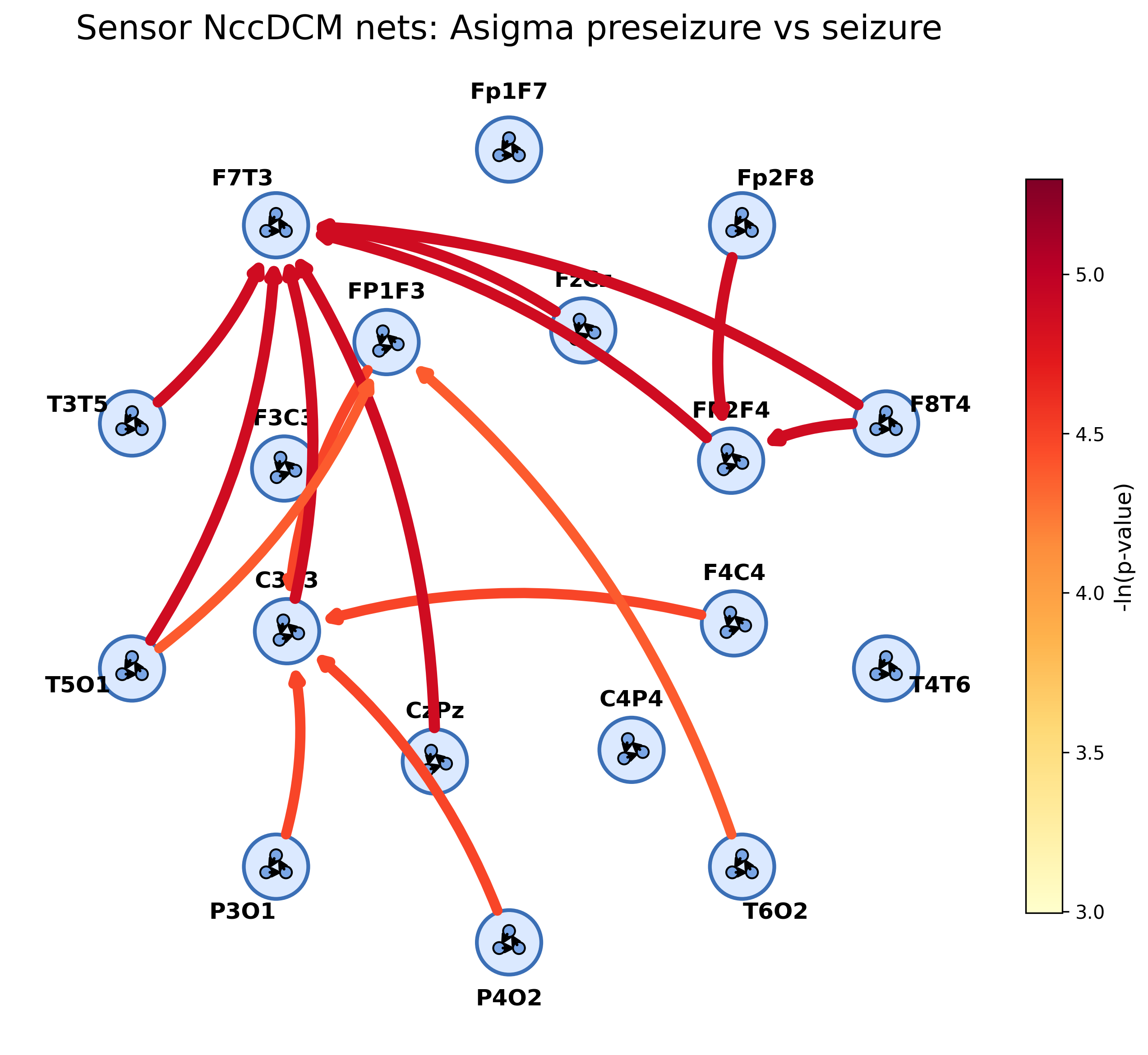}
        \tiny (d)
    \end{minipage}

    \caption{Sensor differential dynamic causal nets of parameters for pre-seizure/interictal v.s. seizure/ictal. (a) $ C_{A}$ , (b) $ K_{A} $, (c) $\nu_0 $ and (d) $\sigma_A$.
The colour bars are in the scale of $-\ln(\mbox{p-value})$, where $\mbox{p-value}$ is BH-adjusted.}
    \label{Fig:effetivenetworkpre}
\end{figure}

 {\bf Sensor level analysis}. We fitted a NccDCM to the EEG recordings for each subject in the preictal period and the ictal period respectively, followed by the pair-matched  post Wilcoxon signed rank test for the period differences. The results illustrated in Figure \ref{Fig:effetivenetworkpre} suggest that, in patients with epilepsy, the NccDCM parameters exhibited noticeable increases from the preictal period to the ictal period. These changes were associated with seizure onset.
 \begin{itemize}
\item In the $\sigma_A$ sensor differential dynamic causal net, channels F7T3, C3P3 and FP2F4, with relatively high degree centralities, imply a spatially distributed preictal-to-ictal EEG activity involving bilateral frontotemporal, centro-parietal, and right frontal regions.
\item In the $C_A$ sensor differential dynamic causal net, the significant change in the local connection parameter was identified due to the input-output relationship between channels FP2F8 and F3C3.
\item  In the $K_A$ sensor differential dynamic causal net, the significant changes of input-output coefficients were identified in channel pairs (T6O2,P3O1), (FP1F3, FP2F4) and (CZPZ, F4C4), related to the connections between the frontal and temporal regions.
\item In the $\nu_0$ sensor differential dynamic causal net, the significant changes of input-output coefficients were identified in channel pairs (FP2F4,CZPZ) and (P3O1, T6O2), related to the connections between the frontal, temporal and posterior regions.
\end{itemize}

\begin{figure}[htbp]
    \centering
    \begin{minipage}[b]{0.45\textwidth}
        \centering
        \includegraphics[height = 6cm,width=\linewidth]{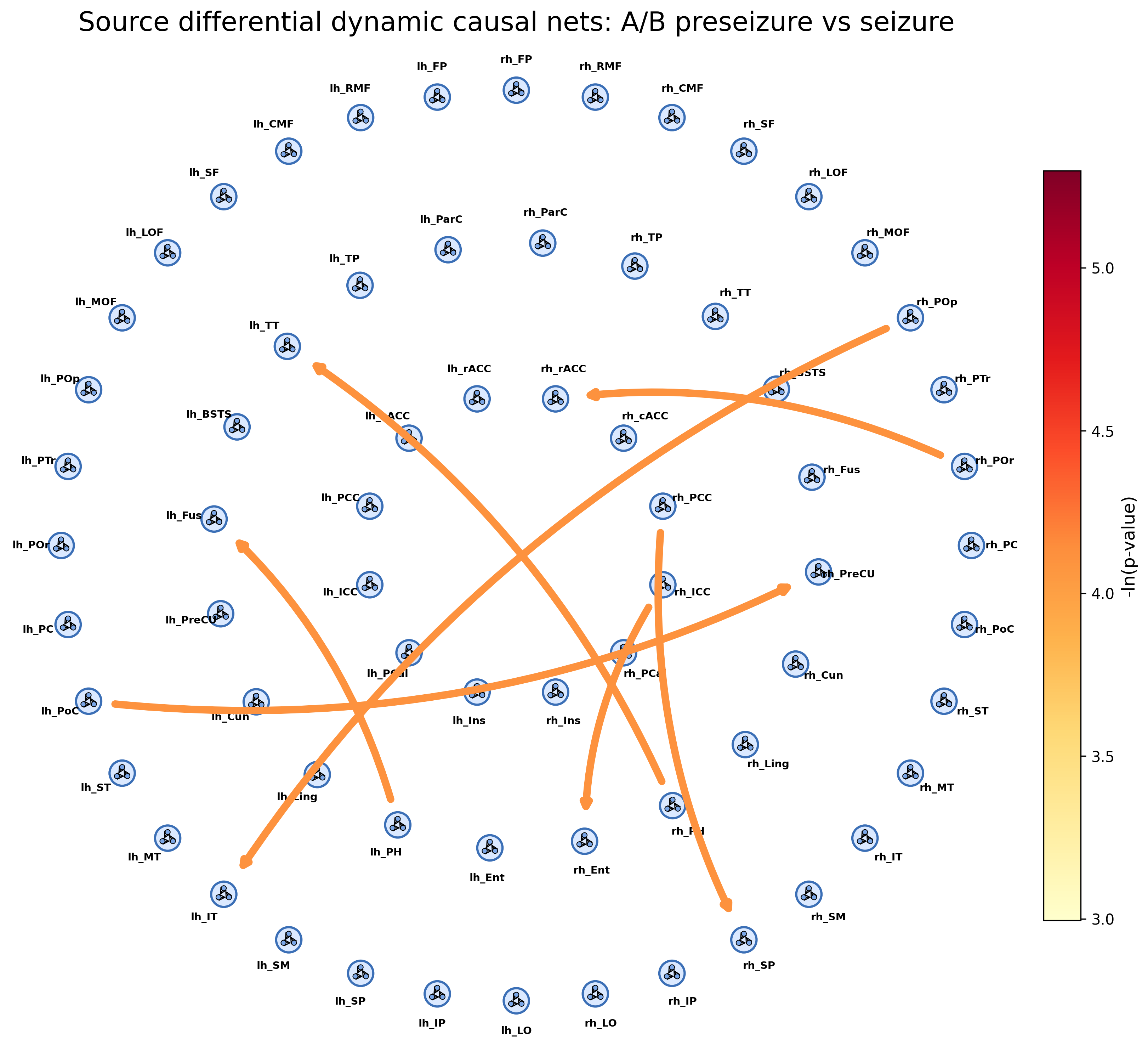}
        \tiny (a)
    \end{minipage}
    \hfill
    \begin{minipage}[b]{0.45\textwidth}
        \centering
        \includegraphics[height = 6cm,width=\linewidth]{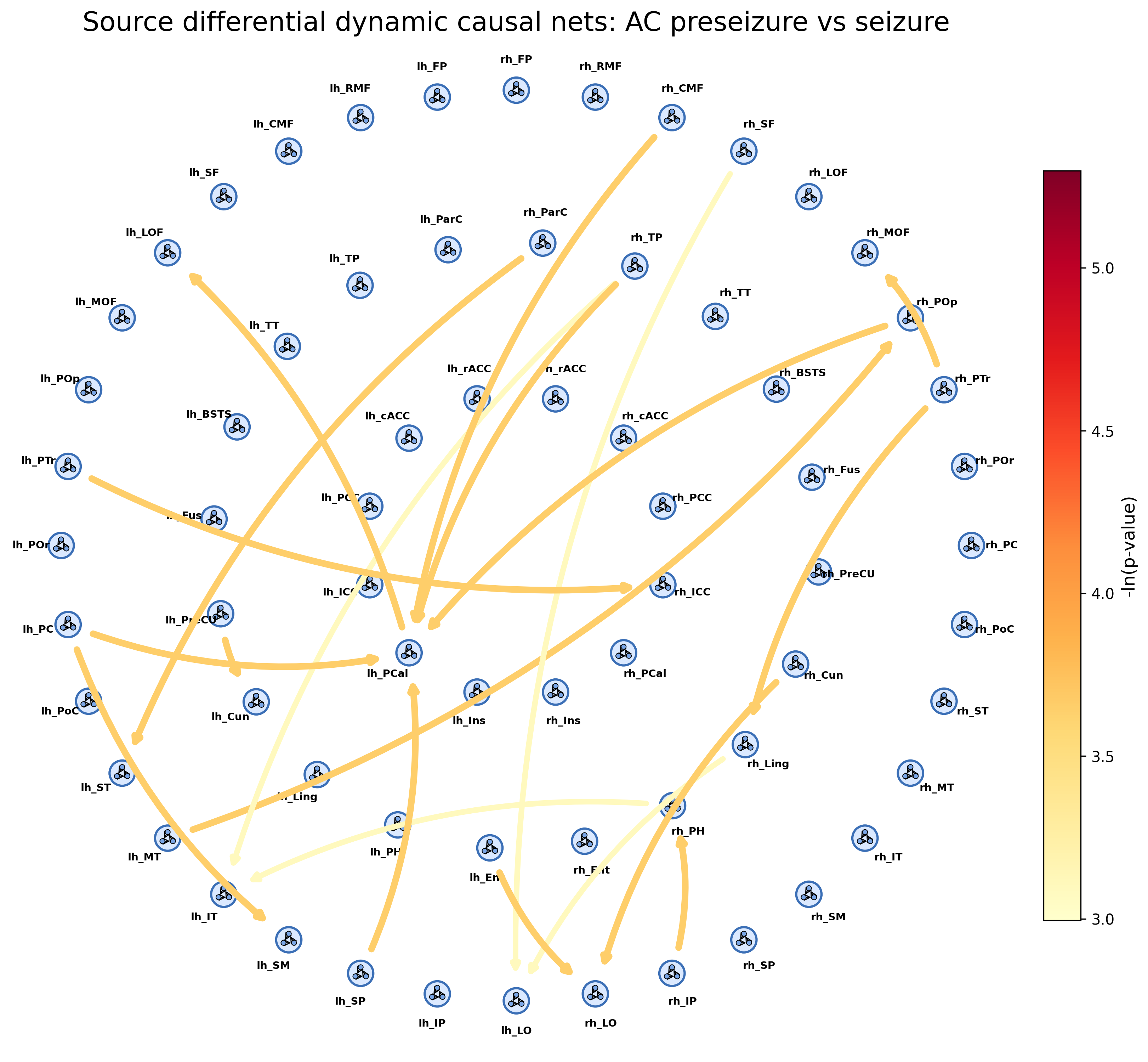}
        \tiny (b)
    \end{minipage}
    \vspace{0.4cm}

    \begin{minipage}[b]{0.45\textwidth}
        \centering
        \includegraphics[height = 6cm,width=\linewidth]{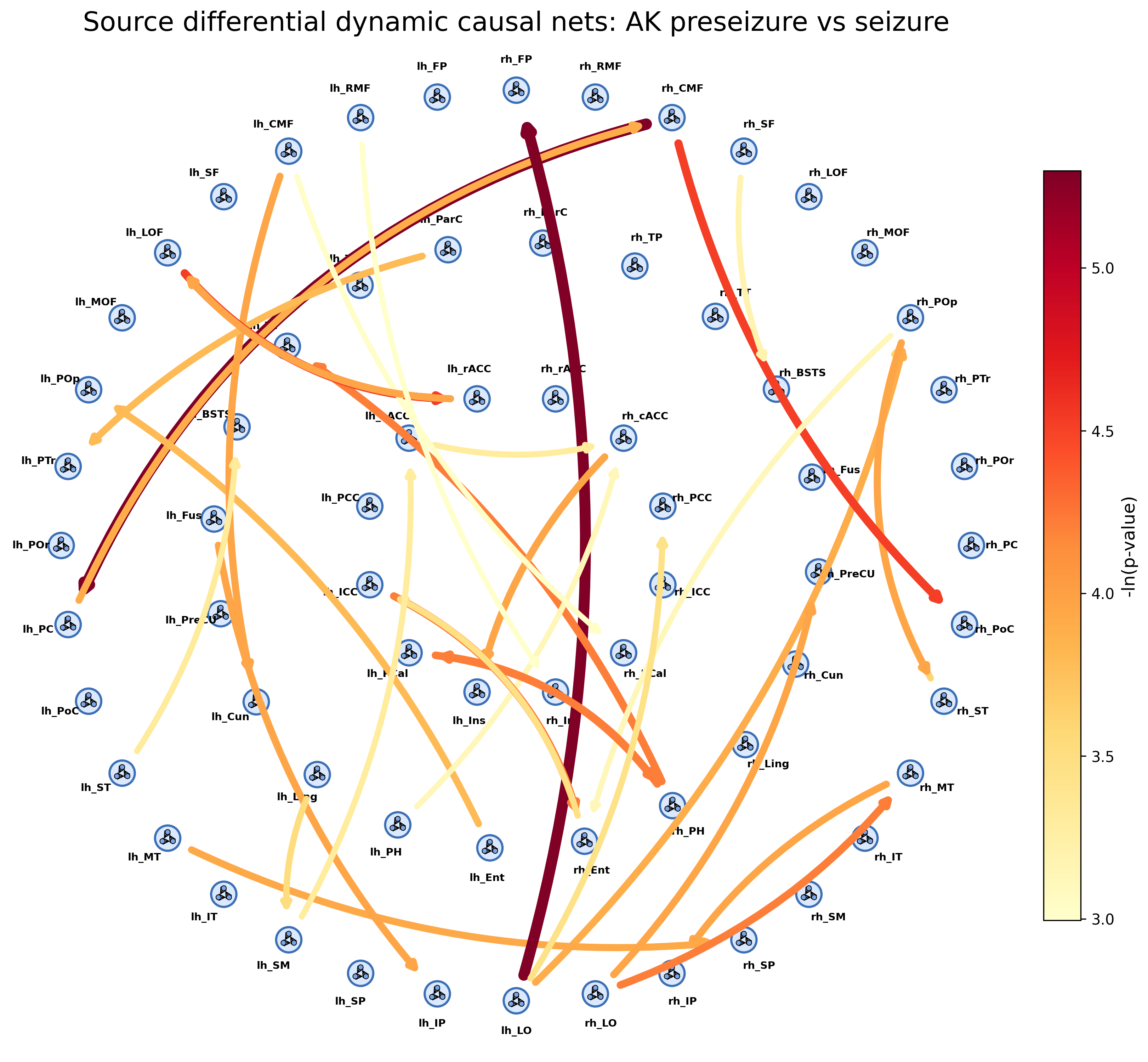}
        \tiny (c)
    \end{minipage}
    \hfill
    \begin{minipage}[b]{0.45\textwidth}
        \centering
        \includegraphics[height = 6cm,width=\linewidth]{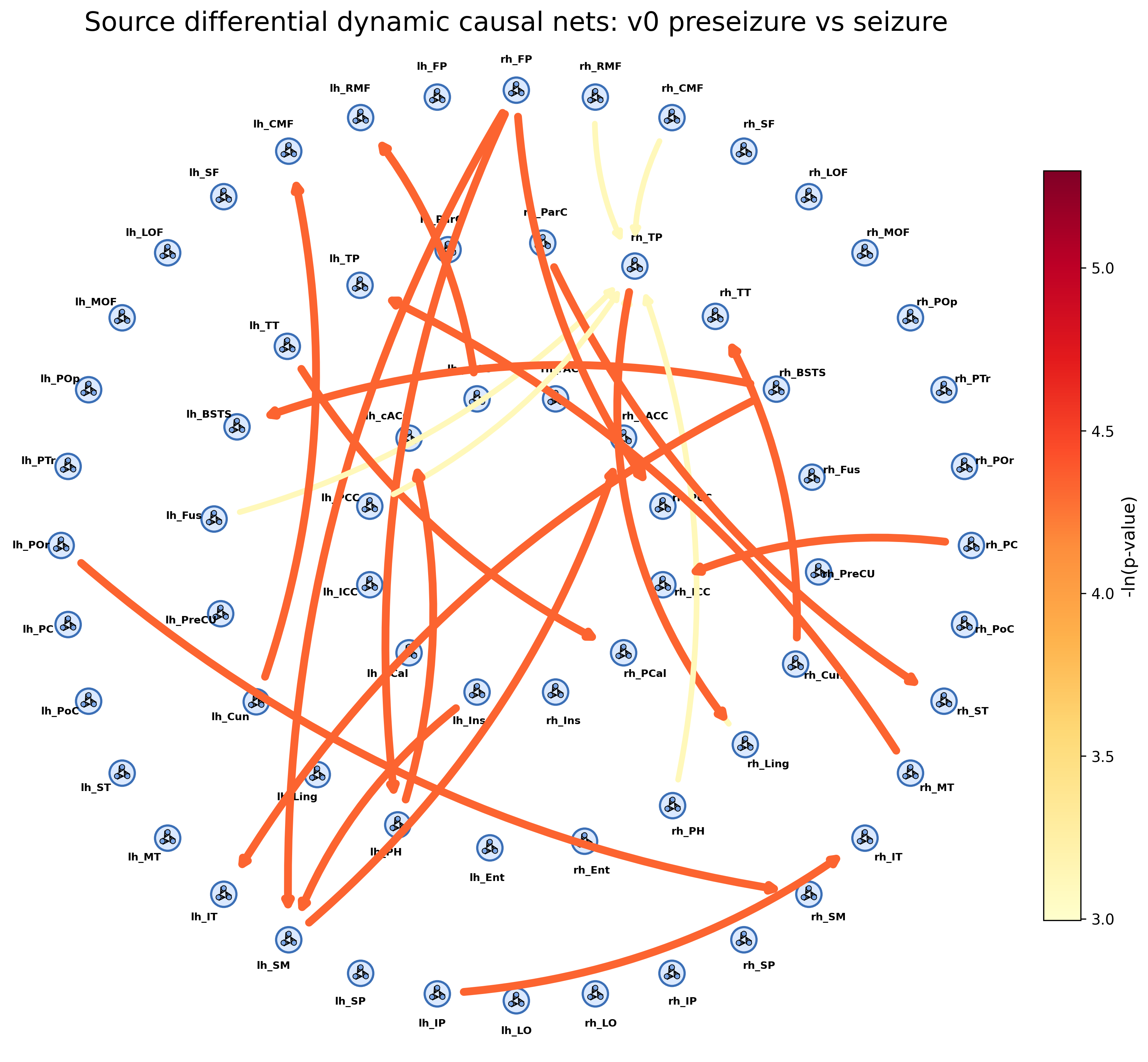}
        \tiny (d)
    \end{minipage}
 \vspace{0.4cm}

    \begin{minipage}[b]{0.45\textwidth}
        \centering
        \includegraphics[height = 6cm,width=\linewidth]{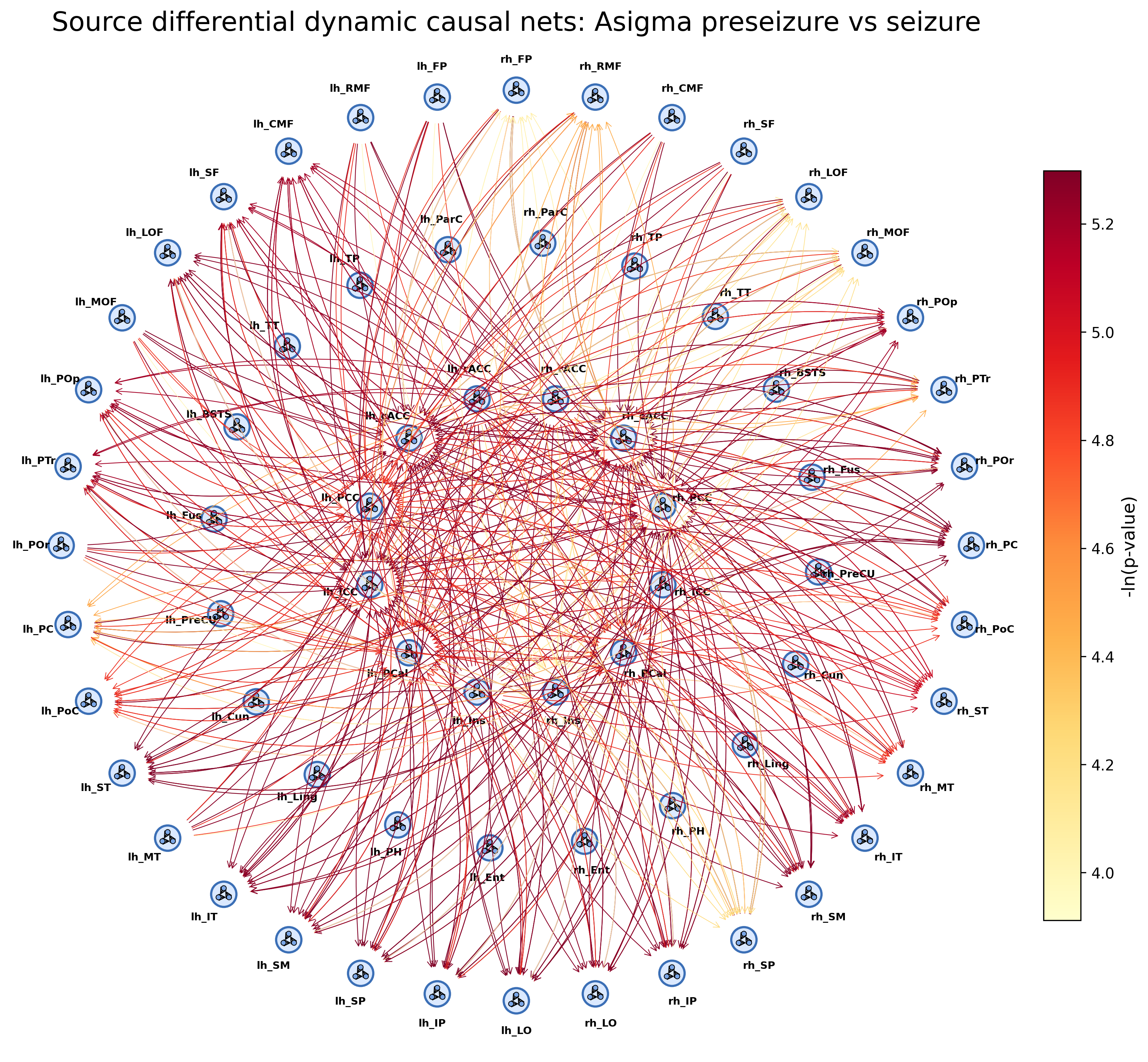}
        \tiny (e)
  \end{minipage}
    \caption{Source differential causal nets of parameters for pre-seizure/interictal v.s. seizure/ictal. (a) $B_{A},$ (b) $ C_{A}$ , (c) $K_{A} $,  (d) $\nu_0 $ and (e) $\sigma_A$.
The colour bars are in the scale of $-\ln(\mbox{p-value})$, where $\mbox{p-value}$ is BH-adjusted.}
    \label{Fig:effetivenetworkpreSource}
\end{figure}

{\bf Source level analysis}. We fitted a NccDCM to the estimated source time series for each subject  in the preictal period and the ictal period respectively, followed by the pair-matched  post Wilcoxon signed rank test for the period differences. Unlike the sensor-level analysis, source-level analysis revealed more significant changes in parameters, even in $B_A$. The results, displayed in Figure~\ref{Fig:effetivenetworkpreSource} below  can be highlighted as follows.

 \begin{itemize}
\item The $B_A$ differential dynamic causal net demonstrates a state-dependent alteration of the input-output network topology during epileptic discharges, particularly involving mesial temporal, posterior cingulate, precuneus and superior parietal regions. The results were partially confirmed by previous EEG/MEG functional connectivity studies \citep{Kumar2023,Sun2023, Liu2024}. In addition, resting-state MEG/fMRI and structural-connectome studies demonstrate that several frontal and temporal regions—including fusiform, entorhinal, inferior temporal, transverse temporal, pars opercularis/orbitalis and anterior cingulate cortices—form a part of altered  epileptic network \citep{Liu2024,Schwarzer2026}. Thus, these regions provide biologically plausible candidate nodes for seizure-state–dependent changes.

\item The $\sigma_A$ differential dynamic causal net shows that a few cortical regions are of high degree centrality, including the insula (lh-Insula and rh-Insula),
anterior cingulate (rh-Caudalanteriorcingulate), posterior cingulate (lh-posteriorcingulate, rh-posteriorcingulate) and isthmus cingulate (lh-isthmuscingulate). Here, degree centrality defines the number of node's first-order neighbors in the network. This is consistent with the recent finding in a cortical thickness study \citep{Ogren2018} that  increased cortical thickness involving the insula, anterior cingulate, posterior cingulate and isthmus cingulate, as well as the pericalcarine cortex. \cite{ Singh2020} using stereo-EEG found heterogeneous but focal seizure-onset patterns within the insula. The anterior cingulate cortex was identified as a potential seizure-onset and epileptogenic region in focal epilepsy \citep{Yang2023}.

\end{itemize}

\subsection{Simulation studies}

\subsubsection{Estimating a single node }

 Consider the following conditionally coupled Jansen-Rit model:
\begin{eqnarray}\label{first}
  d{x}_0(t) &=& x_3(t)dt, \quad
  d{x}_1(t) = x_4(t)dt, \quad
  d{x}_2(t) = x_5(t)dt, \nonumber\\
  d{x}_3(t) &=& \left [ Aa \mathrm{S}\left(x_1(t)-x_2(t)\right)-2ax_3(t)-a^2x_0(t) \right ]dt , \\
  d{x}_4(t) &=&\left [  Aa\left\{C_2\mathrm{S}\left(C_1x_0(t)\right)+K\mathrm{S}\left(y_2(t)\right)\right\}-2ax_4(t)-a^2x_1(t) \right ]dt +Aa dw(t), \nonumber\\
  d{x}_5(t) &=&\left [  BbC_4\mathrm{S}\left(C_3x_0(t)\right)-2bx_5(t)-b^2x_2(t) \right ]dt\nonumber,
\end{eqnarray}
with $C_1=C,$ $C_2=0.8C, C_3=0.25C=C_4,$  state vector $\x(t)=(x_0(t),x_1(t),x_2(t), x_3(t),x_4(t),x_5(t)) $ and initial value $\boldsymbol x_0=(x_0(0),...,x_5(0))^{T}\in\mathbb{R}^{6}$ and the input $y_2(t)$ from the external channel. We use the  Strang splitting method (see S12, 
 the Online Supplementary Material) to generate the samples of $\x(t)$ from the above model. See \cite{Milstein2003, Leimkuhler2016} for more details.
We can convert the above model to a canonical Jansen-Rit model/ccDCM by reparametrisation.
 As ccDCMs play a key role in shaping and influencing the structure and behavior of the NccDCM, we first showed the bifurcation  behavior of ccDCM in S12, 
 the Online Supplementary Material. Then, in the following, we evaluated the accuracy of the proposed estimation method. We measured the accuracy of the proposed model estimation by absolute relative error (ARE):
\[
\mathrm{ARE}(\hat{\theta}) = {|\hat{\theta}-\theta^{(0)}|}/{|\theta^{(0)}|}
\]
with $\theta^{(0)}=(B_A,C_A,K_A,\nu_0,\sigma_A)$ as the ground truth. We calculated the median and the normalized median absolute deviation (NMAD) defined as $1.48\times\mbox{MAD}$ of AREs.

We assessed the ARE of the proposed estimation for the ccDCM model~(\ref{first})  in the following settings :\\
 Setting I. \( A = 3.25 \), \( B = 22 \), \( C = 135 \), \( K = 50 \), \( v_0 = 6 \)  and $\sigma = 10$.\\
 Setting II. \( A = 4 \), \( B = 14 \), and the remaining parameters are kept the same as in Setting I.\\
 Setting III.  $A\sim U(3,6)$, $B\sim U(17.6,30)$, $C\sim U(108,160)$, $K\sim U(0,50)$, $v_0\sim U(3.12,9)$ and $\sigma\sim U(5,10)$, where
$U(a,b)$ is a uniform distribution over the interval $[a,b]$.

\begin{figure}[htbp]
  \centering
  \begin{minipage}{0.32\textwidth}
    \centering
    \includegraphics[width=\linewidth, height=5cm]{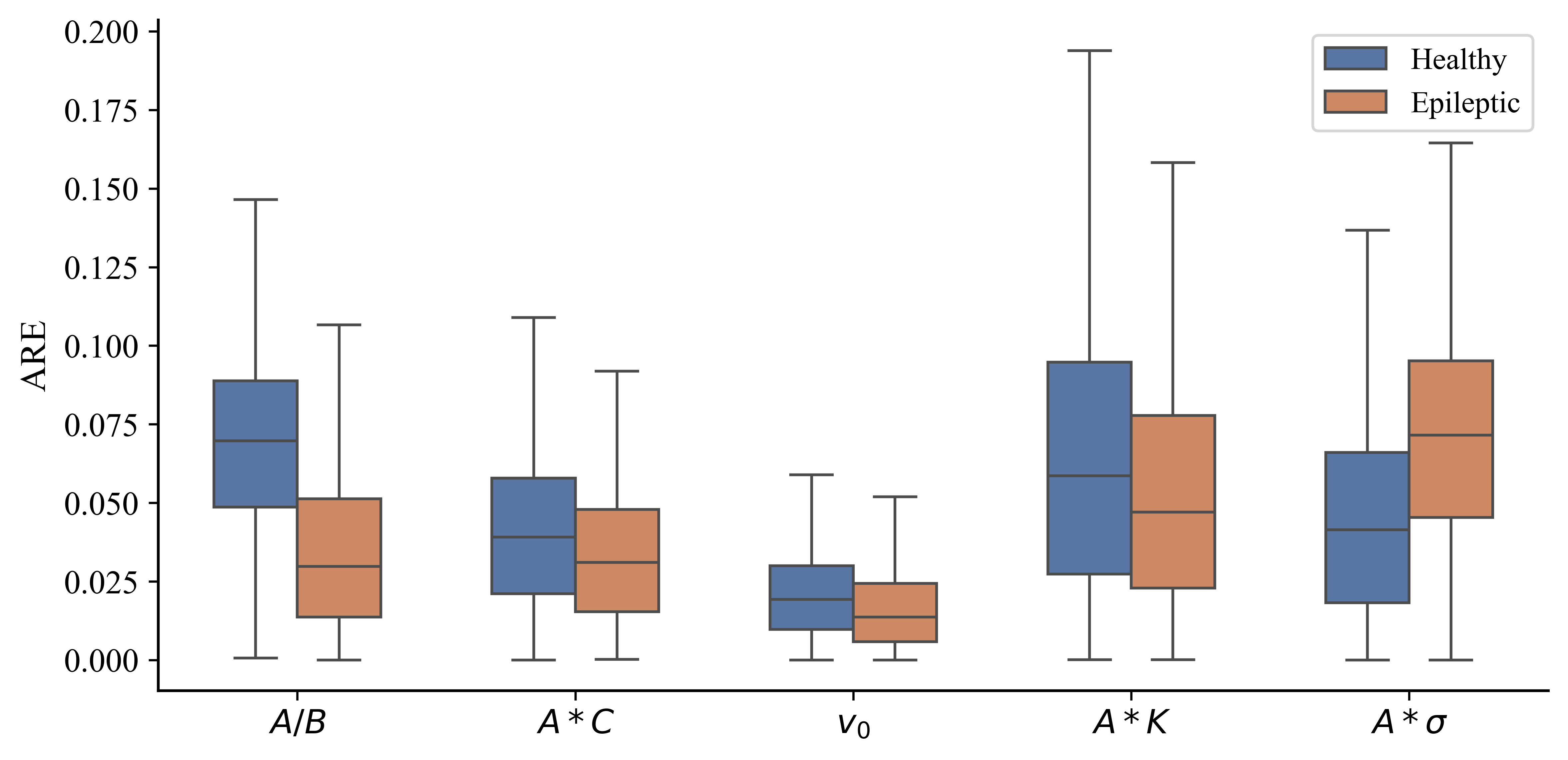}
    \par\vspace{2mm}
{\small Setting I\par}
  \end{minipage}
  \hfill
  \begin{minipage}{0.32\textwidth}
    \centering
    \includegraphics[width=\linewidth, height=5cm]{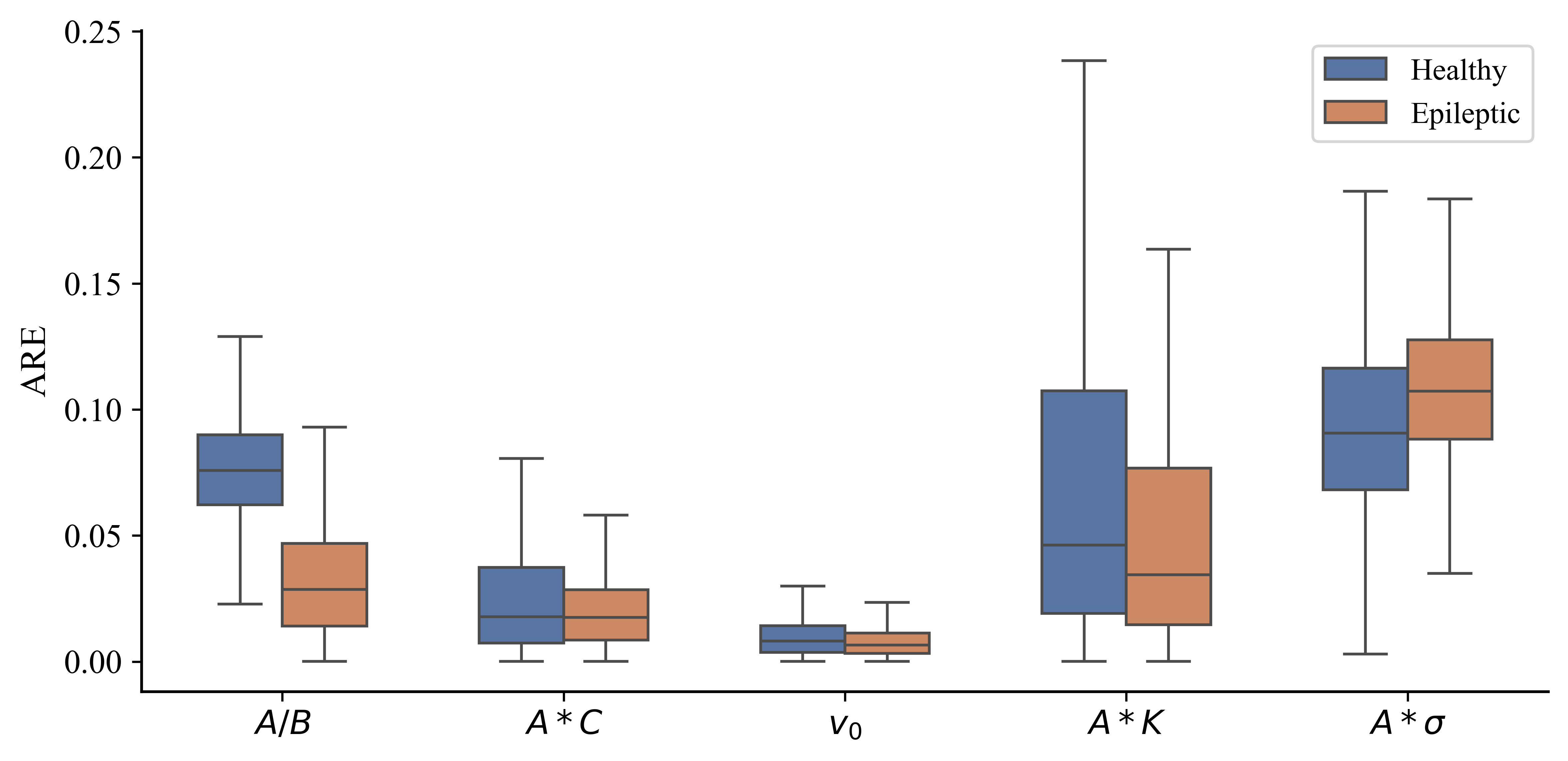}
    \par\vspace{2mm}
{\small Setting II\par}
  \end{minipage}
  \hfill
  \begin{minipage}{0.32\textwidth}
    \centering
    \includegraphics[width=\linewidth, height=5cm]{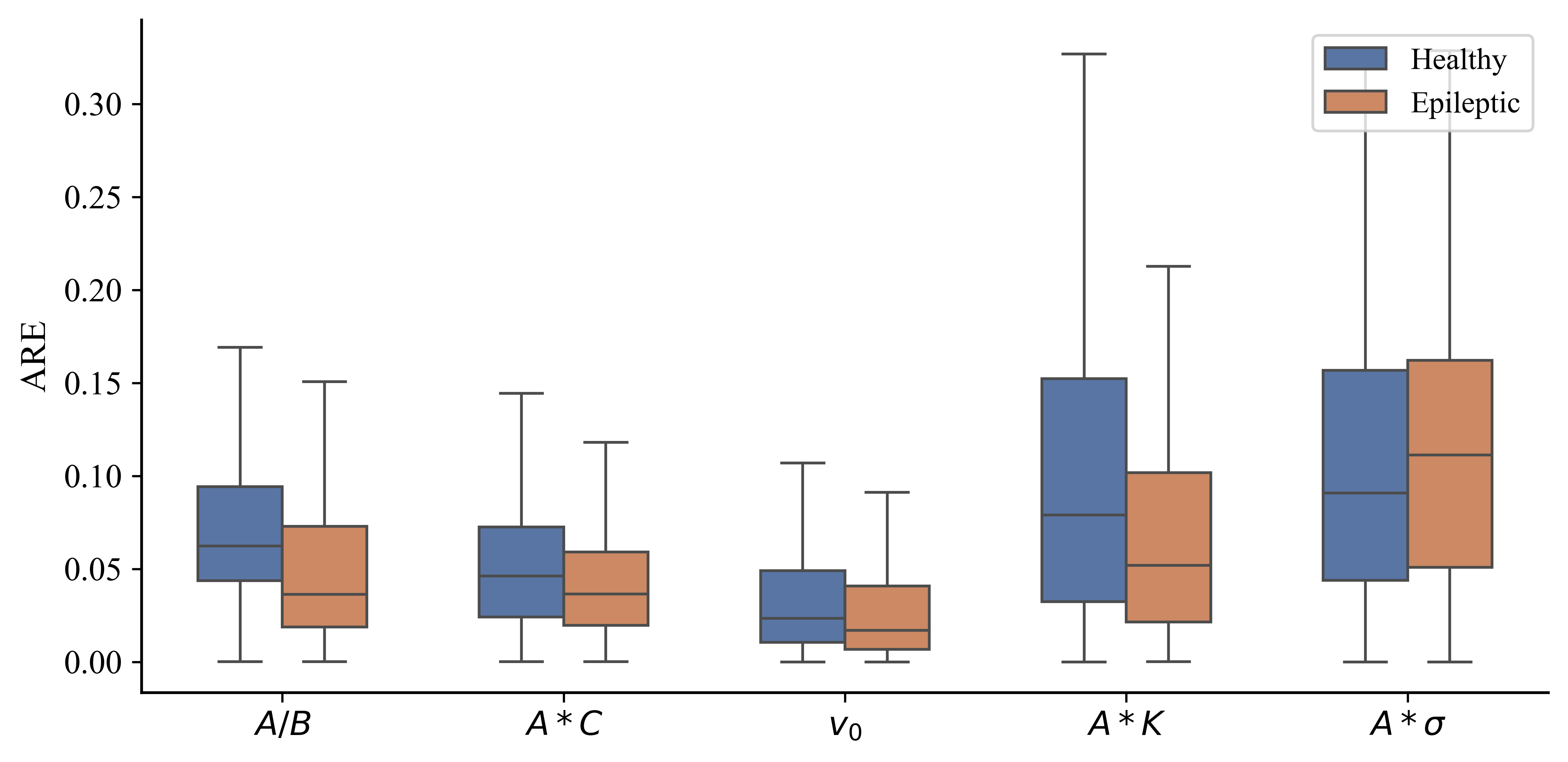}
    \par\vspace{2mm}
{\small Setting III\par}
  \end{minipage}
  \caption{Box-plots of ARE values of the estimated $B_A=A/B$, $C_A=AC$, $v_0$, $K_A=AK$ and $\sigma_A=A\sigma$ compared to the ground truths in Settings I, II, III.}
  \label{ARE_boxplot}
\end{figure}

In these settings, input time series $y_2(t)$ were collected from epileptic patients (P1, P3, and P5 to P10) and healthy subjects (H1 to H10) in the channels FP2F4, F4C4, C4P4, P4O2, FP1F3, F3C3, C3P3, P3O1, FP2F8, F8T4, T4T6, T6O2, FP1F7, F7T3, T3T5, T5O1, FZCZ, and CZPZ respectively.  There were $10\times 18$ time series from healthy subjects and $8\times 18$ time series from epileptic subjects. Each of these time series were made of $6$ segments,  with the duration of $20$ seconds each.  We set the simulation step size  \( h^{\mathrm{sim}} = 1/256 \) and the block length of $2$ (i.e., two time instances) to ensure that each time block was short enough so that the Chen--Fliess series approximation was accurate enough. For each input time series segment, we generated an output time series segment from the model~(\ref{first}). Totally,  we simulated $M_H=1080$ healthy input-output time series segment pairs and $M_P=864$ epileptic input-output time series segment pairs. For each input-output time series segment pair, we fitted it by ccDCM, resulting in estimated parameter values of $(B_A, C_A, K_A, \nu_0, \sigma_A)$. Compared them to the ground truth, we then evaluated the median and NMADs of AREs for simulated healthy segment pairs and for simulated epileptic segment pairs respectively.
The results, displayed by boxplots in  Figure~ \ref{ARE_boxplot} and by Table \ref{tab:MDF-NMM-MEAN}, showed that the proposed estimation is promising.

\begin{table}[htbp]
\centering
\caption{\label{tab:MDF-NMM-MEAN}
Median ARE values with corresponding NMADs (in parentheses) for ccDCM estimation. The left and right blocks present results obtained using EEG time series from epileptic and healthy subjects as external inputs, respectively.}
\renewcommand{\arraystretch}{1.1}
\setlength{\tabcolsep}{6pt}
\resizebox{\textwidth}{!}{
\begin{tabular}{c c c c | c c c c c}
\hline
\multicolumn{4}{c|}{\textbf{Epileptic}} & \multicolumn{5}{c}{\textbf{Healthy}} \\
\textbf{Parameter} & \textbf{Setting I} & \textbf{Setting II} & \textbf{Setting III}
& \textbf{Parameter} & \textbf{Setting I} & \textbf{Setting II} & \textbf{Setting III} & \\
\hline
$B_A$      & 0.03 (0.03) & 0.03 (0.02) & 0.04 (0.03)        & $B_A$       & 0.07 (0.03)  & 0.08 (0.02) & 0.06 (0.03) & \\
$C_A$      & 0.03 (0.02) & 0.02 (0.01) & 0.04(0.03)         & $C_A$       & 0.04 (0.03)  & 0.02 (0.02) & 0.05 (0.04)& \\
$\nu_0$      & 0.01 (0.01) & 0.01 (0.01) & 0.02 (0.02)        & $\nu_0$       & 0.02  (0.01) & 0.01 (0.01) & 0.02 (0.02)& \\
$K_A$      & 0.05 (0.04) & 0.03 (0.04) & 0.05 (0.05)        & $K_A$       & 0.06 (0.05)  & 0.05 (0.05) & 0.08 (0.08)& \\
$\sigma_A$ & 0.07 (0.04) & 0.11 (0.03) & 0.11 (0.08)        & $\sigma_A$  & 0.04 (0.03)  & 0.09 (0.04) & 0.09 (0.08)& \\
\hline
\end{tabular}
}
\end{table}

\subsubsection {Joint estimation of multiple nodes.} \label{joint_es}
\begin{figure}[htbp]
  \centering
  \begin{minipage}[b]{0.45\textwidth}
    \centering
    \includegraphics[
      width=\linewidth,
      height=5cm,
      keepaspectratio
    ]{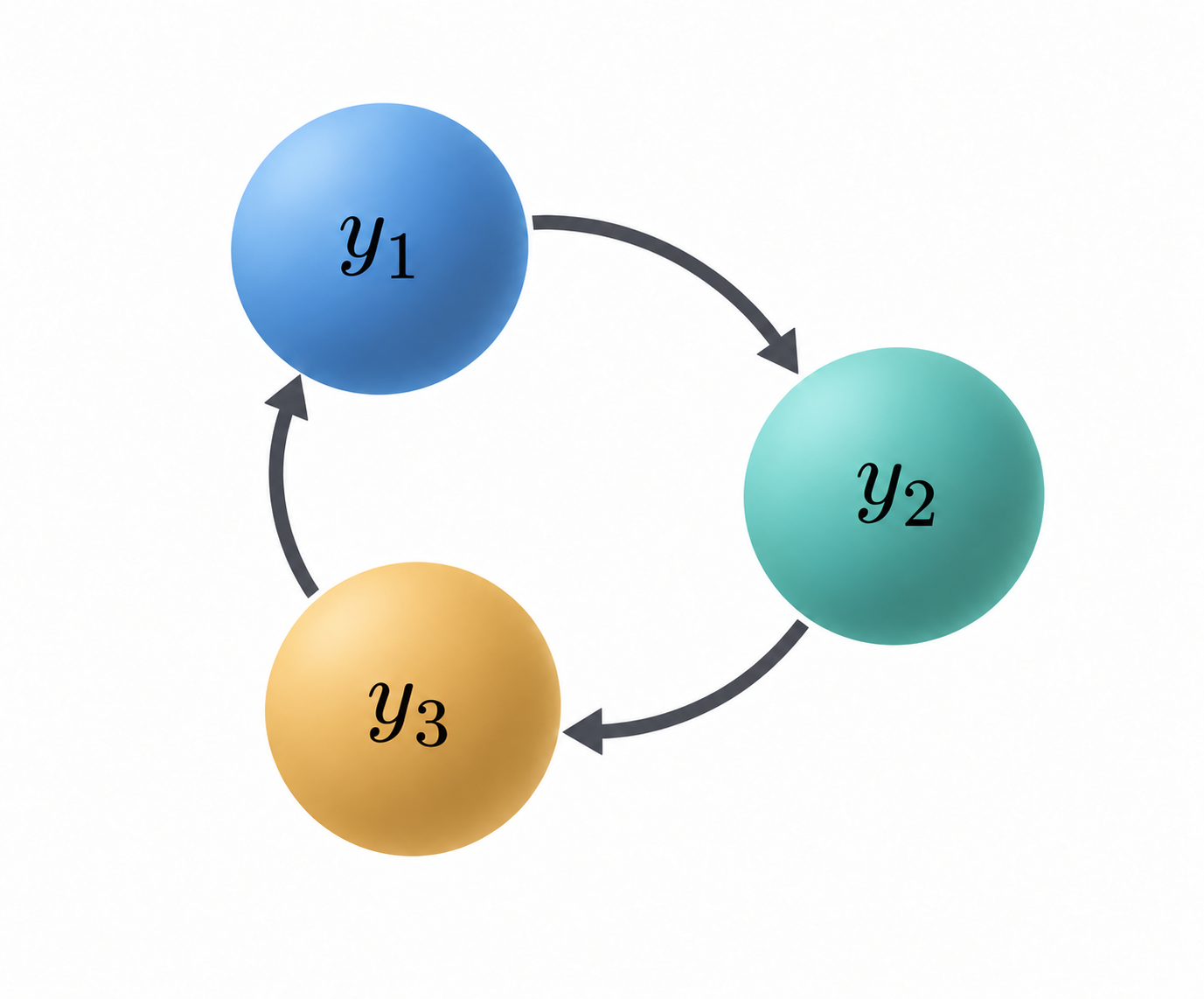}
    \par\vspace{2mm}
    {\small (a) Unidirectional coupling\par}
  \end{minipage}
  \hspace{0.04\textwidth}
  \begin{minipage}[b]{0.45\textwidth}
    \centering
    \includegraphics[
      width=\linewidth,
      height=5cm,
      keepaspectratio
    ]{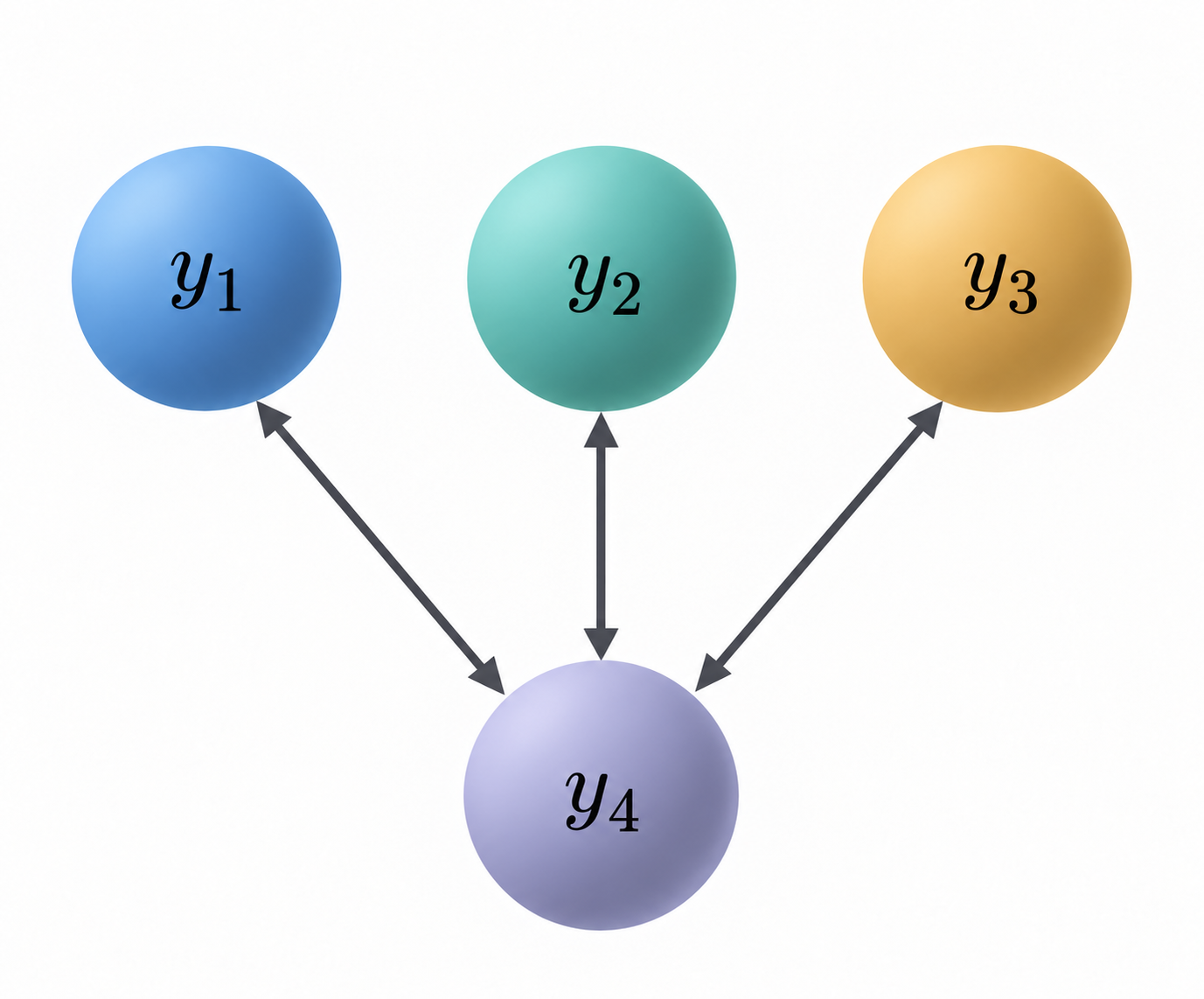}
    \par\vspace{2mm}
    {\small (b) Bidirectional coupling\par}
  \end{minipage}

  \caption{Schematic illustrations of two multi-node
    configurations. $y_i$ is the output of node $i$ while arrows show the direction of coupling between these nodes. (a) Unidirectional coupling configuration.
    (b) Bidirectional coupling configuration.}
  \label{fig:network_configurations}
\end{figure}

In the following, we assessed the ARE of the proposed estimation for multiple nodes, where each node stands for a Jansen-Rit neural mass model.
 Note that the proposed estimation of NccDCM depends on the methods of solving the Jansen-Rit equations.  In this paper, we proposed using Chen-Fliess method, instead of  standard numerical solvers such as the RK and EM schemes  to solve the Jansen-Rit differential equations. We consider two distinct multi-node configurations: one with unidirectional coupling and the other with bidirectional coupling, as illustrated in Figure~\ref{fig:network_configurations}. The detailed mathematical forms can be found in S13, 
the Online Supplementary Material.

 For unidirectional coupling setting, we assessed the proposed method against RK and EM standard solver-based methods. Note that the RK approximation for the stochastic Jansen--Rit model is computed using the function \texttt{stratSRS2} from the Python library \texttt{sdeint}.
The parameters in unidirectional coupling setting were set as follows:
$A^{(1)}= 3.25$, $B^{(1)}_A =3.25/ 22$, $C^{(1)}_A =3.25 \times 135$, $K^{(1|3)}_{A} = 3.25\times 50$,  $\nu^{(1)}_{0} = 6$, $\sigma^{(1)}_{A} =3.25\times 10$;
$A^{(2)} = 4$, $B^{(2)}_A = 4/14$, $C^{(2)}_A = 4\times135$, $K^{(2|1)} =4\times 70$, $\nu^{(2)}_{0} = 4$, $\sigma^{(2)}_{A} =4\times 10$;
$A^{(3)} = 3$, $B^{(3)} = 3/15$, $C^{(3)}_A = 3\times 135$, $K^{(3|2)} =3\times 50$, $\nu^{(3)}_{0} = 5$, $\sigma^{(3)}_{A} = 3\times 10$,
where node~1 receives input $y^{(3)}(t)$ from node 3, node~2 receives input $y^{(1)}(t)$ from node 1, and node~3 receives input $y^{(2)}(t)$ from node 2.
The simulation time interval was $[0,T]=[0,20]$ seconds with the sampling frequency to $256$ Hz.  To reduce the influence of the initial conditions on the loss function, the first second simulated time series was discarded and the remaining time series was used to calculate the loss function. To further examine the robustness of the proposed method with respect to the choice of initial conditions, the following three scenarios were considered:
(i) zero initial conditions $(0,0,0,0,0,0)$;
(ii) constant initial conditions $(0,0.5,1,1.5/A^{(i)},2,2.5)$; and
(iii) random initial conditions drawn from normal distributions, namely, $x^{(i)}_{A0}(0)\sim N(0,1/A^{(i)}),x^{(i)}_{1}(0)\sim N(0,1) , x^{(i)}_{2}(0)\sim N(0,1),x^{(i)}_{A3}(0)\sim N(0,1/A^{(i)}),x^{(i)}_{4}(0)\sim N(0,1) , x^{(i)}_{5}(0)\sim N(0,1)$ .
We repeatedly simulated the triple time series $(y^{(i)}(t): i=1,2,3), t\in [0,T]$ $100$ times from the above model. The AREs of estimated parameters in NccDCM were calculated for the Chen-Fliess-, RK- and EM-based schemes respectively.  The simulation results were summarised in Table \ref{tab:ARE_all_scenarios} and Table \ref{tab:ARE_methods_initial_conditions}, the Online Supplementary Material.
The results demonstrated that the Chen-Fliess method outperformed the RK- and EM- schemes in terms of ARE. The results in Table \ref{tab:ARE_methods_initial_conditions}, the Online Supplementary Material, also suggested  that the Chen-Fliess method was robust against different initial state conditions. Furthermore, Table \ref{tab:time} indicated that the Chen-Fliess method had much lower CPU time costs than the RK-based method when we ran these procedures in the High-Performance Computing Service provided by the University of  Kent. The EM-scheme had lower CPU time costs than the Chen-Fliess (the details are omitted).

For  the bidirectional coupling configuration,  the simulation results, displayed in Tables~\ref{tab:4channel_parameter} and \ref{tab:4channel_cross}, the Online Supplementary Material, suggested that the multiple-input based parameter estimation outperformed the pairwise parameter estimation in terms of ARE.

\begin{table}[htbp]
\centering
\caption{
Median and NMAD (in parenthesis) of ARE values for the EM-, RK- and CF-based methods  with zero initial state values. EM, RK and CF stand for  Euler-Maruyama,  Runge-Kutta and Chen-Fliess.
The bold face entries are the minimum values in their corresponding rows.
}
\label{tab:ARE_all_scenarios}

\renewcommand{\arraystretch}{1.15}
\setlength{\tabcolsep}{12pt}

\resizebox{\textwidth}{!}{
\begin{tabular}{c c c c c}
\hline
\textbf{Parameter} & \textbf{Node} & \textbf{EM-based} & \textbf{RK-based} & \textbf{CF-based} \\
\hline

\multirow{3}{*}{$B_A$}
& 1   & 0.40 (0.18) & 0.55 (0.17) & \textbf{0.05} (0.04) \\
& 2  & 0.80 (0.85) & 0.77 (0.90) & \textbf{0.02} (0.02) \\
& 3 & 0.18 (0.19) & 0.16 (0.17) & \textbf{0.03} (0.02) \\
\hline

\multirow{3}{*}{$C_A$}
& 1  & 0.39 (0.09) & 0.42 (0.13) & \textbf{0.03} (0.02) \\
& 2  & 0.97 (0.04) & 0.96 (0.05) & \textbf{0.04} (0.03) \\
& 3 & 0.34 (0.08) & 0.34 (0.16) & \textbf{0.03} (0.03) \\
\hline

\multirow{3}{*}{$v_0$}
& 1   & 0.09 (0.06) & 0.08 (0.07) & \textbf{0.03} (0.02) \\
& 2  & 0.20 (0.19) & 0.26 (0.20) & \textbf{0.02} (0.02) \\
& 3 & 0.20 (0.10) & 0.19 (0.09) & \textbf{0.02} (0.02) \\
\hline

\multirow{3}{*}{$K_A$}
& 1  & 0.34 (0.18) & 0.55 (0.21) & \textbf{0.09} (0.07) \\
& 2  & 0.32 (0.06) & 0.31 (0.06) & \textbf{0.11} (0.12) \\
& 3 & 0.10 (0.07) & 0.11 (0.12) & \textbf{0.05} (0.04) \\
\hline

\multirow{3}{*}{$\sigma_A$}
& 1  & 0.84 (0.01) & 0.83 (0.02) & \textbf{0.02} (0.02) \\
& 2  & 0.86 (0.02) & 0.85 (0.03) & \textbf{0.10} (0.04) \\
& 3 & 0.82 (0.01) & 0.82 (0.02) & \textbf{0.03} (0.03) \\
\hline

\end{tabular}
}
\end{table}

\begin{table}[htbp]
\centering
\caption{
Mean computation time and corresponding standard deviation (Std) for the Chen-Fliess (CF)-based and RK-based methods based on 200 replications.
}
\label{tab:time}

\renewcommand{\arraystretch}{1.15}
\setlength{\tabcolsep}{8pt}

\begin{tabular}{lcc}
\toprule
\textbf{Method} & \textbf{Mean Time (sec)} & \textbf{Std (sec)} \\
\midrule
CF-based   & \textbf{570.507}  & 54.693  \\
RK-based   & 3894.649 & 356.709 \\
\bottomrule
\end{tabular}
\end{table}

\section {Discussion and conclusions}

\paragraph {Enabling data-driven composite network modelling at whole brain scales.}
The question as to how large-scale dynamic causal patterns emerge from underlying neuronal interactions is one that has attracted much interest in neurological disease studies \citep{Friston2019, Forrester2020, Pereira2021, Forrester2024, Singh2025} and in statistics \citep{Pokern2009, Ditlevsen2025}. The existing imaging studies often focused on a single subject study, either with models for a small number of cortical columns or with models in a large scale but involving only excitatory/inhibitory neuron populations or with models in a large scale but for homogeneous brain regions, where all nodes in the network models are identical. In this paper, to relax the above limitations, we have proposed a heterogeneous composite network model, NccDCM, for groups of EEG scans respectively.
In the NccDCM, each node represents a local Jansen-Rit neural mass model for describing synaptic interactions in excitatory/inhibitory and pyramidal neuron populations, which consists of six-dimensional stochastic differential equations, while directed edges stand for coupling/transmissions between these local neuron populations across different channels. To account for individual subject-variability, we have introduced a mixed-effects model for network parameters. We have developed the Chen-Fliess expansion-based evolutionary optimisation algorithm for estimating NccDCMs. For ease of exposition, we have adopted a pairwise screening strategy to estimate edges in the proposed composite network. To identify dynamic causal nets that differs cases from controls, we have carried out a post-analysis of variance for estimated parameters in cases and controls. 
In the theoretical study, we have revealed that the excitation and inhibition parameters are not structurally identifiable although their ratio is locally structurally identifiable. A similar phenomenon has been found in conductance-based DCM by
\cite{Hauke2026}. This has given rise to the approach of reparametrisation of the original parameters that the new parameters, including the excitation/inhibition balance ratio, are locally structurally identifiable except the excitation parameter.  We have assessed the absolute relative errors of the model estimation, demonstrating that the performance of the proposed estimates for  the node and edge parameters is relatively accurate.

\paragraph {Detecting and quantifying differential epileptic network.}
Epilepsy is considered a network disease that affects the brain across multiple levels of spatial and temporal scales. The epileptic network, however, is not static but evolves in time which requires novel approaches for an in-depth characterisation \citep {Brohl2024}. The proposed biophysics-informed network of neural mass models aims to detect and quantify such a network.
In particular, we have pursued a computational approach to elucidating the difference of brain dynamic connectivity patterns called dynamic causal nets between cases and controls or between the preictal period and the ictal period.
We first fitted a NccDCM to the EEG scan for each subject in the case and control groups respectively. Following that, a mixed-effects model-based post-analysis of variance has been conducted to contrast estimated case-control parameters. This has given rise to various differential causal nets that differ the case-networks from the control-networks. The same analysis has also been performed on the segments of time series in preictal and ictal periods. We  have found similar differential causal nets that differ the preictal period from the ictal period. These nets are associated with epileptic seizures even after the Benjamini-Hochberg correction for multiple testing.

 It is widely recognised that EEG scans are inherently noisy with weak signals due to the minuscule amplitude of brain-generated electrical activity recorded at the scalp and the susceptibility to various sources of contamination. Only the synchronous firing of millions of neurons produces a measurable signal at the scalp.
 Despite this,  we have demonstrated the advantage of the NccDCM analysis over the conventional temporal correlation-based analysis in detecting links between regions. We have identified a list of dynamic causal nets to characterise the spatial-temporal varying epileptic network.  These differential dynamic causal nets partially covered default mode network (DMN), a collection of distributed and interconnected brain regions.  DMN is typically suppressed when the brain receives external stimuli; however, in the absence of attention to external stimuli, the DMN switches to internally focused thought processes. Therefore, the DMN activity can be observed in subject's resting states.  It follows from \cite{Menon2023} that the DMN nodes include posterior cingulate cortex (PCC) and retrosplenial cortex (RSC) in posterior medial parietal cortex; medial prefrontal cortex PFC (mPFC) with its dorsomedial (dmPFC) and ventromedial (vmPFC) subdivisions; anterior temporal cortex (ATC); middle temporal gyrus (MTG) in lateral temporal cortex; medial temporal lobe (MTL); and angular gyrus (AG) in lateral parietal cortex; and caudate nucleus (CAU).
Activities in PCC and RSC are captured by electrodes FP2, P3, P4 and O2; AG is captured by P3, P4, T5 and T6;
 PFC is captured by FP1. FP2, Fz, F3 and F4; ATC is captured by T3 and T4; MTG is captured by T5 and T6; MTL is captured by F3, T4, T5, and T6.  EEG channels do not directly cover subcortical structures like the thalamus (THAL) and caudate nucleus (CAU) because these deep brain regions do not produce sufficient electrical signals that reach the scalp surface to be localised to a single channel. In Figures~\ref{Fig:effetivenetwork} and \ref{Fig:effetivenetworkvariation},  the identified dynamic causal nets differ cases from controls in the means and variations of the estimated NccDCM parameters. These nets imply epilepsy-predisposed changes in the DMNs: (a) Edge changes in the excitatory/inhibitory synaptic ratio: AG/PCC points to PFC. (b) Many edge changes in the local structure parameters $C_A$. (c) Changes in the transmission parameter $K_A$: ATC points to PFC; AG points to PFC. (d) Edge changes in $\nu_0$:
AG points to PFC; ATC points to PFC.  Figure \ref{Fig:effetivenetworkpre},  compared the ictal period to the preictal period, yields many dynamic causal nets that have been related to the estimated region-transmission parameters. This finding implies that the changes in the region-transmission parameters can be used for chracterising onset of seizures.

Our findings have provided  a few epileptic network configurations than the DMN, highlighting the contribution of dysfunctions in the DMN to epileptic activities
and cognitive behavior in patients\citep[\, e.g.,][]{Jiang2022}. As epileptic seizures are associated with abnormal neuronal hyperexcitability, we hypothesize that alterations in the excitation–inhibition balance, local connectivity, or interregional connectivity within the patients’ default mode network (DMN) are associated with epilepsy.

\paragraph{Source level analysis.} Although the cortical regions identified in the $B_A$ differential causal net have previously
been implicated in epilepsy-related structural and functional abnormalities.
Epilepsy is increasingly
conceptualized as a disorder involving distributed structural and functional
brain networks that can extend beyond the putative seizure-onset region
\citep{Whelan2018}. However, the patients have heterogeneous and differently lateralised epilepsies, so group-level left- or right-hemisphere findings need to align the side of onset.

The regions identified in the present analysis included the left
banks of the superior temporal sulcus (bankssts), left lateral
orbitofrontal cortex, bilateral lingual cortex, left parahippocampal
cortex, left postcentral cortex, bilateral superior temporal cortex,
left transverse temporal cortex, right cuneus, bilateral insula,
right inferior temporal cortex, right inferior parietal cortex,
right precuneus, right isthmus cingulate cortex, bilateral
paracentral cortex, right medial orbitofrontal cortex, right
pars opercularis, and left lateral occipital cortex.

The temporal and limbic regions identified in the present analysis
are also consistent with the network-level alterations frequently
reported in temporal lobe epilepsy. In particular, the
parahippocampal and superior temporal regions form part of
distributed temporal networks that may show structural and
functional abnormalities in epilepsy.
The insular cortex warrants particular consideration. The insula
has extensive anatomical and functional connections with frontal,
temporal, parietal, and subcortical regions and can participate in
both seizure generation and seizure propagation
\citep{Solanki2023}. Insular seizures can also produce clinical
manifestations that resemble seizures originating from temporal,
frontal, or parietal regions. Consequently, increased connectivity
involving the insula may reflect its role as part of a distributed
epileptic network rather than necessarily indicating an isolated
insular seizure-onset zone \citep{Ryvlin2021,Solanki2023}.

Overall, the presence of these regions among the areas showing
altered directed connectivity is broadly consistent with previous
evidence that epilepsy involves distributed cortico-cortical and
cortico-subcortical networks. Nevertheless, the present connectivity
results alone cannot establish that any individual region is
epileptogenic. Rather, they indicate that these regions participate
in the pattern of directed functional interactions observed in the
studied population. This distinction is particularly important
when interpreting regions showing high in-degree or high out-degree.
A high in-degree indicates that a region receives directed
connections from many other regions, whereas a high out-degree
indicates that a region sends directed connections to many other
regions. These network measures therefore describe the organization
of directed connectivity and should not be interpreted directly as
measures of seizure onset or epileptogenicity.

\paragraph {Limitations.} We view our present model as having the following limitations: 1) algorithms for estimating the NccDCM and 2) the omission of subjects' demographic information.
We have developed a marginal screening approach to estimating the node-wise connection parameters. Marginal screening is a method used to estimate the coefficients of multiple external node inputs by identifying the most significant coefficients for each node. This approach is particularly useful when a large number of nodes involved. The method involves constructing a model that includes all external nodes and then selecting the most significant nodes to include in the final model. Our simulation studies have demonstrated that this method helps to avoid overfitting and ensures that the model is generalisable to new data. However, the method is at the cost of estimation efficiency.  The penalisation regression or forward/backward variable selection has been explored in another paper.
The second limitation potentially arise from the neglect of subjects' demographic information such as age and gender. However, the numbers of sujects in the case and control groups are relatively small. A further work is needed to take age and gender as additional predictors in each ccDCM for increased sizes of case-control gropus.

\paragraph {Conclusions and future work.} In summary, the proposed NccDCM is a novel and useful tool for detecting disease-predisposed effective connectivity in case-control study using EEG data. We have shown that the NccDCM can be used to characterise epileptic brain network at the whole brain scale in both sensor and source levels. The proposed method also demonstrates promising performance on synthetic data. This work can be extended to other neural mass models in \cite{Friston2019}.

\section*{Appendix A}
\paragraph{Proposition 1.} The canonical Jansen-Rit model (\ref{JR1}) is locally structurally identifiable.

{\bf Proof}: 
Note that the canonical Jansen-Rit model has non-rational vector field functions while the existing software SIAN for structural identifiability analysis is only applicable to neural mass differential equations with rational vector field functions. In order to use the software SIAN to analyse the identifiability of the canonical Jansen-Rit model with zero noises, we first bring the canonical Jansen-Rit model to the form with rational vector field functions described in \cite{Hong2019}  by introducing new states $x_i(t), i= A3, 4,5,6,7,8$:
\begin{eqnarray}\label{rational:JR}
\dot{x}_{A0}&=&x_{A3}, \quad \dot{x}_1(t)=x_4(t), \quad  \dot{x}_2(t)=x_5(t),\nonumber\\
 \dot{x}_{A3}(t)&=&ax_6(t)-2ax_{A3}(t)-a^2x_{A0}(t),\nonumber\\
 \dot{x}_4(t)&=&a[P(t)+0.8C_Ax_7(t)]-2ax_4(t)-a^2x_1(t),\nonumber\\
 \dot{x}_5(t)&=&B_A^{-1}b0.25C_Ax_8(t)-2bx_5(t)-b^2x_2(t),\\
 \dot{x}_6(t)&=&rx_6(t)(1-x_6(t))(x_4(t)-x_5(t)),\nonumber\\
 \dot{x}_7(t)&=&rx_7(t)(1-x_7(t))C_Ax_{A3}(t),\nonumber\\
 \dot{x}_8(t)&=&rx_8(t)(1-x_8(t))0.25C_Ax_{A3}(t),\nonumber\\
\end{eqnarray}
where
\begin{eqnarray*}
x_6(t)&=&1/(1+\exp(r(\nu_0-x_1(t)+x_2(t))),\quad x_7(t)=1/(1+\exp(r(\nu_0-C_Ax_{A0}(t))),\\
x_8(t)&=&1/(1+\exp(r(\nu_0-0.25C_Ax_{A0}(t))).\quad P(t)=\sum_{j\not=k} K^{(k|j)}_Ay_j(t)
\end{eqnarray*}
with the output $(y_j(t): j\in I_k).$
The SIAN analysis shows that the parameters $B_A, C_A, P(0),$ $ x_{A0}(0),x_1(0), x_2(0), x_{A3}(0)$, $x_i(t)$, $i=4,5,6,7,8$ in equations (\ref{rational:JR}) are locally structurally identifiable, but has not identified any globally structurally identifiable parameters. 

It follows from the definition that 
\begin{eqnarray*}
\nu_0&=&r^{-1}\log\left(\frac{1-x_7(0)}{x_7(0)}\right),\\
r&=&\frac 1{\nu_0-0.25C_Ax_{A0}(0)}\log\left(\frac {1-x_8(0)}{x_8(0)}\right)
\end{eqnarray*}
which imply that given the external input $z(t)$, parameters $\nu_0$ and $r$ are also locally structurally identifiable. Note that $P(t)$ is identifiable as $P(0)$ is identifiable.  Given time instances $t_1<\cdots<t_m $ with $m$ being larger than the size $|I_k|$ of $I_k$, we have
\begin{eqnarray*}
\begin{pmatrix} P(t_1)\\
                     \vdots\\
                     P(t_m) 
\end{pmatrix}=\begin{pmatrix} y_1(t_1) &\cdots &y_{|I_k|}(t_1)\\
                            \vdots&          &\vdots\\
                       y_1(t_m) &\cdots &y_{|I_k|}(t_m)
\end{pmatrix}
\begin{pmatrix} K^{(k|1)}\\
                     \vdots\\
                     K^{(k||I_k)|}
\end{pmatrix} 
\end{eqnarray*}
which yields unique values $K^{(k|j)}, j\in I_k$ under the condition ($\bf C0$).
The proof is completed.

\section*{Appendix B}

Consider the following set of differential system of the form
\begin{eqnarray}\label{cf1}
  \dot{z}(t) &=& g_0(z(t))+\sum\limits_{i=1}^{m}g_i(z(t))u_i(t),~~z(0)=z_0; ~~~ y(t) = h(z(t)),
\end{eqnarray}
where each $g_i$ is an analytic vector field expressed in some neighborhood of $z_0$, and the
 $h$ is an analytic output function. Assume that (\ref{cf1}) has a well defined solution $z(t)$, $t\in [t_0,T + t_0]$  for any given input $u_i(t)$, and $y(t)
=h(z(t))$, $t\in [t_0,T + t_0]$ .

\paragraph{Lemma}
  The solution to equation $(\ref{cf1})$ can be expressed as a Chen-Fliess series of the  form
  \begin{eqnarray}\label{cf2}
 y(t)&=&\sum\limits_{1\leq i_1,...,i_k\leq m,k\geq0}\left(\int_0^t\int_0^{\tau_k}\cdots\int_0^{\tau_2} u_{i_k}(\tau_k)\cdots u_{i_1}(\tau_1)d\tau_1\cdots d\tau_{k-1}d\tau_k\right)\non\\
 &&\times\left(L_{g_{i_1}}\circ\cdots \circ L_{g_{i_k}}h(z_0)\right),
\end{eqnarray}
where $L_{g_{i_1}}\circ\cdots \circ L_{g_{i_k}}h(z_0)$ is a Lie-derivative in the direction of the vector field. See \cite{Lyons1998}.

\paragraph{ Chen-Fliess expansion of states in model (\ref{first})).}  For a small increment $h$ and $t\in[t_0, t_0+h]$, truncating the Chen-Fliess expansion of $x_0(t)$ and $x_2(t)$ at the 3rd order, we have
\begin{eqnarray}\label{x0}
 x_{A0}(t)
 &=&x_{A0}(t_0)+x_{A3}(t_0)(t-t_0)\non\\
&&+\left(\frac{\nu_{\max}a}{1+e^{r(v_0-y(t_0))}}-a^2x_{A0}(t_0)-2ax_{A3}(t_0)\right)\frac{(t-t_0)^2}{2}\non\\
 &&+\left(\frac{r\nu_{\max}ae^{r(v_0-y(t_0))}\dot{y}(t_0)}{\left(1+e^{r(v_0-y(t_0))}\right)^2}-a^2x_{A3}(t_0)\right.\non\\
 &&\left. -2a\left(\frac{\nu_{\max}a}{1+e^{r(v_0-y(t_0)))}}-a^2x_{A0}(t_0)-2ax_{A3}(t_0)\right)
\right)\frac{(t-t_0)^3}{3!}\non\\
&&+O_p((t-t_0)^4),
\end{eqnarray}
\begin{eqnarray}\label{x2}
 x_2(t)
 &=&x_2(t_0)+x_5(t_0)(t-t_0)\non\\
&&+\left(\frac{\nu_{\max}B_A^{-1}b0.25C_{A}}{1+e^{r(v_0-C_{A}x_{A0}(t_0))}}-2bx_5(t_0)-b^2x_2(t_0)\right)\frac{(t-t_0)^2}{2}\non\\
&&+\left(\frac{r\nu_{\max}B_A^{-1}b(0.25C_{A})^2x_{A3}(t_0)e^{r(v_0-0.25C_{A}x_{A0}(t_0))}}{(1+e^{r(v_0-0.25C_{A}x_{A0}(t_0))})^2}
-b^2x_5(t_0)\right.\non\\
&&\left.-2b\left(\frac{\nu_{\max}B_A^{-1}b0.25C_{A}}{1+e^{r(v_0-0.25C_{A}x_{A0}(t_0))}}-2bx_5(t_0)-b^2x_2(t_0) \right)\right)\frac{(t-t_0)^3}{3!}\non\\
&&+O_p((t-t_0)^4).
\end{eqnarray}
Similarly, given the input $y^{(j)}(t)$ from channel $j$,  truncating the Chen-Fliess expansion of $x^(t)$ at the 2nd order, we have
\begin{eqnarray}\label{x1}
 x_1(t)
 &=&x_1(t_0)+x_4(t_0)(t-t_0)\non\\
&&+\left(\frac{\nu_{\max}a0.8C_{A}}{1+e^{r(v_0-C_{A}x_{A0}(t_0))}}
 +\frac{\nu_{\max}aK_A}{1+e^{r(v_0-y^{(j)}(t_0))}}-a^2x_1(t_0)-2ax_4(t_0)\right)\frac{(t-t_0)^2}{2}\non\\
 &&+\sigma_A a\int_{t_0}^{t}(t-s)d\tilde{w}_{s}+o_p((t-t_0)^3),
\end{eqnarray}
It follows from equations (\ref{x2}) and (\ref{x1}) that the truncated Chen-Fliess expansion of $y(t)$ at the 2nd order can be written as
\begin{eqnarray}\label{yg}
  y(t)&\approx&g(t|x_{-\{3\}}(t_0),y(t_0),\dot{y}(t_0),y^{(j)}(t_0),\bthe+\sigma_Aa\int_{t_0}^{t}(t-s)d\tilde{w}_{s},
\end{eqnarray}
where
\begin{eqnarray*}
&& g(t|x_{-\{3\}}(t_0),y(t_0),\dot{y}(t_0),y^{(j)}(t_0),\bthe)\\
 & &\quad= y(t_0)+\dot{y}(t_0)(t-t_0)\non\\
  &&\qquad+\left(\frac{\nu_{\max}a0.8C_{A}}{1+e^{r(v_0-C_{A}x_{A0}(t_0))}}
 +\frac{\nu_{\max}aK_A}{1+e^{r(v_0-y^{(j)}(t_0))}}-a^2x_1(t_0)-2ax_4(t_0)\right)\frac{(t-t_0)^2}{2}\non\\
 &&\qquad-\left(\frac{\nu_{\max}B^{-1}_Ab0.25C_{A}}{1+e^{r(v_0-0.25C_{A}x_{A0}(t_0))}}-2bx_5(t_0)-b^2x_2(t_0)\right)\frac{(t-t_0)^2}{2}.
\end{eqnarray*}
 It is clear that $g$ function depends on the initial values of states $x_{-\{A3\}}(t_0)$.

Let
\begin{eqnarray*}
L_A(t_0)&=&\frac{\nu_{\max}a}{1+e^{r(v_0-y(t_0))}}-a^2x_{A0}(t_0)-2ax_{A3}(t_0)\\
L_B(t_0)&=&\frac{\nu_{\max}B^{-1}_Ab0.25C_{A}}{1+e^{r(v_0-0.25C_{A}x_{A0}(t_0))}}-2bx_5(t_0)-b^2x_2(t_0).
\end{eqnarray*}
The 3rd-order Chen-Fliess series of $x_{A0}$ and $x_2$ are
\begin{eqnarray}\label{x0chen}
 &&g_0(t|x_{\{A0,A3\}}(t_0),y(t_0),\dot{y}(t_0))\non\\
 &=&x_{A0}(t_0)+x_{A3}(t_0)(t-t_0)+L_A(t_0)\frac{(t-t_0)^2}{2}\non\\
 &+&\left(\frac{r\nu_{\max}ae^{r(v_0-y(t_0))}\dot{y}(t_0)}{\left(1+e^{r(v_0-y(t_0))}\right)^2}-a^2x_{A3}(t_0)-2aL_A(t_0)\right)\frac{(t-t_0)^3}{3!},
\end{eqnarray}
and
\begin{eqnarray}\label{x2chen}
g_2(t|x_{-\{1,4\}}(t_0))
 &=&x_2(t_0)
+x_5(t_0)(t-t_0)+L_B(t_0)\frac{(t-t_0)^2}{2}\non\\
&&
 +\left(\frac{r\nu_{\max}B^{-1}_Ab0.25C_{A}0.25C_{A}x_{A3}(t_0)e^{r(v_0-0.25C_{A}x_{A0}(t_0))}}{(1+e^{r(v_0-0.25C_{A}x_{A0}(t_0))})^2}\right.\non\\
&&\left.-b^2x_5(t_0)-2bL_B(t_0)\right)\frac{(t-t_0)^3}{3!},
\end{eqnarray}
respectively.

 The 2nd-order Chen-Fliess series of $x_3$ and $x_5$ are
\begin{eqnarray}\label{x3chen}
&&g_3(t|x_{\{A0,A3\}}(t_0), y(t_0),\dot{y}(t_0))
 =x_{A3}(t_0)+L_A(t_0)(t-t_0)\non\\
& &\qquad+\left(\frac{r\nu_{\max}ae^{r(v_0-y(t_0))}\dot{y}(t_0)}{\left(1+e^{r(v_0-y(t_0))}\right)^2}-a^2x_{A3}(t_0)-2aL_A(t_0)\right)
\frac{(t-t_0)^2}{2},
\end{eqnarray}
and
\begin{eqnarray}\label{x5chen}
 g_5(t|x_{-\{1,4\}}(t_0))
&= &x_5(t_0)+L_B(t_0)(t-t_0)\non\\
&&
 +\left(\frac{r\nu_{\max}B^{-1}_Ab(0.25C_{A})^2x_{A3}(t_0)e^{r(v_0-0.25C_{A}x_{A0}(t_0))}}{(1+e^{r(v_0-0.25C_{A}x_{A0}(t_0))})^2}\right.\non\\
&&\left.-b^2x_5(t_0)-2bL_B(t_0)\right)\frac{(t-t_0)^2}{2},
\end{eqnarray}
respectively.

\section*{Author Contributions}
K.Y. conceptualisation, methodology, coding, data analysis, writing \& editing of original draft;
G.G. conceptualisation, methodology, review \& editing of original draft;
J.Z. conceptualisation, methodology, coding, data analysis, writing and editing of original draft, supervision,  funding acquisition and project administration.

\section*{Data and Code Availability}

Codes to reproduce simulation results are in https://github.com/zhangjsib/NccDCM. The EEG data on epilepsy is publicly available in \cite{OK2021}.

\section*{Funding}
The research of K.Y. and J.Z. is supported by the Engineering and Physical Sciences Research Council (EPSRC) grant (EP/X038297/1).

\section*{Declaration of Competing Interests}
K.Y.,  J.Z. and G.G. have no competing interests.

\section*{Acknowledgments}
We are grateful to the University of Kent for letting us use the University HPC service throughout the project.
We are grateful to the Senior Editor and reviewers for their specific, detailed, and valuable comments that have helped to improve the presentation of the paper. The use of chatGPT in polishing English wording is acknowledged. 

\section*{Supplementary material}

  Supplementary material in a pdf file includes additional information as follows.

S1:  The identifiability analysis of the canonical Jansen--Rit model.

S2: An introduction on the optimisation  algorithm JADE.

S3: Multiple permutation-based Wilcoxon signed-rank tests for significant NccDCM changes from the preictal period to the ictal period.

S4: Bootstrap bias correction.

S5: Lagged correlation-based functional connectivity.

S6: EEG electrode allocation on the scalp.

S7: Post-ANOVA tables.

S8: More results on subject-heterogeneity analysis.

S9: Bipolar EEG source localisation.

S10: Region of Interest Desikan-Killiany atlas.


S11: Bifurcation behavior of ccDCMs.

S12: Unidirectional circle network of three nodes.

S13: Investigate the more general case where one channel receives more than one inputs from other channels.

The pdf file of the Supplementary Material is available in https://github.com/zhangjsib/NccDCM.

\nolinenumbers

\bibliographystyle{chicago}
\bibliography{library}

\end{document}


\vspace*{0.35in}

\begin{flushleft}
{\Large
\textbf\newline{Differential Dynamic Causal Nets: Model Construction, Identification and Group Comparisons}
\\
\vspace{0.2cm}
\hspace{0.1cm}---Supplementary Material}
\newline
\\
Kang You\textsuperscript{1},
Gary Green\textsuperscript{2},
Jian Zhang\textsuperscript{1,*}
\\
\bigskip
\bf{1} School of Engineering, Mathematics and Physics, University of Kent, Canterbury, Kent CT2 7NF, UK.
\\
\bf{2} York Neuroimaging Centre, The Biocentre, York Science Park, University of York, Heslington, York YO10 5NY, UK.
\\
\bigskip
* Correponding author: Jian Zhang (J.Zhang-79@kent.ac.uk)

\end{flushleft}

\section*{Supporting information}
The supplementary material contains: The identifiability analysis of the canonical Jansen--Rit model (S1), an introduction to the JADE optimisation algorithm (S2), and permutation-based Wilcoxon signed-rank tests of NccDCM changes from the preictal to the ictal period (S3). It also describes bootstrap bias correction (S4), lagged correlation-based functional connectivity (S5), and EEG electrode allocation (S6). Further materials include post-ANOVA tables (S7), additional subject heterogeneity results (S8), bipolar EEG source localisation (S9), and regions of interest in the Desikan--Killiany atlas (S10). Finally, it presents the bifurcation behaviour of ccDCMs (S11), a three-node unidirectional ring network (S12), and the case in which a channel receives inputs from multiple other channels (S13).
\newpage

\section{The identifiability analysis of the cananical Jansen-Rit model}\label{Section:S1}
a) Lie symmetry analysis

We introduce the reparameterisation
\begin{eqnarray*}
    x_{A0}={x_0}/{A}, B_A=A/B, C_A=AC, K_A=AK, \nu_0=\nu_0.
\end{eqnarray*}
The resulting model is
\begin{eqnarray}\label{eq:reparam_model}
L_a x_{A0} &=&a S(y), \nonumber\\
L_a x_1
&=&a\left( K_Az+0.8C_A S(C_A x_{A0})\right),\nonumber\\
L_b x_2
&=&0.25{C_A}b{B_{A}}^{-1}
  S\!\left(0.25C_A x_{A0}\right),\\
y&=&x_1-x_2,\nonumber
\end{eqnarray}
where
\begin{eqnarray*}
L_a&=&\partial^2/\partial t^2+2a\partial /\partial t^2+a^2,
\qquad
L_b=\partial^2/\partial t^2+2b\partial /\partial t^2+b^2,\\
S(v)&=&
\frac{\nu_{\max}}
{1+\exp\left(r(v_0-v)\right)}.
\label{eq:sigmoid}
\end{eqnarray*}
The constants $a$, $b$, $e_0$ and $r$ and external input $z$ are assumed to be fixed,
whereas $x_{A0},x_1,x_2$ are hidden states and
$
\boldsymbol{\theta}
=
(B_{A},C_A,K_A,\nu_0)
$
are the unknown parameters.

For any continuous transformation $\phi(\cdot,\epsilon)$, if the output $y$ does not change when the parameter in (\ref{JR1}) take values $\phi((B_A,C_A,K_A,\nu_0),\epsilon), 0\le \epsilon\le\delta.$ Then, we can find a small ball in the parametric space such that the output $y$ also does not change under the following transformation in a sufficiently small ball:
\begin{eqnarray*}
(x^*_{A0},x^*_1,x^*_2,B^*_A,C^*_A,K^*_A,\nu^*_0)&=&(\psi(x_{A0},\epsilon_0),\psi(x_1,\epsilon_1),\psi(x_2,\epsilon_2),\psi(B_A,\epsilon_3),\\
&&\psi(C_A,\epsilon_4),\psi(K_A,\epsilon_5),\psi(\nu_0,\epsilon_6)),
\end{eqnarray*}
where
\begin{eqnarray*}
\psi(w,\epsilon)=\begin{cases}w+\epsilon, & w=0\\
                                                 \pm e^{\epsilon}w, &w\not=0
                               \end{cases}
\end{eqnarray*}
and $|\epsilon_i|\le \delta_i, i=0,1,...,6$. Let
\begin{eqnarray*}
\eta(w)=\partial\psi(w,\epsilon)/\partial\epsilon|_{\epsilon=0}=\begin{cases} 1, &w=0\\
                                                                                                                           \pm w, &w\not=0
                                                                                                                        \end{cases}.
\end{eqnarray*}
In general, suppose there exists a small parametric ball under which the output keeps the same for a time period. Then, we want to prove that the ball has zero radius.

For any inner point  $(\theta,x_{A0},x_1,x_2)$  in the parametric and state space, suppose that there exists a small ball in which for any $(x^*_{A0},x^*_1,x^*_2,\theta^*) $, the output is preserved. We assume that there exist nonegative constants $\delta_i, i=0,1,...,6$, in which we can define a scaling map within the ball:
$ x^*_{A0}=e^{\epsilon_0}x_{A0},$ $ x^*_{1}=e^{\epsilon_1}x_{1}$ and  $ x^*_{2}=e^{\epsilon_2}x_{2},$
 $\theta^*_i=e^{\epsilon_{2+i}}\theta_i, i=1,2,3,4,$  $|\epsilon_i|\le \delta_i$, $i=0,1,...,6.$
 To prove the local weak identifiability, it is sufficient to show that for the above Jansen-Rit model, only output-preserving scaling transformation is the identity mapping, that is, $\delta_i=0, 1,...,6$.

 Note that under the scaling transformations, the differential coefficients of $x_0,x_1,x_2, B_A,C_A, K_A,\nu_0$ are
$\eta(x_{A0})=\partial x^*_{A0}/\partial {\epsilon}|_{\epsilon=0}=x_{A0},$ $\eta(x_1)= x_1,$ $\eta(x_2)=x_2$,
$\eta(B_A)=B_A,$ $\eta(C_A)=C_A,$ $\eta(K_A)=K_A$ and $\eta(\nu_0)=\nu_0.$  Following the Lie-symmetry formulation in Merkt et al. (2015),  denote
 its infinitesimal generator by
\begin{eqnarray*}
X&=&
x_{A0}I(\delta_0>0)
  \frac{\partial}{\partial x_{A0}}
+x_1I(\delta_1>0)
  \frac{\partial}{\partial x_1}
+x_2I(\delta_2>0)
  \frac{\partial}{\partial x_2}
+{B_A}I(\delta_3>0)
  \frac{\partial}{\partial {B_A}}\nonumber\\
&&+{C_A}I(\delta_4>0)
  \frac{\partial}{\partial  C_A}
+K_AI(\delta_5>0)
  \frac{\partial}{\partial K_A}
+\nu_0I(\delta_6>0)
  \frac{\partial}{\partial \nu_0},
\end{eqnarray*}

According to Theorem 2 in \cite{Merkt2015}, the Lie symmetry preserves the output and, therefore, satisfies
$
    X(y)=0
$
which  gives
\begin{eqnarray}\label{eq:output_condition}
    x_1I(\delta_1>0)-x_2I(\delta_2>0)=0.
\end{eqnarray}


Following \cite{Yates2009}, define the first prolonged function
\begin{eqnarray*}
F_1=L_ax_{A0}-aS(y),
\end{eqnarray*}
with the infinitesimal invariance condition
\begin{eqnarray*}
X(F_1)=0
\end{eqnarray*}
which gives
\begin{equation}
L_a(x_{A0})I(\delta_0>0)
-S_v(y)(x_1I(\delta_1>0)-x_2I(\delta_2>0))
-aS_{v_0}(y){\nu_0}I(\delta_6>0)=0.
\label{eq:det1}
\end{equation}

Using the output condition
\eqref{eq:output_condition}, this reduces to
\begin{eqnarray*}
L_a(x_{A0})I(\delta_0>0)-aS_{v_0}(y){\nu_0}I(\delta_6>0)=0.
\end{eqnarray*}
Combining this with model (\ref{eq:reparam_model}) yields
\begin{eqnarray*}
aS(y)I(\delta_0>0)=aS_{v_0}(y)\nu_0I(\delta_6>0)=-ar\nu_{\max}S(y)
\left(1-S(y)/\nu_{\max}\right){\nu_0}I(\delta_6>0)
\end{eqnarray*}
as the derivatives of the sigmoid  $S$ with respect to $\nu$ and $\nu_0$ satisfy
\begin{eqnarray*}
S_v(v)\hat{=}\partial S(v)/\partial v=
r\nu_{\max}S(v)
\left(1-S(v)/\nu_{\max}\right)=-\partial S(v)/\partial \nu_{0}\hat{=}-S_{\nu_0}(\nu).
\end{eqnarray*}
We have
\begin{eqnarray*}
aI(\delta_0>0)=-ar\nu_{\max}\left(1-S(y)/\nu_{\max}\right)\nu_0I(\delta_6>0)
\end{eqnarray*}
which implies $a(\delta_0>0)=0$, $a\nu_0I(\delta_6>0)=0$ and
  the above equation implies
\begin{equation}
L_a(x_{A0})I(\delta_0>0)=0.
\label{eq:etaq_homogeneous}
\end{equation}


Define the second prolonged function
\begin{eqnarray*}
F_2
=
L_ax_1-a\left(K_A z
-0.8C_A S(C_Ax_{A0})\right).
\end{eqnarray*}
Note that
\begin{eqnarray*}
X(C_Ax_{A0})
=
x_{A0}{C_A}I(\delta_4>0)+C_Ax_{A0}I(\delta_0>0).
\end{eqnarray*}

Since $\nu_0I(\delta_6>0)=0$, the determining equation on the prolonged function
\begin{eqnarray*}
X(F_2)=0
\end{eqnarray*}
becomes
\begin{eqnarray}
&&L_a(x_1)I(\delta_1>0)-a\left({K_A}zI(\delta_5>0)-0.8C_A S(C_Ax_{A0}){C_A}I(\delta_4>0)\right.\nonumber\\
&&\left.-0.8C_A S_v(C_Ax_{A0})
\left(x_{A0}{C_A}I(\delta_4>0)+C_Ax_{A0}I(\delta_0>0)\right)\right)
=0.
\label{eq:det2}
\end{eqnarray}
Since $\delta_0=0,$ combining the above determing equation with the model (\ref{eq:reparam_model}) yields
\begin{eqnarray*}
0&=&a\left\{(K_Az+0.8C_AS(C_Ax_{A0}))I(\delta_1>0)\right.\\
&&\left.-K_AzI(\delta_5>0)+0.8C^2_AS(C_Ax_{A0})I(\delta_4>0)\right.\\
&&\left.+0.8C_AS_{\nu}(C_Ax_{A0})(x_{A0}C_AI(\delta_4>0)
+C_Ax_{A0}I(\delta_0>0))\right\}\\
&=&a\left\{(K_Az+0.8C_AS(C_Ax_{A0}))I(\delta_1>0)\right.\\
&&\left.-K_AzI(\delta_5>0)+0.8C^2_AS(C_Ax_{A0})I(\delta_4>0)\right.\\
&&\left.+0.8C_A^2x_{A0}S_{\nu}(C_Ax_{A0})(I(\delta_4>0)+I(\delta_0>0))\right\}
\end{eqnarray*}
which holds only if
\begin{eqnarray*}
aI(\delta_1>0)=0, aC_AI(\delta_4>0)=0, aK_AI(\delta_5>0)=0.
\end{eqnarray*}


Define the third prolonged function
\begin{eqnarray*}
F_3
=
L_bx_2
-
0.25b\frac{C_A}{B_A}
S(0.25C_Ax_{A0})
\end{eqnarray*}

\begin{eqnarray*}
X(F_3)=0.
\end{eqnarray*}

Note that
\begin{eqnarray*}
X\left(
\frac{C_A}{B_A}
\right)
&=&
\frac{C_A}{B_{A}}I(\delta_4>0)
-
\frac{C_A}{B_{A}^2}B_AI(\delta_3>0)=-\frac{C_A}{B_{A}^2}B_AI(\delta_3>0),\\
X(0.25C_Ax_{A0})
&=&
0.25(x_{A0}C_AI(\delta_4>0)+C_Ax_{A0}I(\delta_0>0)).
\end{eqnarray*}

Thus the determining equation on the prolonged function $F_3$ becomes
\begin{eqnarray*}
0&=&
L_b(x_2)I(\delta_2>0)+0.25b\frac{C_A}{B_{A}^2}B_AI(\delta_3>0)S(0.25C_Ax_{A0})\nonumber\\
&&-0.25b{C_A}{B_{A}}^{-1}
  S(0.25C_A x_{A0})0.25(x_{A0}C_AI(\delta_4>0)+C_Ax_{A0}I(\delta_0>0))\nonumber\\
&=&0.25b{C_A}{B_{A}}^{-1}
  S(0.25C_A x_{A0})(I(\delta_2>0)+I(\delta_3>0))\nonumber\\
&&-0.25^2b{C_A}{B_{A}}^{-1}
  S(0.25C_A x_{A0})(x_{A0}C_AI(\delta_4>0)+C_Ax_{A0}I(\delta_0>0))\nonumber\\
&=&0.25b{C_A}{B_{A}}^{-1}
  S(0.25C_A x_{A0})\nonumber\\
&&\times \left( I(\delta_2>0)+I(\delta_3>0)-0.25x_{A0}C_A(I(\delta_4>0)+I(\delta_0>0))\right)
\end{eqnarray*}
which yields
\begin{eqnarray}\label{eq:det3}
0=bC_A(I(\delta_2>0), 0=bC_A(I(\delta_3>0)=0,  bC_AI(\delta_4>0))=0, bC_AI(\delta_0>0)=0.
\end{eqnarray}

Combining the above two equations give
\begin{equation}
L_b(x_2)I(\delta_2>0)=0.
\end{equation}

The output-preserving condition
\[
x_1I(\delta_1>0)=x_2I(\delta_2>0).
\]
gives $L_a(x_2)I(\delta_2>0)=L_a(x_1)I(\delta_1>0)=0.$

Let $D_t=\partial/\partial t$. subtracting $L_b$ from $L_a$, on $\delta_2>0,$ we have
\begin{eqnarray*}
0=(La-L_b)(x_2)=2(a-b)D_tx_2+(a^2-b^2)x_2.
\end{eqnarray*}
For $a\neq b$,
\begin{eqnarray*}
2D_tx_2+(a+b)x_2=0
\end{eqnarray*}
which yields
\[
x_2(t)=c\exp(-(a+b)t/2).
\]
Substitution $x_2(t)$ into $L_a(x_2)=0$ gives
\begin{eqnarray*}
\frac{(a-b)^2}{4}c=0.
\end{eqnarray*}
Hence, for $a\neq b$,
\begin{equation}
x_2(t)=0
\end{equation}
which implies
\[
0=L_b(x_)2=0.25{C_A}b{B_{A}}^{-1}
  S\!\left(0.25C_A x_{A0}\right).
\]
On $(\delta_2>0)$, $bC_A=0.$ Similarly, on $(\delta_1>0),$ $aC_A=0$ and $x_1(t)=0.$

In summary, for $a\not= b$ and non-zero $B_A, C_A, K_A,\nu_0$, we have $\delta_i=0, i=0,1,...,6.$
Therefore, there is no nontrivial continuous Lie transformation
acting on the parameter vector
\[
(B_{A},C_A, K_A, v_0)
\]
while preserving the output.

For the cases that $a\not-b$ or $a=0$, the differential equations in (\ref{eq:reparam_model}) have no solutions.

Consequently, parameters
$
(B_{A},
C_A,
K_A,
v_0)$
have no non-trival Lie symmetry for the canonical Jansen-Rit model.

\vskip 0.2in
b) The output of structural identifiability analysis (SIAN) of the rationalised Jansen-Rit model defined in the proof of
Proposition 1.
\begin{verbatim}
=======================================================
0. Extracting states, inputs, outputs, and parameters from the system
=======================================================

=== Input info ===
State variables:          [x0(t), x1(t), x2(t), x3(t), x4(t), x5(t), 
                          x6(t), x7(t), x8(t)]
Output variables:         [y(t)]
Input variables:          []
Parameters in equations:  [B, C, K]
===================


=======================================================
1. Constructing the maximal polynomial system
=======================================================

=======================================================
2. Truncating the polynomial system based on the Jacobian condition
=======================================================

=======================================================
3. Assessing local identifiability
=======================================================
Locally identifiable paramters:  [B, C, K, x0(0), x1(0),
 x2(0), x3(0), x4(0), x5(0), x6(0), x7(0), x8(0)]
Nonidentifiable parameter:  []

=======================================================
4. Randomizing the truncated system
=======================================================

=======================================================
Applying Weighted Ordering
=======================================================

=======================================================
5. Assessing global identifiability
=======================================================

NULL;

\end{verbatim}

\section{The optimisation algorithm JADE}\label{Section:S2}

To minimize the forward loss function, we employed an adaptive differential evolution with optional external archive (JADE), a population-based global optimization method proposed by \cite{Zhang2009} and implemented in the \texttt{mealpy} library.
 JADE is well suited for non-convex, multi-modal, and non-differentiable optimization problems, as it does not rely on gradient information. Instead, it evolves a population of candidate solutions through mutation, crossover, and selection, with the aim of converging to the global minimum. The method is particularly advantageous in our context because the objective function may be non-smooth or contain multiple local minima due to the models nonlinear structure.

As a metaheuristic optimization algorithm, the performance of JADE is sensitive to the choice of hyperparameters,  the initial population size and epoch size. These two hyperparameters substantially affect both the convergence speed and the quality of the final solution.

Figure~\ref{population_real_2} illustrates how the value of the loss function evolves under different settings of population size and iteration number in simulation studies, emphasizing the role of these hyperparameters in shaping optimisation performance.
It is evident that almost algorithm JADE   eventually converges to the same optimal value when the number of iterations exceeds 150 for almost all population sizes. Intuitively, smaller populations may force the algorithm to converge to a local minimum, limiting the coverage of its search space. In contrast, larger populations generally facilitate faster convergence owing to enhanced diversity, though further increases in population size yield diminishing returns in performance while incurring substantially higher computational costs.
To strike a balance between convergence accuracy and computational efficiency, we set the population size and the maximum number of iterations to $15$ and $250$, respectively.

We assessed the sensitivity of the JADE-based optimisation by fitting the NccDCM to the EEG data from subject P3, obtaining the parameter estimates $(\widehat{B}_A,\widehat{C}_A,\widehat{K}_A,\widehat{v}_0,\widehat{\sigma}_A)$. For each parameter in turn, we evaluated the loss function while holding the other four parameters fixed at their estimated values. The resulting one-dimensional loss profiles (Figure~\ref{fig:objective-curves}) appear locally convex in the vicinity of the estimated parameter values, although several local minima are also evident.

\begin{figure}[htbp]
  \centering

  \begin{subfigure}[t]{0.31\linewidth}
    \centering
    \includegraphics[width=\linewidth]{figures/objective_curve_AB.png}
    \caption{$B_A$}
    \label{fig:objective-BA}
  \end{subfigure}\hfill
  \begin{subfigure}[t]{0.31\linewidth}
    \centering
    \includegraphics[width=\linewidth]{figures/objective_curve_AC.png}
    \caption{$C_A$}
    \label{fig:objective-CA}
  \end{subfigure}\hfill
  \begin{subfigure}[t]{0.31\linewidth}
    \centering
    \includegraphics[width=\linewidth]{figures/objective_curve_AK.png}
    \caption{$K_A$}
    \label{fig:objective-KA}
  \end{subfigure}

  \par\medskip

  \begin{subfigure}[t]{0.31\linewidth}
    \centering
    \includegraphics[width=\linewidth]{figures/objective_curve_v0.png}
    \caption{$v_0$}
    \label{fig:objective-v0}
  \end{subfigure}\hspace{0.04\linewidth}%
  \begin{subfigure}[t]{0.31\linewidth}
    \centering
    \includegraphics[width=\linewidth]{figures/objective_curve_Asigma.png}
    \caption{$\sigma_A$}
    \label{fig:objective-sigmaA}
  \end{subfigure}

  \caption{Plots of loss function in $B_A$, $C_A$, $K_A$, $v_0$,
  and $\sigma_A$.}
  \label{fig:objective-curves}
\end{figure}
\begin{figure}[h!]
  \centering
  \begin{minipage}{0.48\textwidth}
    \centering
    \includegraphics[width=\linewidth, height=6cm]{figures/JADEconvergence/Dif_P3_population_difference_plot.png}

    {\small P3 }
  \end{minipage}
  \hfill
  \begin{minipage}{0.48\textwidth}
    \centering
    \includegraphics[width=\linewidth, height=6cm]{figures/JADEconvergence/Dif_P6_population_difference_plot.png}

    {\small P6}
  \end{minipage}

  \begin{minipage}{0.48\textwidth}
    \centering
    \includegraphics[width=\linewidth, height=6cm]{figures/JADEconvergence/Dif_P8_population_difference_plot.png}

    {\small P8 }
  \end{minipage}
  \hfill
  \begin{minipage}{0.48\textwidth}
    \centering
    \includegraphics[width=\linewidth, height=6cm]{figures/JADEconvergence/Dif_H2_population_difference_plot.png}

    {\small H2}
  \end{minipage}

  \begin{minipage}{0.48\textwidth}
    \centering
    \includegraphics[width=\linewidth, height=6cm]{figures/JADEconvergence/Dif_H5_population_difference_plot.png}

    {\small H5 }
  \end{minipage}
  \hfill
  \begin{minipage}{0.48\textwidth}
    \centering
    \includegraphics[width=\linewidth, height=6cm]{figures/JADEconvergence/Dif_H6_population_difference_plot.png}

    {\small H6}
  \end{minipage}

  \caption{Examples of trace plots of the loss function as the number of iterations is increasing for subjects P3, P6, P8, H2, H5 and  H6: L-Shape under different population sizes.}
  \label{population_real_2}
\end{figure}

\linenumbers

\section{Multiple permutation-based Wilcoxon signed-rank tests for significant NccDCM changes from the preictal period to the ictal period  }\label{Section:S3}

For each of parameters in \((B^{(ilj)k}_{Anq}, C^{(i|j)k}_{Anq}, K^{(i|j)k}_{Anq}, v^{(i|j)k}_{0nq}, \sigma^{(i)k}_{Anq})\) and for each node pair, let \(\Pi_{\mathrm{pre}}^{(\cdot)}\) and \(\Pi_{\mathrm{on}}^{(\cdot)}\) denote the matrices of estimates derived from the NccDCM in the preictal and ictal periods respectively. Column of \(\Pi_{\mathrm{pre}}^{(\cdot)}\) and \(\Pi_{\mathrm{on}}^{(\cdot)}\) stand for the estimators of  different case-subjects indexed by $1\le n\le N_1 $ while rows correspond to different time segments indexed by $1\le q\le 6$. We performed multiple permutation-based Wilcoxon signed-rank tests for significant NccDCM changes from the preictal period to the ictal period  They are implemented in four steps as follows.

\begin{itemize}
  \item \textbf{Compute the test statistics \(W = (W_1, W_2, \ldots, W_{N_1})^{\top}\).}
  For subject \(n\), denote the \(n\)-th columns of \(\Pi_{\mathrm{pre}}^{(\cdot)}\) and \(\Pi_{\mathrm{on}}^{(\cdot)}\) as \(\pi_{\mathrm{pre}}^{(\cdot,j)}\) and \(\pi_{\mathrm{on}}^{(\cdot,j)}\), respectively.
  Compute the paired differences \(\pi_{\mathrm{on}}^{(\cdot,n)} - \pi_{\mathrm{pre}}^{(\cdot,n)}\).
  After removing zero differences, rank the absolute values of the remaining differences in ascending order, assigning average ranks in the case of ties.
  The statistic \(W_n\) is then obtained by summing the ranks corresponding to positive differences, i.e., those instances where \(\pi_{\mathrm{on}}^{(\cdot,n)} > \pi_{\mathrm{pre}}^{(\cdot,n)}\).

  \item \textbf{Pair-matched permutation procedure.}
 For each column in  \(\Pi_{\mathrm{pre}}^{(\cdot)}\), find the corresponding column in \(\Pi_{\mathrm{on}}^{(\cdot)}\). We randomly swap the corresponding entries in the two columns, yielding \(2^6\) distinct permutations.
  For each permutation \(\alpha\),  we denote the resulted permuted matrices by \(\Pi_{\mathrm{pre}}^{(\cdot),\alpha}\) and \(\Pi_{\mathrm{on}}^{(\cdot),\alpha}\).
  \(W^{\alpha}\) is reduced to the observed statistic \(W\) when $\alpha $ is an identity permutation. The procedure is implemented as follows.

  \item \textbf{Compute the empirical \(p\)-values.}
  For the \(n\)-th subject, the empirical \(p\)-value is computed as
  \begin{align*}
    p^{(n)} = \frac{\#\{ W_n^{\alpha} \geq W_n \}}{\text{Total number of permutations}} \,,
  \end{align*}
  where the numerator counts the number of permuted statistics at least as extreme as the observed \(W_n\).
  This yields the individual \(p\)-value for each subject.

  \item \textbf{Combine the \(p\)-values using Fisher's method.}
  Combining these individual \(p\)-values using Fisher's method, we have
  the combined test statistic
  \begin{align*}
    \chi^2_{\mathrm{Fisher}} = -2 \sum_{n=1}^{N_1} \ln p^{(n)}.
  \end{align*}
  The above statistic follows a chi-square distribution with \(2 \times N_1\) degrees of freedom under the null hypothesis that there are no changes in the parameters from the preictal period to the ictal period.
  The overall significance can derived from the above chi-square distribution.
\end{itemize}
Analogous to the case-control group comparison, adjustment for multiple input effects can be also implemented. The details are omitted.

\section{Bootstrap bias correction}\label{Section:S4}

 The detailed parameter estimation procedure is described in the following steps.
\begin{description}
  \item[Step 1] Apply the Chen--Fliess method to construct the objective function (\textbf{see Section 2.2}).
  \item[Step 2] The parameters are estimated using the genetic evolutionary algorithm JADE, yielding the estimator
$
\widehat{\theta} = (\widehat{B}_A,\; \widehat{C}_A,\; \widehat{K}_A,\; \widehat{v}_0,\; \widehat{\sigma}_A).
$
  \item[Step 3] Based on the estimated parameter values $\widehat{\theta}$ and the canonical Jansen--Rit model \textbf{(4)}, 
  we generate $n_{\mathrm{boot}}$ synthetic EEG signals and use these signals to obtain $n_{\mathrm{boot}}$ bootstrap estimators
$
\widehat{\theta}^{*,b},  b = 1,\ldots,n_{\mathrm{boot}}.
$
\end{description}
Define the bootstrap mean estimator by
$
\widehat{\theta}^{*}
=
\frac{1}{n_{\mathrm{boot}}}
\sum_{b=1}^{n_{\mathrm{boot}}}
\widehat{\theta}^{*,b}.
$
Then, the bias-corrected estimator is given by
$
\widehat{\theta}_{B}
=
2\widehat{\theta}
-
\widehat{\theta}^{*}.
$

Note that, during the bootstrap procedure, since the true parameter values used to generate the synthetic EEG signals are known, we employ the \texttt{minimize} function provided in \texttt{scipy} for parameter estimation. Moreover, the initial value of the optimization algorithm is set to $\widehat{\theta}$ to improve computational efficiency.

\section{Lagged correlation-based functional connectivity}\label{Section:S5}
We calculate the functional connectivity based on the lagged correlation coefficient.
The EEG signals used are extracted from the same time block across all subjects and from all channels.
For each subject, we compute pairwise cross-correlations between all channel pairs across time lags ranging from $-50$ to $+50$.
For each channel pair, the maximum absolute correlation across these time lags is calculated, which allows us to construct correlation matrices for all subjects.

To normalise the correlation values, we apply Fisher's $z$-transform, yielding the matrices
$F_{P_i}$ ($1 \leq i \leq n_0$) and $F_{H_j}$ ($1 \leq j \leq n_1$).
For the control and case groups, the mean matrices are then computed as
$
\overline{F}_P = \frac{1}{n_0}\sum_{i=1}^{n_0} F_{P_i},
$
$
\overline{F}_H = \frac{1}{n_1}\sum_{j=1}^{n_1} F_{H_j}.
$
Then, the corresponding variance matrices are defined as
$
\overline{V}_P = \frac{1}{n_0}\sum_{i=1}^{n_0} \bigl(F_{P_i}-\overline{F}_P \bigr)^2
$
and
$
\overline{V}_H = \frac{1}{n_1}\sum_{i=1}^{n_1} \bigl(F_{H_i}-\overline{F}_H \bigr)^2.
$
Let $\overline{\rho}_{ij}^{(0)}$, $\overline{\rho}_{ij}^{(1)}$,
$\overline{\varrho}_{ij}^{(0)}$, and $\overline{\varrho}_{ij}^{(1)}$
denote the $(i,j)$-th elements of $\overline{F}_P$, $\overline{F}_H$,
$\overline{V}_P$, and $\overline{V}_H$, respectively.
For each channel pair, we compute the z-score as
\[
f_{ij} = \frac{\overline{\rho}_{ij}^{(0)}-\overline{\rho}_{ij}^{(1)}}%
{\sqrt{\dfrac{(n_0-1)\,\overline{\varrho}_{ij}^{(0)}+(n_1-1)\,\overline{\varrho}_{ij}^{(1)}}%
{n_0+n_1-2}}}.
\]
Finally, only the values of $|f_{ij}|$ greater than $0.5$ are retained, and the results are  based on the absolute z-scores.

In addition, we apply the following permutation test procedure  to compute the $p$-values.

\begin{enumerate}
  \item \textbf{Permutation procedure.}
  Let \(\mathcal{D} = (H_1, \dots, H_{10}, P_1, P_3, P_5, \dots, P_{10})\) denote the combined dataset.
  To assess the statistical significance of the observed \(F=(f_{ij})\), we apply a permutation test: the labels ``case'' and ``control'' are randomly reassigned by permuting \(\mathcal{D}\).
  In each permutation, the first 10 subjects in the permuted dataset \(\mathcal{D}^{\pi}\) are designated as the control group and the remaining subjects as the case group.
  Then, we recompute the test statistic  \(f_{ij}^{\pi}\).

  \item \textbf{Significance assessment.}
  This permutation procedure is repeated 999 times.
  The observed statistic \(f_{ij}\) is included as part of the permutation distribution \(\{f_{ij}^{\pi}\}\).
  The empirical $p$-value is then given by
  \begin{align*}
    p = \frac{\#\{|f_{ij}^{\pi}| \geq |f_{ij}|\}}{\text{Total number of permutations}} \,,
  \end{align*}
  where the numerator counts the number of permuted statistics at least as extreme as the observed \(f_{ij}\).
\end{enumerate}

\section{EEG electrode allocation on the scalp}\label{Section:S6}
Table \ref{tab:eeg_positions} shows the positions of the electrodes and the brain regions that the channels cover.

\begin{table}[htbp]
\centering
\caption{EEG electrode pairs and their brain regions}
\begin{tabular}{lll}
\hline
\textbf{Electrode pair} & \textbf{ Position} & \textbf{Brain Region} \\
\hline
FP2F4   & Right frontopolar             & Right prefrontal cortex \\
P4O2    & Right parieto-occipital   & Visual association \\
F4C4    & Right fronto-central      & Right frontal $\rightarrow$ motor cortex \\
F3C3    & Left fronto-central       & Left frontal $\rightarrow$ motor cortex \\
C4P4    & Right centro-parietal     & Right sensorimotor $\rightarrow$ parietal \\
C3P3    & Left centro-parietal      & Left sensorimotor $\rightarrow$ parietal \\
FP1F3   & Left frontopolar             & Left prefrontal cortex \\
P3O1    & Left parieto to occipital    & Visual cortex \\
FP2F8    & Right frontopolar    &Anterior inferior frontal  \\
F8T4    &Right inferior frontal     &Broca homolog \\
T4T6    &Right mid to posterior temporal    &Auditory association  \\
T6O2    &Right posterior temporal    &Visual association  \\
FP1F7   & Left frontopolar   & Anterior inferior frontal cortex \\
F7T3    &Left inferior frontal      & Broca area vicinity \\
T3T5   &Left mid to posterior temporal lobe   &Auditory and language \\
T5O1    &Left posterior temporal to occipital   & Visual association \\
FZCZ    & Medial frontal cortex    &  Supplementary motor\\
CZPZ    & Central midline to parietal midline   & Primary motor \\
BP3REF    & A reference electrode   &Not a cortical region  \\
\hline
\end{tabular}
\label{tab:eeg_positions}
\end{table}

\newpage
\section{Post-ANOVA tables}\label{Section:S7}

 The results of the post ANOVA are summarised in Tables \ref{tab:permanova_AB_SENSOR_STANDARD}---\ref{tab:permanova_Asigma_SOURCE_STANDARD}. The results indicated that the identified dynamic causal nets were highly significant for each parameter.
\begin{table}
\caption{Post ANOVA: $B_A.$}
\label{tab:permanova_AB_SENSOR_STANDARD}
\begin{tabular}{lrrrrr}
\toprule
\textbf{Factor} & \textbf{Df} & \textbf{Sum of Sq.} & $\mathbf{R^2}$ & \textbf{F} & \textbf{p-value} \\
\midrule
Case-control & 1 & 361.3590 & 0.0471 & 268.4587 & 0.0002 \\
Channel & 305 & 405.1509 & 0.0528 & 0.9869 & 0.5942 \\
Case-control:Channel & 305 & 311.8932 & 0.0407 & 0.7597 & 1.0000 \\
Residual & 4896 & 6590.2635 & 0.8594 & -- & -- \\
Total & 5507 & 7668.6666 & 1.0000 & -- & -- \\
\bottomrule
\end{tabular}
\end{table}

\begin{table}
\caption{Post ANOVA: $C_A.$}
\label{tab:permanova_AC_SENSOR_STANDARD}
\begin{tabular}{lrrrrr}
\toprule
\textbf{Factor} & \textbf{Df} & \textbf{Sum of Sq.} & $\mathbf{R^2}$ & \textbf{F} & \textbf{p-value} \\
\midrule
Case-control & 1 & 505865346.0343 & 0.0191 & 108.8468 & 0.0002 \\
Channel & 305 & 2003473690.6040 & 0.0757 & 1.4134 & 0.0002 \\
Case-control:Channel & 305 & 1186482323.8943 & 0.0449 & 0.8370 & 1.0000 \\
Residual & 4896 & 22754144867.7666 & 0.8603 & -- & -- \\
Total & 5507 & 26449966228.2993 & 1.0000 & -- & -- \\
\bottomrule
\end{tabular}
\end{table}

\begin{table}
\caption{Post ANOVA: $K_A.$}
\label{tab:permanova_AK_SENSOR_STANDARD}
\begin{tabular}{lrrrrr}
\toprule
\textbf{Factor} & \textbf{Df} & \textbf{Sum of Sq.} & $\mathbf{R^2}$ & \textbf{F} & \textbf{p-value} \\
\midrule
Case-control & 1 & 102982330.7693 & 0.0150 & 104.0527 & 0.0002 \\
Channel & 305 & 1469172393.7312 & 0.2144 & 4.8670 & 0.0002 \\
Case-control:Channel & 305 & 434145625.5117 & 0.0634 & 1.4382 & 0.0002 \\
Residual & 4896 & 4845636908.6169 & 0.7072 & -- & -- \\
Total & 5507 & 6851937258.6290 & 1.0000 & -- & -- \\
\bottomrule
\end{tabular}
\end{table}

\begin{table}
\caption{Post ANOVA: $\nu_0$.}
\label{tab:permanova_v0_SENSOR_STANDARD}
\begin{tabular}{lrrrrr}
\toprule
\textbf{Factor} & \textbf{Df} & \textbf{Sum of Sq.} & $\mathbf{R^2}$ & \textbf{F} & \textbf{p-value} \\
\midrule
Case-control & 1 & 184766.4927 & 0.0296 & 168.7859 & 0.0002 \\
Channel & 305 & 426044.4232 & 0.0684 & 1.2761 & 0.0002 \\
Case-control:Channel & 305 & 262557.1241 & 0.0421 & 0.7864 & 1.0000 \\
Residual & 4896 & 5359552.0694 & 0.8599 & -- & -- \\
Total & 5507 & 6232920.1094 & 1.0000 & -- & -- \\
\bottomrule
\end{tabular}
\end{table}

\begin{table}
\caption{Post ANOVA: $\sigma_A$.}
\label{tab:permanova_Asigma_SENSOR_STANDARD}
\begin{tabular}{lrrrrr}
\toprule
\textbf{Factor} & \textbf{Df} & \textbf{Sum of Sq.} & $\mathbf{R^2}$ & \textbf{F} & \textbf{p-value} \\
\midrule
Case-control & 1 & 15530065.7022 & 0.1430 & 873.5486 & 0.0002 \\
Channel & 305 & 4757543.9657 & 0.0438 & 0.8774 & 0.9592 \\
Case-control:Channel & 305 & 1287117.7224 & 0.0119 & 0.2374 & 1.0000 \\
Residual & 4896 & 87041755.7707 & 0.8014 & -- & -- \\
Total & 5507 & 108616483.1611 & 1.0000 & -- & -- \\
\bottomrule
\end{tabular}
\end{table}

\begin{table}
\caption{Source Post ANOVA: $B_A$.}
\label{tab:permanova_AB_SOURCE_STANDARD}
\begin{tabular}{lrrrrr}
\toprule
\textbf{Factor} & \textbf{Df} & \textbf{Sum of Sq.} & $\mathbf{R^2}$ & \textbf{F} & \textbf{p-value} \\
\midrule
Case-control & 1 & 181.7473 & 0.0226 & 99.5304 & 0.0002 \\
Cortical region & 235 & 595.5941 & 0.0740 & 1.3879 & 0.0002 \\
Case-control:Cortical region & 235 & 377.3340 & 0.0469 & 0.8793 & 0.9837 \\
Residual & 3776 & 6895.1545 & 0.8566 & -- & -- \\
Total & 4247 & 8049.8298 & 1.0000 & -- & -- \\
\bottomrule
\end{tabular}
\end{table}

\begin{table}
\caption{Source Post ANOVA: $C_A$.}
\label{tab:permanova_AC_SOURCE_STANDARD}
\begin{tabular}{lrrrrr}
\toprule
\textbf{Factor} & \textbf{Df} & \textbf{Sum of Sq.} & $\mathbf{R^2}$ & \textbf{F} & \textbf{p-value} \\
\midrule
Case-control & 1 & 640100846.1766 & 0.0535 & 258.3829 & 0.0002 \\
Cortical region & 235 & 1228835491.5024 & 0.1026 & 2.1108 & 0.0002 \\
Case-control:Cortical region & 235 & 750632429.7321 & 0.0627 & 1.2894 & 0.0002 \\
Residual & 3776 & 9354414664.0195 & 0.7812 & -- & -- \\
Total & 4247 & 11973983431.4305 & 1.0000 & -- & -- \\
\bottomrule
\end{tabular}
\end{table}

\begin{table}
\caption{Source Post ANOVA: $K_A$.}
\label{tab:permanova_AK_SOURCE_STANDARD}
\begin{tabular}{lrrrrr}
\toprule
\textbf{Factor} & \textbf{Df} & \textbf{Sum of Sq.} & $\mathbf{R^2}$ & \textbf{F} & \textbf{p-value} \\
\midrule
Case-control & 1 & 7227433.4849 & 0.0018 & 12.2902 & 0.0003 \\
Cortical region & 235 & 1539002062.1353 & 0.3869 & 11.1364 & 0.0002 \\
Case-control:Cortical region & 235 & 211223240.4413 & 0.0531 & 1.5284 & 0.0002 \\
Residual & 3776 & 2220533084.5214 & 0.5582 & -- & -- \\
Total & 4247 & 3977985820.5830 & 1.0000 & -- & -- \\
\bottomrule
\end{tabular}
\end{table}

\begin{table}
\caption{Source Post ANOVA: $\nu_0$.}
\label{tab:permanova_v0_SOURCE_STANDARD}
\begin{tabular}{lrrrrr}
\toprule
\textbf{Factor} & \textbf{Df} & \textbf{Sum of Sq.} & $\mathbf{R^2}$ & \textbf{F} & \textbf{p-value} \\
\midrule
Case-control & 1 & 132730.9279 & 0.0267 & 122.2962 & 0.0002 \\
Cortical region & 235 & 420670.2229 & 0.0847 & 1.6494 & 0.0002 \\
Case-control:Cortical region & 235 & 315820.1184 & 0.0636 & 1.2383 & 0.0002 \\
Residual & 3776 & 4098179.8733 & 0.8250 & -- & -- \\
Total & 4247 & 4967401.1424 & 1.0000 & -- & -- \\
\bottomrule
\end{tabular}
\end{table}

\begin{table}
\caption{Source Post ANOVA: $\sigma_A$.}
\label{tab:permanova_Asigma_SOURCE_STANDARD}
\begin{tabular}{lrrrrr}
\toprule
\textbf{Factor} & \textbf{Df} & \textbf{Sum of Sq.} & $\mathbf{R^2}$ & \textbf{F} & \textbf{p-value} \\
\midrule
Case-control & 1 & 7298611.6045 & 0.0185 & 103.2300 & 0.0002 \\
Cortical region & 235 & 114634103.3495 & 0.2901 & 6.8994 & 0.0002 \\
Case-control:Cortical region & 235 & 6276374.4906 & 0.0159 & 0.3778 & 1.0000 \\
Residual & 3776 & 266972371.2363 & 0.6756 & -- & -- \\
Total & 4247 & 395181460.6809 & 1.0000 & -- & -- \\
\bottomrule
\end{tabular}
\end{table}

\section{More results on subject-heterogeneity analysis}\label{Section:S8}

We present results of subject-heterogeneity analysis for FP1F7, C4P4, T6O2, T5O1 and P4O2. The results are displayed in Figure \ref{Heterogeneity:controlb}.

\begin{figure}[htbp]
  \centering
  \begin{minipage}{0.4\textwidth}
    \centering
    \includegraphics[width=\linewidth, height=3cm]{figures/dendrogram_autocorr/H_FP1F7_dendrogram_all.png}
  \end{minipage}
  \hspace{0.01\textwidth}
  \begin{minipage}{0.4\textwidth}
    \centering
    \includegraphics[width=\linewidth, height=3cm]{figures/dendrogram_autocorr/P_FP1F7_dendrogram_all.png}
  \end{minipage}
  \begin{minipage}{0.4\textwidth}
    \centering
    \includegraphics[width=\linewidth, height=3cm]{figures/dendrogram_autocorr/H_C4P4_dendrogram_all.png}
  \end{minipage}
  \hspace{0.01\textwidth}
  \begin{minipage}{0.4\textwidth}
    \centering
    \includegraphics[width=\linewidth, height=3cm]{figures/dendrogram_autocorr/P_C4P4_dendrogram_all.png}
  \end{minipage}
  \begin{minipage}{0.4\textwidth}
    \centering
    \includegraphics[width=\linewidth, height=3cm]{figures/dendrogram_autocorr/H_T6O2_dendrogram_all.png}
  \end{minipage}
  \hspace{0.01\textwidth}
  \begin{minipage}{0.4\textwidth}
    \centering
    \includegraphics[width=\linewidth, height=3cm]{figures/dendrogram_autocorr/P_T6O2_dendrogram_all.png}
  \end{minipage}
  \begin{minipage}{0.4\textwidth}
    \centering
    \includegraphics[width=\linewidth, height=3cm]{figures/dendrogram_autocorr/H_T5O1_dendrogram_all.png}
  \end{minipage}
  \hspace{0.01\textwidth}
  \begin{minipage}{0.4\textwidth}
    \centering
    \includegraphics[width=\linewidth, height=3cm]{figures/dendrogram_autocorr/P_T5O1_dendrogram_all.png}
  \end{minipage}
  \begin{minipage}{0.4\textwidth}
    \centering
    \includegraphics[width=\linewidth, height=3cm]{figures/dendrogram_autocorr/H_P4O2_dendrogram_all.png}
    \caption*{Controls}
  \end{minipage}
  \hspace{0.01\textwidth}
  \begin{minipage}{0.4\textwidth}
    \centering
    \includegraphics[width=\linewidth, height=3cm]{figures/dendrogram_autocorr/P_P4O2_dendrogram_all.png}
    \caption*{Cases}
  \end{minipage}
  \caption{Dendrograms for the channels  FP1F7, C4P4, T6O2, T5O1 and P4O2. The left panel is for the control group (the left panel) while the right panel is for the case group.}
  \label{Heterogeneity:controlb}
\end{figure}

\section{Bipolar EEG source localisation}\label{Section:S9}

{\it EEG Data Acquisition and Preprocessing.}
Electroencephalography (EEG) recordings were obtained for 11 subjects in European Data Format (EDF). Signals were recorded using a clinical bipolar montage comprising 18 derivations (Fp2–F4, F4–C4, C4–P4, P4–O2, Fp1–F3, F3–C3, C3–P3, P3–O1, Fp2–F8, F8–T4, T4–T6, T6–O2, Fp1–F7, F7–T3, T3–T5, T5–O1, Fz–Cz, Cz–Pz). A non-EEG channel BP3–REF was excluded from further analysis.
EEG signals were band-pass filtered between 1–40 Hz and re-referenced implicitly through the bipolar montage. No additional re-referencing was applied to preserve the original spatial derivations. Independent component analysis was applied and six components were chosen. Artefacts due to eye-movements and blink were removed. Segmentation information provided in accompanying metadata identified seizure (ictal) intervals via start and end times; all remaining intervals were treated as non-seizure (interictal) periods.

{\it Segmentation of EEG Data.}
For each subject, seizure segments were extracted based on annotated start–end times. Non-seizure segments were defined as the complementary intervals of the continuous recording. Segments shorter than 5 seconds were excluded to ensure reliable estimation of temporal statistics. To mitigate boundary artifacts, 0.5 seconds were trimmed from the beginning and end of each segment.

{\it Forward Model and Source Space.}
EEG source localisation was performed using a template-based approach implemented in MNE-Python. A standard anatomical template (fsaverage) was used in the absence of subject-specific MRI data. The cortical source space was defined using a surface-based discretization ($~$5,000–10,000 vertices per hemisphere). A three-layer boundary element model (BEM) was constructed to model volume conduction.
The standard forward model is defined as:
\[
B = LQ + E
\]
where $ B$ denotes monopolar EEG measurements, $ L$ the leadfield matrix, $ Q $ the cortical source activity, and $ E $ measurement noise.

{\it Bipolar Montage Transformation.}
Because EEG data were recorded in a bipolar montage, the measured signals $A$ were related to the underlying monopolar potentials $ B$ through a linear transformation:
\[
A = M B
\]
where $ M$  is a binary-valued link matrix encoding electrode differences (entries in $\{-1, 0, 1\}$). Substituting the forward model yields:
\[
A = (M L) Q + M E
\]
Thus, source localisation was performed by replacing the standard leadfield $ L$ with the transformed leadfield $ M L$, ensuring consistency between the measurement model and the recorded data. The noise covariance matrix was transformed accordingly. The marginal covariances of $A$, $B$, $Q$ and $E$ satisfy:
\[
\cov(A) = M \cov(B) M^T=ML\cov(Q) L^TM^T+M\cov(E)M^T.
\]

{\it Source Reconstruction.}
Distributed source reconstruction was performed using minimum-norm estimation (MNE) with depth weighting and noise normalization (dSPM). The inverse operator was computed using the transformed leadfield $M L$ and corresponding noise covariance. Source time series were estimated for each segment independently. Let $M=UDV^T$ denote the singular value decomposition and $M^+=VD^+U^T$ the pseudoinverse (i.e., the Moore–Penrose inverse) of $M.$ Let $\Sigma$ denote $M\cov(E)M^T $. The minimum norm estimate is defined by
\begin{eqnarray*}
\hat{Q}&=&\mbox{arg}\min_Q(A-MLQ)^T(A-MLQ)+\lambda \Sigma^{1/2}QQ^T\lambda \Sigma^{1/2} \\
            &=& ((L^TM^T\Sigma^+ML+\lambda I)^{-1}L^TM^T\Sigma^+A.
\end{eqnarray*}

{\it Cortical Parcellation.}
Reconstructed source activity was mapped onto 68 cortical regions defined by the Desikan–Killiany atlas. For each region, vertex-wise time series were aggregated using sign-flipped means to average the estimated source time series over the grid points within each region, yielding one representative time series per region.

{\it Accuracy of source localisation.} Although source reconstructions are not recommended for 18-channel data in the literature, the focus of our study is to understand the neuronal dynamics that give rise to the EEG signals. Our source localisation analysis is a sanity check to confirm whether we can indeed explain sufficient variance with the selected modular networks.

\newpage
\section{Region of Interest (ROI):  Desikan-Killiany atlas}\label{Section:S10}


\begin{longtable}{
    >{\raggedright\arraybackslash}p{0.10\textwidth}
    >{\raggedright\arraybackslash}p{0.60\textwidth}
    >{\raggedright\arraybackslash}p{0.20\textwidth}
}

\caption{Cortical region indices, region names, and abbreviations.}
\label{tab:cortical_regions} \\

\toprule
\textbf{Index} &
\textbf{Region Name} &
\textbf{Abbreviation} \\
\midrule
\endfirsthead

\toprule
\textbf{Index} &
\textbf{Region Name} &
\textbf{Abbreviation} \\
\midrule
\endhead

\midrule
\multicolumn{3}{r}{Continued on next page} \\
\endfoot

\bottomrule
\endlastfoot

1  & ctx-lh-bankssts                 & lh-BSTS \\
2  & ctx-lh-caudalanteriorcingulate  & lh-cACC \\
3  & ctx-lh-caudalmiddlefrontal      & lh-CMF \\
4  & ctx-lh-cuneus                   & lh-CUN \\
5  & ctx-lh-entorhinal               & lh-Ent \\
6  & ctx-lh-frontalpole              & lh-FP \\
7  & ctx-lh-fusiform                 & lh-Fus \\
8  & ctx-lh-inferiorparietal         & lh-IP \\
9  & ctx-lh-inferiortemporal         & lh-IT \\
10 & ctx-lh-insula                   & lh-Ins \\
11 & ctx-lh-isthmuscingulate         & lh-ICC \\
12 & ctx-lh-lateraloccipital         & lh-LO \\
13 & ctx-lh-lateralorbitofrontal     & lh-LOF \\
14 & ctx-lh-lingual                  & lh-Ling \\
15 & ctx-lh-medialorbitofrontal      & lh-MOF \\
16 & ctx-lh-middletemporal            & lh-MT \\
17 & ctx-lh-paracentral              & lh-ParC \\
18 & ctx-lh-parahippocampal           & lh-PH \\
19 & ctx-lh-parsopercularis           & lh-POp \\
20 & ctx-lh-parsorbitalis             & lh-POr \\
21 & ctx-lh-parstriangularis          & lh-PTr \\
22 & ctx-lh-pericalcarine             & lh-Cal \\
23 & ctx-lh-postcentral               & lh-PoC \\
24 & ctx-lh-posteriorcingulate        & lh-PCC \\
25 & ctx-lh-precentral                & lh-PC \\
26 & ctx-lh-precuneus                 & lh-PreCu \\
27 & ctx-lh-rostralanteriorcingulate  & lh-rACC \\
28 & ctx-lh-rostralmiddlefrontal      & lh-RMF \\
29 & ctx-lh-superiorfrontal           & lh-SF \\
30 & ctx-lh-superiorparietal          & lh-SP \\
31 & ctx-lh-superiortemporal          & lh-ST \\
32 & ctx-lh-supramarginal             & lh-SM \\
33 & ctx-lh-temporalpole              & lh-TP \\
34 & ctx-lh-transversetemporal        & lh-TT \\

35 & ctx-rh-bankssts                 & rh-BSTS \\
36 & ctx-rh-caudalanteriorcingulate  & rh-cACC \\
37 & ctx-rh-caudalmiddlefrontal      & rh-CMF \\
38 & ctx-rh-cuneus                   & rh-CUN \\
39 & ctx-rh-entorhinal               & rh-Ent \\
40 & ctx-rh-frontalpole              & rh-FP \\
41 & ctx-rh-fusiform                 & rh-Fus \\
42 & ctx-rh-inferiorparietal         & rh-IP \\
43 & ctx-rh-inferiortemporal         & rh-IT \\
44 & ctx-rh-insula                   & rh-Ins \\
45 & ctx-rh-isthmuscingulate         & rh-ICC \\
46 & ctx-rh-lateraloccipital         & rh-LO \\
47 & ctx-rh-lateralorbitofrontal     & rh-LOF \\
48 & ctx-rh-lingual                  & rh-Ling \\
49 & ctx-rh-medialorbitofrontal      & rh-MOF \\
50 & ctx-rh-middletemporal            & rh-MT \\
51 & ctx-rh-paracentral              & rh-ParC \\
52 & ctx-rh-parahippocampal           & rh-PH \\
53 & ctx-rh-parsopercularis           & rh-POp \\
54 & ctx-rh-parsorbitalis             & rh-POr \\
55 & ctx-rh-parstriangularis          & rh-PTr \\
56 & ctx-rh-pericalcarine             & rh-PCal \\
57 & ctx-rh-postcentral               & rh-PoC \\
58 & ctx-rh-posteriorcingulate        & rh-PCC \\
59 & ctx-rh-precentral                & rh-PC \\
60 & ctx-rh-precuneus                 & rh-PreC \\
61 & ctx-rh-rostralanteriorcingulate  & rh-rACC \\
62 & ctx-rh-rostralmiddlefrontal      & rh-RMF \\
63 & ctx-rh-superiorfrontal           & rh-SF \\
64 & ctx-rh-superiorparietal          & rh-SP \\
65 & ctx-rh-superiortemporal          & rh-ST \\
66 & ctx-rh-supramarginal             & rh-SM \\
67 & ctx-rh-temporalpole              & rh-TP \\
68 & ctx-rh-transversetemporal        & rh-TT \\

\end{longtable}

\newpage
\section{Bifurcation behavior of ccDCMs}\label{Section:S12}

As ccDCMs play a key role in shaping and influencing the structure and behavior of the NccDCM, in the following we exam the behavior of ccDCM and evaluate the accuracy of the proposed parameter estimation. For this purpose, suppressing channel and group subscripts in the model (\ref{sc}),  consider the following ccDCM with states
$\x(t)=(x_0(t),x_1(t),x_2(t),x_3(t),x_4(t),x_5(t)) $:
\begin{eqnarray}\label{first}
  d{x}_0(t) &=& x_3(t)dt, \quad
  d{x}_1(t) = x_4(t)dt, \quad
  d{x}_2(t) = x_5(t)dt, \nonumber\\
  d{x}_3(t) &=& \left [ Aa \mathrm{S}\left(x_1(t)-x_2(t)\right)-2ax_3(t)-a^2x_0(t) \right ]dt , \\
  d{x}_4(t) &=&\left [  Aa\left\{0.8C\mathrm{S}\left(Cx_0(t)\right)+K\mathrm{S}\left(y_2(t)\right)\right\}-2ax_4(t)-a^2x_1(t) \right ]dt +Aa dw(t), \nonumber\\
  d{x}_5(t) &=&\left [  Bb0.25C\mathrm{S}\left(0.25Cx_0(t)\right)-2bx_5(t)-b^2x_2(t) \right ]dt\nonumber,
\end{eqnarray}
with initial value $\boldsymbol x_0=(x_0(0),...,x_5(0))^{T}\in\mathbb{R}^{6}$ and the input $y_2(t)$ from the external channel. We use the following Strang splitting method to generate the samples of $\x(t)$ from the above model. See \citep{Milstein2003, Leimkuhler2016} for more details.
To this end, we first divide the system (\ref{first}) into two subsystems below:
\begin{eqnarray}\label{sc:1}
  dx_0^{[1]}(t) &=& x_3^{[1]}(t)dt, \quad
  dx_1^{[1]}(t) = x_4^{[1]}(t)dt, \quad
  dx_2^{[1]}(t) = x_5^{[1]}(t)dt, \nonumber\\
  dx_3^{[1]}(t) &=& [-2ax_3^{[1]}(t)-a^2x_0^{[1]}(t)]dt, \\
  dx_4^{[1]}(t) &=& \left[-2ax_4^{[1]}(t)-a^2x_1^{[1]}(t)\right]dt, \nonumber\\
  dx_5^{[1]}(t) &=& [-2bx_5^{[1]}(t)-b^2x_2^{[1]}(t)]dt\nonumber,
\end{eqnarray}
and
\begin{eqnarray}\label{sc:2}
  dx_0^{[2]}(t) &=& 0, \quad
  dx_1^{[2]}(t) = 0, \quad
  dx_2^{[2]}(t) = 0, \nonumber\\
  dx_3^{[2]}(t) &=& Aa \mathrm{S}\left(x_1^{[2]}(t)-x_2^{[2]}(t)\right)dt, \\
  dx_4^{[2]}(t) &=& Aa\left[0.8C\mathrm{S}\left(Cx_0^{[2]}(t)\right)+K\mathrm{S}\left(y_2(t)\right)\right]dt+Aa dw(t), \nonumber\\
  dx_5^{[2]}(t) &=& Bb0.25C\mathrm{S}\left(0.25Cx_0^{[2]}(t)\right) dt\nonumber.
\end{eqnarray}
Let $\varphi_{t}^{[1]}$ and $\varphi_{t}^{[2]}$ denote the exact flows of subsystems (\ref{sc:1}) and (\ref{sc:2}). Then, for $\x\in\mathbb{R}^{6}$,  the so-called Strang splitting for system (\ref{first}) can be written as
\begin{align}\label{strang}
  \psi_{\Delta t}(\x)=(\varphi_{\Delta t/2}^{[1]}\circ\varphi_{\Delta t}^{[2]}\circ \varphi_{\Delta t/2}^{[1]})(\x).
\end{align}
 We independently generate the initial values for the states from the standard normal distribution.  We choose the input signal $y_2(t)$  from the channel F4C4  of the healthy subject H9 in the above EEG dataset which contains the EEG measurements  for $18$ pediatric subjects.
Using the discretisation step size of $h^{\mathrm{sim}}=1/256$, we obtain the approximate solution paths for subsystems (\ref{sc:1}) and (\ref{sc:2}) over time-length $T = 20$ seconds.
We then calculate the output signal $y_1(t)=x_1(t)-x_2(t).$

Under the above simulation setting, we first demonstrate that the ccDCM does have bifurcation behavior when $A/B$, $C$, $K$, and $v_{0}$ take values in some ranges. In the analysis, we change the value of one parameter, fixed other parameters to their typical values as specified in Table~\ref{paramer}.
 Figure~ \ref{Fig:output} respectively display the input time series from the channel F4C4 of the healthy subject H9 and the corresponding output time series under varying values of $A/B$, $C$, $K$, and $\nu_{0}$.

As shown in Figure~\ref{Fig:output}(a), the output $y_1(t)$ remains stable when $A/B = 2/25$ and $3.25/22$. However, when $A/B$ increases to $4/14$, the output $y_1(t)$ exhibits unstable oscillatory behavior after approximately 15\,seconds. Note that, as depicted in Figure~\ref{Fig:output}(a), the input signal reaches its maximum around this time. In the remaining panels of Figure~\ref{Fig:output}, no similar phenomenon has been observed for the other parameters. Furthermore, the  higher values of   $B_A=A/B$, $K_A=AK$, $1/C_A=1/(AC)$ and $\nu_0$, the  higher output signals will result.

\begin{figure}[htbp]
    \centering
    \begin{minipage}[b]{0.45\textwidth}
   \caption*{\tiny (a) $B_A=A/B$}
        \centering
        \includegraphics[height = 4cm,width=\linewidth]{figures/Bifurcation_AB.png}
    \end{minipage}
\hfill
    \begin{minipage}[b]{0.45\textwidth}
 \caption*{\tiny (b) $C$}
        \centering
        \includegraphics[height = 4cm,width=\linewidth]{figures/Bifurcation_C.png}
    \end{minipage}
    \vspace{0.4cm}
\\
    \begin{minipage}[b]{0.45\textwidth}
   \caption*{\tiny (c) K}
        \centering
        \includegraphics[height = 4cm,width=\linewidth]{figures/Bifurcation_K.png}
    \end{minipage}
\hfill
    \begin{minipage}[b]{0.45\textwidth}
   \caption*{\tiny (d) $v_0$ }
        \centering
        \includegraphics[height = 4cm,width=\linewidth]{figures/Bifurcation_v0.png}
    \end{minipage}
    \caption{Given the input time series in Figure~Output time-series for varying values of $(A,B)\in\{(2,15), (3.25,22),(4, 14)\}$, $C\in \{68, 135, 675\}$, $K\in\{10, 50, 100 \}$ and $v_{0}\in\{2 ,6, 12\}.$
Output time-series exhibit bifurcation behavior only when $A/B$ increases to 0.2857. The magnitude of output signal increases quickly in $A/B$, $v_0$ and $1/C$ but slowly in $K$.}
    \label{Fig:output}
\end{figure}

\section{Unidirectional circle network of three nodes}\label{Section:S13}
This section contains the mathematical expression of multi-coupled Jansen-Rit model, the simulation results for Chen-Fliess, RK and EM methods based on different initial conditions, and the the analysis of the bidirectional coupling configuration.

First, we give the mathematical expression of multi-coupled Jansen-Rit model:
\begin{eqnarray}\label{three1}
  dx^{(i)}_{A0}(t) &=& x^{(i)}_{A3}(t)\,dt, ~~~ dx^{(i)}_{1}(t) = x^{(i)}_{4}(t)\,dt, ~~~ dx^{(i)}_{2}(t) = x^{(i)}_{5}(t)\,dt, \nonumber\\
  dx^{(i)}_{A3}(t) &=& \left[a\, \mathrm{S}^{(i)}\left(y^{(i)}(t)\right) - 2a\,x^{(i)}_{A3}(t) - a^2\,x^{(i)}_{A0}(t)\right]dt,\nonumber \\
  dx^{(i)}_{4}(t) &=&  \left\{a\left[\sum\limits_{i\neq j}K^{(i|j)}_A\,\mathrm{S}^{(i)}\left(y^{(j)}(t)\right) + 0.8C^{(i)}_{A}\,\mathrm{S}^{(i)}\left(C^{(i)}_{A}\,x^{(i)}_{A0}(t)\right)\right]\right.\\
  &&\left. - 2a\,x^{(i)}_{4}(t) - a^2\,x^{(i)}_{1}(t)\right\}dt +a\sigma^{(i|j)}_{A} d\tilde{\textbf{w}}^{(i|j)}(t), \nonumber\\
  dx^{(i)}_{5}(t) &=& \left[(1/B^{(i)}_A)b0.25\,C^{(i|j)}_{A}\,\mathrm{S}^{(i)}\left(0.25C^{(i|j)}_{A}\,x^{(i)}_{A0}(t)\right) - 2b\,x^{(i)}_{5}(t) - b^2\,x^{(i)}_{2}(t)\right]dt, \nonumber\\
y^{(i)}(t)&=&x^{(i)}_1(t)-x^{(i)}_2(t).\nonumber
\end{eqnarray}
with
\begin{eqnarray*}
  \mathrm{S}_i(x(t)) = \frac{\nu_{\max}}{1 + \exp\!\left(r(\nu^{(i)}_{0} - x(t))\right)},
  \quad t \in [0,T], \quad i = 1,2,3.
\end{eqnarray*}
For the unidirectional configuration shown in the Figure \ref{fig:network_configurations}, all connections other than \(K^{(2\mid1)}\), \(K^{(3\mid2)}\), and \(K^{(1\mid3)}\) were set to zero.

The simulation results for Chen-Fliess, RK and EM methods based on different initial conditions can be found in Table \ref{tab:ARE_methods_initial_conditions}

\begin{sidewaystable}[p]
\centering

\caption{Median ARE values for CF, EM,
and RK based methods under different initial conditions.
NMADs are reported in parentheses.}
\label{tab:ARE_methods_initial_conditions}

\begingroup
\renewcommand{\arraystretch}{1.25}
\setlength{\tabcolsep}{4pt}

\resizebox{\linewidth}{!}{%
\begin{tabular}{@{}cc*{9}{c}@{}}
\toprule
\multirow{2}{*}{\textbf{Parameter}}
& \multirow{2}{*}{\textbf{Node}}
& \multicolumn{3}{c}{\textbf{CF-based}}
& \multicolumn{3}{c}{\textbf{EM-based}}
& \multicolumn{3}{c}{\textbf{RK-based}} \\
\cmidrule(lr){3-5}
\cmidrule(lr){6-8}
\cmidrule(l){9-11}
& & Zero & Constant & Normal
  & Zero & Constant & Normal
  & Zero & Constant & Normal \\
\midrule

\multirow{3}{*}{$B_A$}
& 1
& 0.05 (0.04) & 0.05 (0.04) & 0.04 (0.03)
& 0.40 (0.18) & 0.44 (0.19) & 0.46 (0.22)
& 0.55 (0.17) & 0.56 (0.15) & 0.52 (0.20) \\
& 2
& 0.02 (0.02) & 0.02 (0.02) & 0.02 (0.02)
& 0.80 (0.85) & 0.75 (0.38) & 0.76 (0.76)
& 0.77 (0.90) & 0.71 (0.56) & 0.79 (0.84) \\
& 3
& 0.03 (0.02) & 0.02 (0.02) & 0.03 (0.03)
& 0.18 (0.19) & 0.16 (0.14) & 0.14 (0.14)
& 0.16 (0.17) & 0.20 (0.24) & 0.13 (0.13) \\
\addlinespace[5pt]

\multirow{3}{*}{$C_A$}
& 1
& 0.03 (0.02) & 0.03 (0.02) & 0.03 (0.03)
& 0.39 (0.09) & 0.43 (0.10) & 0.42 (0.11)
& 0.42 (0.13) & 0.45 (0.13) & 0.42 (0.12) \\
& 2
& 0.04 (0.03) & 0.04 (0.03) & 0.04 (0.03)
& 0.97 (0.04) & 0.96 (0.05) & 0.97 (0.04)
& 0.96 (0.05) & 0.96 (0.05) & 0.96 (0.05) \\
& 3
& 0.03 (0.03) & 0.02 (0.02) & 0.02 (0.02)
& 0.34 (0.08) & 0.33 (0.07) & 0.32 (0.09)
& 0.34 (0.16) & 0.33 (0.17) & 0.31 (0.14) \\
\addlinespace[5pt]

\multirow{3}{*}{$\nu_0$}
& 1
& 0.03 (0.02) & 0.02 (0.02) & 0.02 (0.02)
& 0.09 (0.06) & 0.09 (0.07) & 0.11 (0.08)
& 0.08 (0.07) & 0.08 (0.07) & 0.08 (0.07) \\
& 2
& 0.02 (0.02) & 0.02 (0.01) & 0.02 (0.02)
& 0.20 (0.19) & 0.25 (0.23) & 0.17 (0.17)
& 0.26 (0.20) & 0.33 (0.23) & 0.23 (0.25) \\
& 3
& 0.02 (0.02) & 0.01 (0.01) & 0.02 (0.02)
& 0.20 (0.10) & 0.21 (0.07) & 0.22 (0.09)
& 0.19 (0.09) & 0.20 (0.08) & 0.18 (0.09) \\
\addlinespace[5pt]

\multirow{3}{*}{$K_A$}
& 1
& 0.09 (0.07) & 0.07 (0.06) & 0.07 (0.04)
& 0.34 (0.18) & 0.38 (0.18) & 0.39 (0.21)
& 0.55 (0.21) & 0.58 (0.25) & 0.54 (0.25) \\
& 2
& 0.11 (0.12) & 0.13 (0.12) & 0.13 (0.09)
& 0.32 (0.06) & 0.32 (0.05) & 0.31 (0.07)
& 0.31 (0.06) & 0.31 (0.06) & 0.31 (0.07) \\
& 3
& 0.05 (0.04) & 0.05 (0.04) & 0.05 (0.04)
& 0.10 (0.07) & 0.09 (0.07) & 0.10 (0.07)
& 0.11 (0.12) & 0.11 (0.11) & 0.10 (0.09) \\
\addlinespace[5pt]

\multirow{3}{*}{$\sigma_A$}
& 1
& 0.02 (0.02) & 0.03 (0.02) & 0.02 (0.02)
& 0.84 (0.01) & 0.84 (0.01) & 0.84 (0.01)
& 0.83 (0.02) & 0.84 (0.01) & 0.84 (0.02) \\
& 2
& 0.10 (0.04) & 0.11 (0.04) & 0.11 (0.05)
& 0.86 (0.02) & 0.86 (0.02) & 0.86 (0.02)
& 0.85 (0.03) & 0.84 (0.05) & 0.85 (0.03) \\
& 3
& 0.03 (0.03) & 0.03 (0.03) & 0.04 (0.02)
& 0.82 (0.01) & 0.82 (0.01) & 0.82 (0.02)
& 0.82 (0.02) & 0.82 (0.03) & 0.82 (0.02) \\
\bottomrule
\end{tabular}%
}
\endgroup

\par\smallskip
\begin{minipage}{\linewidth}
\small
\textit{Note.}
Zero denotes all-zero initial conditions;
Constant denotes the fixed initial condition
$[0,\,0.5,\,1,\,1.5,\,2,\,2.5]$;
Normal denotes initial conditions drawn from
$\mathcal{N}(0,1)$.
\end{minipage}

\end{sidewaystable}

\section{Bidirectionally coupled system with multiple inputs }

In this section, we consider a bidirectionally coupled configuration in which a single Jansen--Rit node receives inputs from multiple nodes. The mathematical formulation of this configuration is identical to that in Equation~\eqref{three1}, and its schematic illustration is provided in Figure \ref{fig:network_configurations} in the main text.

The loss function for the bidirectionally coupled system differs from that for the single-input system because it incorporates multiple inputs. For the output signal $y(t)=x_1(t)-x_2(t)$, we construct the likelihood using the same approach as in the single-input case. For subject $n$ in group $k$, the resulting negative log-likelihood at node $i$ during period $q$, with inputs from nodes $j=1,\ldots,J$, is given by
\begin{eqnarray*}
L^{(i|j)k}_{nq}( \bthe^{(i|j)k}_{nq})=\frac{D}{2}\log 2\pi+\frac 12
\sum\limits_{d=1}^{D}\log G^{(i|j)k}_{nqd}
+\frac{1}{2}\sum\limits_{d=1}^D\frac{\left(y^{(i)k}_{nq}(t_{b_q-1}^d)-\hat{y}^{(i)k}_{nq}(t_{b_q-1}^d)\right)^2}{G^{(i|j)k}_{nqd}},
\end{eqnarray*}
where $G^{(i|j)k}_{nqd} = (\sigma^{(i)k}_{Anq}a)^2(t^{d}_{b_q-1}-t^d_0)^3/6$ and
\begin{align*}
\hat{y}^{(i)k}_{nq}(t^d_{b_q-1})\approx g\left(t^{d}_{b_q-1}|x^{(i)k}_{-\{A3 \}nq}(t^d_0), y^{(i)k}_{nq}(t^d_0), \dot{y}^{(i)k}_{nq}(t^d_0), \{y^{(j)k}_{nq}(t^d_0)\}_{j\neq i}^{J},\bthe^{(i)k}_{nq}  \right)
\end{align*}
with
\begin{eqnarray*}
&& g(t|x^{(i)}_{-\{3\}}(t_0),y^{(i)}(t_0),\dot{y}^{(i)}(t_0),\{y^{(j)}(t_0)\}_{j\neq i}^{J},\bthe)\\
 & &\quad= y^{(i)}(t_0)+\dot{y}^{(i)}(t_0)(t-t_0)\non\\
  &&\qquad+\left(\frac{\nu_{\max}a0.8C_{A}}{1+e^{r(v_0-C_{A}x^{(i)}_{A0}(t_0))}}
 +\sum\limits_{i\neq j}^J\frac{\nu_{\max}aK_A^{(i|j)}}{1+e^{r(v_0-y^{(j)}(t_0))}}-a^2x_1^{(i)}(t_0)-2ax_4^{(i)}(t_0)\right)\non\\
&&\qquad\times\frac{(t-t_0)^2}{2}\non\\
 &&\qquad-\left(\frac{\nu_{\max}B^{-1}_Ab0.25C_{A}}{1+e^{r(v_0-0.25C_{A}x^{(i)}_{A0}(t_0))}}-2bx_5^{(i)}(t_0)-b^2x_2^{(i)}(t_0)\right)\frac{(t-t_0)^2}{2}.
\end{eqnarray*}
and $\bthe = (B_A,C_A,v_0,\sigma_A, K_A^{(i|j\neq i)})$.
Furthermore, the hidden states $x_1$, $x_2$, and $x_3$ are recovered using the same procedure as that employed in the single-input case.

To investigate both the influence of multiple inputs on the proposed estimation method originally developed for the single-input case and the performance of the proposed estimation method designed for the multiple-input setting, we conducted a simulation study.

In this study, a four-node network is considered, and the inter-channel coupling structure is defined as follows: node~1 receives input from $y_4(t)$, node~2 receives input from $y_4(t)$, node~3 receives input from $y_4(t)$, and node~4 receives inputs from $y_1(t)$, $y_2(t)$, and $y_3(t)$. Consequently, node~4 acts as an integration node that aggregates information from the other three nodes, thereby forming a more complex coupling structure than that in the single-input case.

Such a configuration enables us to evaluate the robustness and adaptability of the proposed estimation framework under multi-source interactions and to assess its capability in characterising complex neural connectivity patterns commonly observed in large-scale brain networks.
%
%
%
%
%

In the simulation study, we set the parameters as follows:
for node~1, $A_1=3.25$, $B_1=22$, $C_1=135$, $v_{0,1}=6$, $K^{(1|2)}=0$, $K^{(1|3)}=0$, $K^{(1|4)}=100$, and $\sigma_1=10$;
for node~2, $A_2=4$, $B_2=14$, $C_2=135$, $v_{0,2}=4$, $K^{(2|1)}=0$, $K^{(2|3)}=0$, $K^{(2|4)}=50$, and $\sigma_2=10$;
for node~3, $A_3=3.25$, $B_3=18$, $C_3=135$, $v_{0,3}=5$, $K^{(3|1)}=0$, $K^{(3|2)}=0$, $K^{(3|4)}=50$, and $\sigma_3=10$;
and for node~4, $A_4=5$, $B_4=22$, $C_4=135$, $v_{0,4}=6$, $K^{(4|1)}=50$, $K^{(4|2)}=100$, $K^{(4|3)}=50$, and $\sigma_4=10$. We repeated the simulation 200 times.
The simulation results  were summarised in Tables \ref{tab:4channel_parameter} and \ref{tab:4channel_cross} presented the ARE rates of the multiple-inputs adjusted method and the original single input method for estimating these unknown parameters. The results demonstrated that the multiple-inputs adjusted method outperformed the original single-input method.

Note that  when the true $K_A^{(j|\cdot)}=0$, the denominator of ARE would be zero. So, the ARE could not be useful for evaluating the error rate of estimating $K_A^{(j|\cdot)}$. Instead,  for the multiple-input method, we used the mean absolute errors for estimating $K_A(j|\cdot)$ at nodes $j=1, 2,$ and $3,$ with NMAD values in parentheses. We obtained  the mean errors of $4.25$ $(6.3)$ for $K_A^{(1|2)}$ and $1.95$ $(2.89)$ for $K_A^{(1|3)}$;  $7.53$ $(11.12)$ for $K_A^{(2|1)}$ and $0$ $(0)$ for $K_A^{(2|3)}$; $25.55$ $(8.89)$ for $K_A^{(3|1)}$ and $2.68$ $(3.97)$ for $K_A^{(3|2)}$. Similarly, we estimated the errors for the single input method. The further details were omitted but they also demonstrated that the multiple-inputs adjusted method outperformed the original single-input method.

\begin{table}[htbp]
\centering
\caption{
The median (upper row)  and the NMAD (lower row) of ARE for the proposed multiple-input estimation method. 
}
\label{tab:4channel_parameter}
\small
\renewcommand{\arraystretch}{1.15}
\setlength{\tabcolsep}{10pt}

\resizebox{\textwidth}{!}{
\begin{tabular}{c c c c c c c c c}
\hline
\textbf{Node (j)} & $B_A^{(j)}$ & $C_A^{(j)}$ & $v_0^{(j)}$ & $\sigma_A^{(j)}$ & $K_A^{(j|1)}$ & $K_A^{(j|2)}$ & $K_A^{(j|3)}$ & $K_A^{(j|4)}$ \\
\hline

\multirow{2}{*}{1}
& 0.18 & 0.04 & 0.09 & 0.43 & & $\infty$ & $\infty$ & 0.07 \\
& 0.03 & 0.02 & 0.01 & 0.01 && $\infty$ & $\infty$ & 0.02 \\
\hline

\multirow{2}{*}{2}
& 0.06 & 0.04 & 0.11 & 0.45 & $\infty$ & & $\infty$ & 0.04 \\
& 0.01 & 0.01 & 0.01 & 0.01 & $\infty$ & & $\infty$ & 0.04 \\
\hline

\multirow{2}{*}{3}
& 0.15 & 0.04 & 0.08 & 0.45 & $\infty$ & $\infty$&  & 0.03 \\
& 0.03 & 0.02 & 0.01 & 0.01 & $\infty$ & $\infty$&  & 0.03 \\
\hline

\multirow{2}{*}{4}
& 0.18 & 0.04 & 0.08 & 0.43 & 0.11 & 0.13 & 0.08 & \\
& 0.03 & 0.02 & 0.01 & 0.01 & 0.07 & 0.02 & 0.07 &\\
\hline

\end{tabular}
}
\end{table}

\begin{table*}[htbp]
\centering
\caption{Median ARE values for the four-node system using the proposed
single-input estimation method. NMAD values are reported in parentheses.
Rows correspond to output nodes and columns to input nodes.}
\label{tab:4channel_cross}

\small
\renewcommand{\arraystretch}{1.15}
\setlength{\tabcolsep}{5pt}

\begin{tabular}{cccccc}
\toprule
\multirow{2}{*}{\textbf{Parameter}}
& \multirow{2}{*}{\textbf{Output node}}
& \multicolumn{4}{c}{\textbf{Input node}} \\
\cmidrule(lr){3-6}
& & \textbf{1} & \textbf{2} & \textbf{3} & \textbf{4} \\
\midrule

\multirow{4}{*}{$B_A$}
& 1 &  & 0.50 (0.09) & 1.17 (0.08) & 0.04 (0.03) \\
& 2 & 0.12 (0.02) &  & 0.20 (0.02) & 0.03 (0.02) \\
& 3 & 0.11 (0.04) & 0.21 (0.05) &  & 0.03 (0.02) \\
& 4 & 0.28 (0.03) & 0.35 (0.05) & 0.48 (0.05) &  \\
\midrule

\multirow{4}{*}{$C_A$}
& 1 &  & 0.22 (0.06) & 0.30 (0.07) & 0.03 (0.02) \\
& 2 & 0.11 (0.02) &  & 0.09 (0.03) & 0.03 (0.02) \\
& 3 & 0.07 (0.02) & 0.15 (0.04) &  & 0.02 (0.02) \\
& 4 & 0.19 (0) & 0.03 (0.03) & 0.19 (0) &  \\
\midrule

\multirow{4}{*}{$v_0$}
& 1 &  & 0.09 (0.04) & 0.22 (0.05) & 0.02 (0.02) \\
& 2 & 0.05 (0.02) &  & 0.06 (0.02) & 0.01 (0.01) \\
& 3 & 0.02 (0.02) & 0.07 (0.03) &  & 0.01 (0.01) \\
& 4 & 0.07 (0.02) & 0.05 (0.02) & 0.11 (0.03) &  \\
\midrule

\multirow{4}{*}{$K_A$}
& 1 &  & $\infty$ & $\infty$ & 0.07 (0.05) \\
& 2 & $\infty$ &  & $\infty$ & 0.05 (0.04) \\
& 3 & $\infty$ & $\infty$ &  & 0.03 (0.03) \\
& 4 & 1.40 (0) & 0.20 (0) & 1.19 (0.28) &  \\
\midrule

\multirow{4}{*}{$\sigma_A$}
& 1 &  & 0.57 (0.04) & 0.57 (0.06) & 0.19 (0.03) \\
& 2 & 0.05 (0.02) &  & 0.12 (0.03) & 0.08 (0.02) \\
& 3 & 0.09 (0.03) & 0.19 (0.03) &  & 0.04 (0.03) \\
& 4 & 0.46 (0.03) & 0.21 (0.03) & 0.70 (0.02) &  \\
\bottomrule
\end{tabular}
\end{table*}
\nolinenumbers
\newpage
\bibliographystyle{chicago}
\bibliography{library}